\documentclass[aps,prb,superscriptaddress,twocolumn,floatfix,10pt,nofootinbib]{revtex4-2}
\usepackage[utf8]{inputenc}
\usepackage{amsmath,amssymb}
\usepackage{empheq}
\usepackage{lipsum}
\usepackage{mathtools,cuted}
\usepackage{esvect}
\usepackage{multirow}
\usepackage{xcolor}
\usepackage{soul}
\usepackage[export]{adjustbox}
\usepackage{graphicx}
\graphicspath{{Figures/}}
\usepackage{dcolumn}
\usepackage{bm}
\usepackage{hyperref}
\usepackage{physics}
\usepackage[mathlines]{lineno}
\usepackage[capitalise]{cleveref}
\usepackage{multirow}
\usepackage{bbm}
\usepackage{orcidlink}
\usepackage{booktabs}
\usepackage[inkscapeexe=/Applications/Inkscape.app/Contents/MacOS/inkscape,inkscapearea=page]{svg}
\usepackage{nicematrix}
\usepackage{babel}
\usepackage{natbib}

\makeatletter
\def\bbl@set@language#1{%
	\edef\languagename{%
		\ifnum\escapechar=\expandafter`\string#1\@empty
		\else\string#1\@empty\fi}%
	\@ifundefined{babel@language@alias@\languagename}{}{%
		\edef\languagename{\@nameuse{babel@language@alias@\languagename}}%
	}%
	\select@language{\languagename}%
	\expandafter\ifx\csname date\languagename\endcsname\relax\else
	\if@filesw
	\protected@write\@auxout{}{\string\select@language{\languagename}}%
	\bbl@for\bbl@tempa\BabelContentsFiles{%
		\addtocontents{\bbl@tempa}{\xstring\select@language{\languagename}}}%
	\bbl@usehooks{write}{}%
	\fi
	\fi}
\newcommand{\DeclareLanguageAlias}[2]{%
	\global\@namedef{babel@language@alias@#1}{#2}%
}
\makeatother

\DeclareLanguageAlias{en}{english}
\DeclareLanguageAlias{English}{english}
\DeclareLanguageAlias{Englisch}{english}
\DeclareLanguageAlias{EN}{english}
\DeclareLanguageAlias{en-US}{english}

\newcommand{\kk}{\mathbf{k}}

\newcommand{\sbar}[0]{\overline{S}}

\begin{document}
	
	\title{Higher-order topology protected by latent crystalline symmetries}
	
 \author{L. Eek}
 \affiliation{Institute of Theoretical Physics, Utrecht University, Utrecht, 3584 CC, Netherlands}
 \author{M. Röntgen}
 \affiliation{Laboratoire d'Acoustique de l'Université du Mans, Unite Mixte de Recherche 6613, Centre National de la Recherche Scientifique, Avenue O. Messiaen, F-72085 Le Mans Cedex 9, France}
 \author{A. Moustaj}
 \affiliation{Institute of Theoretical Physics, Utrecht University, Utrecht, 3584 CC, Netherlands}
  \author{C. Morais Smith}
 \affiliation{Institute of Theoretical Physics, Utrecht University, Utrecht, 3584 CC, Netherlands}

	\date{\today}
	
	\begin{abstract}
    We demonstrate that rotation symmetry is not a necessary requirement for the existence of fractional corner charges in $C_n$-symmetric higher-order topological crystalline insulators. Instead, it is sufficient to have a latent rotation symmetry, which may be revealed upon performing an isospectral reduction on the system. We introduce the concept of a filling anomaly for latent crystalline symmetric systems, and propose modified topological invariants. The notion of higher-order topology in two dimensions protected by $C_n$ symmetry is thus generalized to a protection by latent symmetry. Our claims are corroborated by concrete examples of models that show non-trivial corner charge in the absence of $C_n$-symmetry. This work extends the classification of topological crystalline insulators to include latent symmetries.
	\end{abstract}
	
	\maketitle{}
	
\section{introduction}
Topological phases of matter have been extensively studied and form a cornerstone of condensed matter physics. These phases were originally understood in the framework of the celebrated Altland-Zirnbauer classification for non-interacting systems\cite{altland_nonstandard_1997}. It considers the presence of time-reversal, particle-hole, and chiral symmetry to obtain 10 symmetry classes for topological insulators (TIs) and superconductors. It was also realized that there are various relationships between the classes in different dimensions, leading to what is called the ten-fold way \cite{ryu_topological_2010}. Later, there have been various extensions that lie outside the original Altland-Zirnbauer classification. Examples are Floquet topological materials \cite{roy_periodic_2017}, disordered materials/topological Anderson insulators \cite{li_topological_2009}, non-Hermitian systems \cite{kawabata_symmetry_2019}, and topological crystalline insulators (TCI's) \cite{fu_topological_2011}, to name a few. The latter are known to host quantized electronic boundary states protected by the geometric symmetries of the crystalline system \cite{fu_topological_2011,fang_new_2015,slager_space_2013,zhang_general_2023}.

Of particular interest are \textit{higher-order topological} (crystalline) \textit{insulators} (HOTIs). While `conventional' TIs in $D$ dimension host states with codimension one on their $D-1$ dimensional boundary, HOTIs host boundary states with a codimension $d$ on their $D-d$ dimensional boundary. One way of realizing HOTIs is by considering materials with quantized higher multipole moments in the bulk. Ref.~\cite{benalcazar_electric_2017} proposed a classification of materials with higher multipole moments as an extension of the modern theory of electric polarization \cite{vanderbilt_electric_1993,king-smith_theory_1993}. These systems are now known to be topological and characterized in terms of \textit{nested Wilson loops}. However, systems without a bulk multipole moment may still give rise to quantized edge or corner states. Such systems are in a so-called obstructed atomic limit, a phase with a weaker topological protection than in the conventional topological materials. This was further investigated in Refs.~\cite{MiertRot, benalcazar_quantization_2019}, where the presence of rotation symmetry leads to the definition of rotation invariants to characterize the band topology. Energy spectra with different rotation invariants cannot be deformed into each other without closing the bulk gap and are, therefore, topologically distinct. However, the protection and classification of these novel topological phases of matter require the presence of a geometric rotation symmetry. Additionally, the emergence of these phases is independent of time-reversal, particle-hole, and chiral symmetries, making them more vulnerable since rotational symmetries are more readily disrupted compared to spectral symmetries. 

Recently, a new type of symmetry, named \emph{latent symmetry}, has been proposed \cite{Smith2019PA514855HiddenSymmetriesRealTheoretical,Rontgen2021GraphtheoreticalAnalysisLocalLatent}.
A latent symmetry only becomes apparent at the reduced level upon performing a suitable dimensional reduction, the so-called \textit{isospectral reduction} (ISR)\cite{Bunimovich2014IsospectralTransformationsNewApproach}.
In particular, seemingly asymmetric Hamiltonians may feature latent geometric symmetries which have a strong impact on the system's eigenstates and eigenenergies \cite{Kempton2020LAIA594226CharacterizingCospectralVerticesIsospectral,Rontgen2021GraphtheoreticalAnalysisLocalLatent}.
This connection has been used, for instance, in the design of flat-band lattices \cite{morfonios_flat_2021}, and to explain ``accidental degeneracies'' in band structures \cite{Rontgen2021PRL126180601LatentSymmetryInducedDegeneracies}.
Very recently, latent symmetries were also used to construct latent versions of the Su-Schrieffer-Heeger (SSH) model \cite{Su1979PRL421698SolitonsPolyacetylene,PhysRevB.110.035106} and of the non-Hermitian SSH model \cite{eek_emergent_2024}.

In this work, we explore the implications of the existence of latent symmetries on the classification of topological phases.
Our main result is that the presence of a geometric rotation symmetry is not a necessary constraint for obtaining the topological phases introduced in Ref.~\cite{benalcazar_quantization_2019}. Indeed, the requirement of a rotation symmetry can be relaxed to the less stringent condition of a \textit{latent rotation symmetry}.

The outline of this paper is the following. In \cref{sec:anomaly}, we introduce the concept of filling anomaly in the context of the SSH model. We then relate this concept to the latent SSH model, proposed in Ref.~\cite{PhysRevB.110.035106}, discuss the ISR in more detail, and formalise the concept of latent crystalline symmetries. In \cref{sec:class}, we review the topological classification of $C_n$-symmetric TCI's introduced in Ref.~\cite{benalcazar_quantization_2019} and argue how the topological invariants should be modified for latent symmetries. In \cref{sec:generators}, the notion of primitive generators is introduced, and the generators proposed in Ref.~\cite{benalcazar_quantization_2019} are generalized. We propose latent primitive generators that have no direct rotation symmetry, but are latently rotation symmetric. The topological behaviour of these generators is characterised in phase diagrams using topological rotation invariants. Using the generators, we construct examples of crystalline latent HOTIs in \cref{sec:corn}. Finally, we conclude our work in \cref{Sec: Conclusion}.
 
\section{Filling anomaly by latent symmetry}\label{sec:anomaly}
Crystalline symmetries impose restrictions on the distribution of electrons in a crystal. As a consequence, it may not be possible to maintain charge neutrality everywhere in the lattice. The simplest example of a model that exhibits this behaviour is the SSH model. Its Bloch-Hamiltonian is given by 
\begin{equation}
    h_\text{SSH}(k) = \begin{pmatrix}
        0 & v + e^{i k} \\
        v + e^{-i k} & 0
    \end{pmatrix},\label{eq:SSH}
\end{equation}
where the intracell hopping is denoted by $v$, the intercell hopping is fixed to $1$, and $k$ is the dimensionless crystal momentum (we set the lattice constant $a=1$). The SSH model exhibits chiral and mirror symmetry, and both can be used fpr the topological classification of the system. The former leads to the definition of a winding number, while the latter makes use of symmetry indicators. Since the classification in terms of the (crystalline) mirror symmetry is more relevant for the remainder of this work, we focus on it. For the SSH model, the mirror symmetry $M$ is given by
\begin{equation}
    M h_\text{SSH}(k) M^{-1} = h_\text{SSH}(-k), \qquad M = \begin{pmatrix}
        0 & 1 \\ 1 & 0 
    \end{pmatrix}. \label{eq:mirror}
\end{equation}
\begin{figure}[b]
    \centering
    \includegraphics[width=\linewidth]{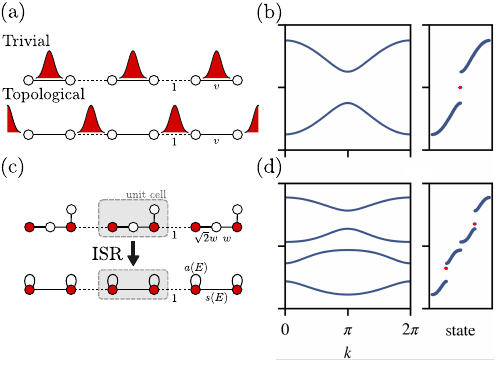}
    \caption{(a) Sketch of the SSH model in its trivial ($|v|>1$) and topological ($|v|<1$) phase. In the trivial (topological) phase, Wannier centers localize in the center (boundary) of the unit cell. (b) Spectrum of the SSH model for PBC (left) and OBC (right) in the topological phase. The OBC spectrum shows in-gap edge localized modes. (c) Latent SSH model consisting of four sites with intra cell hoppings $w$ (thin solid line) and $\sqrt{2}w$ (thick solid line). Upon performing an ISR, an energy dependent SSH model is obtained. (d) Spectrum of the latent SSH model for PBC (left) and OBC (right) in the topological phase. The OBC spectrum shows in-gap edge states for filling $1/4$ and $3/4$ in the topological phase.}
    \label{fig:1DISR}
\end{figure}
The presence of this mirror symmetry gives rise to two gapped phases, separated by a band closing at $|v| = 1$. At half filling, the system is an insulator, and for periodic boundary conditions (PBC), each unit cell hosts one electron. Overall, charge neutrality dictates that each unit cell is composed of an ion with charge $Q = |e|$. For open boundary conditions (OBC), without cutting through unit cells, the distribution of electrons depends on the phase. As a consequence of mirror symmetry, the Wannier centers of the electrons can only be located at either the center or the edge of a unit cell. For $|v|>1$ (trivial phase), the Wannier centers are located at the center of the unit cell and the system can be adiabatically connected to the trivial atomic limit. On the other hand, when $|v|<1$ (topological phase), the Wannier centers sit at the edge of the unit cell and the system can be connected to its obstructed atomic limit. The Wannier centers in both phases of the SSH model are shown in \cref{fig:1DISR}(a). In the trivial phase, $N$ electrons can be distributed symmetrically over $N$ unit cells to yield charge neutrality. However, in the topological phase, this is no longer possible. $N$ unit cells can only be filled in a symmetric manner using $N-1$ or $N+1$ electrons. Since mirror symmetry requires the energies of the edge states to be degenerate [see \cref{fig:1DISR}(b)], increasing the Fermi energy yields an increase of $N-1$ electrons to $N+1$ electrons, skipping $N$. This leads us to the notion of a \textit{filling anomaly} $\eta$,
\begin{equation}
    \eta = \#\text{ions} - \#\text{electrons}\quad \mod 2.
\end{equation}
Mirror symmetry distributes this charge imbalance over the two sides of the system such that there will be a fractional boundary charge of $e/2$ at the ends of the system. Consequently, the dipole moment in the topological phase equals $p=e/2$, while it vanishes in the trivial phase. One can also conclude that in order to respect the symmetries of the system at bulk insulating filling, the system with OBC must feature a filling anomaly $\eta$. These statements also hold for higher-order rotational symmetries $C_n$, which now divide the lattice into $n$ sectors, each spanning an angle of $2\pi/n$ rad, with $n$ being the order of the rotation. In this case, the filling anomaly is expressed as \cite{benalcazar_quantization_2019}
\begin{equation*}
    \eta = \#\text{ions} - \#\text{electrons}\quad \mod n.
\end{equation*}

\subsection{Isospectral reduction and latent symmetry}
Recently, Ref.~\cite{PhysRevB.110.035106} showed that although mirror symmetry leads to topological edge states in the SSH model, its not a necessary constraint. Indeed, the requirement of preserved mirror symmetry can be relaxed to preserved \textit{latent} mirror symmetry. A latent symmetry is hidden and only becomes apparent upon performing what is called an isospectral reduction (ISR)---akin to an effective Hamiltonian---on the system \cite{Smith2019PA514855HiddenSymmetriesRealTheoretical}. Take a Hamiltonian $H$ as a starting point. The Hilbert space on which it acts may be partitioned in a set $S$ and its complement $\overline{S}$. The ISR $\mathcal{R}_S(H,E)$ of $H$ is then defined through
\begin{equation} \label{eq:ISR}
    \mathcal{R}_S(H,E) = H_{SS} - H_{S\overline{S}}\left( H_{\overline{SS}} - E\mathbb{I} \right)^{-1} H_{\overline{S}S},
\end{equation}
where $\mathbb{I}$ is the identity matrix \cite{Bunimovich2014IsospectralTransformationsNewApproach}. The ISR converts the linear eigenvalue problem $H\boldsymbol{\psi} = E \boldsymbol{\psi}$ into a reduced, albeit non-linear, eigenvalue problem $\mathcal{R}_S(H,E) \psi_S~\!=~\!E~\!\psi_S$. For example, consider the Bloch Hamiltonian
\begin{equation}
    h_\text{LSSH}(k) = \begin{pmatrix}
        0 & e^{i k} & \sqrt{2}w & 0 \\
        e^{-i k} & 0 & w & w \\
        \sqrt{2} w & w & 0 & 0 \\
        0 & w & 0 & 0
    \end{pmatrix},\label{eq:latSSH}
\end{equation}
which is also depicted in \cref{fig:1DISR}(c) for OBC. By inspection, one can conclude that this Hamiltonian does not have a mirror symmetry. Nevertheless, upon performing an ISR over the red sites, we obtain
\begin{align}
    \mathcal{R}_S[h_\text{LSSH}(k),E] &= \begin{pmatrix}
        a(E) & s(E) + e^{i k} \\
        s(E) + e^{-i k} & a(E)
    \end{pmatrix}\notag\\
    &\equiv \mathfrak{h}_\text{LSSH}(k), \label{eq:latSSHreduc}
\end{align}
which bears a strong resemblance with the SSH model given in \cref{eq:SSH}, though with energy-dependent on-site potential $a(E)=2w^2/E$ and coupling $s(E)=\sqrt{2}w^2/E$. At the level of the ISR, there is a mirror symmetry: $M \mathfrak{h}_\text{LSSH}(k) M^{-1}=\mathfrak{h}_\text{LSSH}(-k)$ \footnote{For simplicity, from now on we will use $\mathfrak{h}(k)$ instead of $\mathcal{R}_S[h(k),E]$ to denote the ISR of $h(k)$. The energy dependence and reduction to $S$ are implied.}.
In other words, the system given by \cref{eq:latSSH} hosts a \textit{latent mirror symmetry}, which is revealed through the ISR.

The main idea of this work is to employ latent symmetries to construct HOTIs.
Before we do so, let us discuss two important properties of latent symmetries and the ISR. Firstly, it has been recently shown that the presence of latent symmetry also implies a certain (though in general non-geometric) symmetry on the level of the original Hamiltonian \cite{rontgen_symmetries_2021}. Let us define the eigenvalue problem $A\mathbf{v} = \lambda \mathbf{v}$ and let $\mathcal{R}_S (A,\lambda)$ be the ISR of $A$. If there exist a symmetry $T$ that becomes apparent after the ISR, that is
\begin{equation}\label{eq:latent}
    T \mathcal{R}_S (A,\lambda) T^{-1} = \mathcal{R}_S (A,\lambda),
\end{equation}
then there also exists a symmetry on the level of $A$ of the form $Q A Q^{-1} = A$, with $Q = T \oplus \overline{Q}$, where $\overline{Q}$ is a normal matrix that acts on $\overline{S}$.
Equivalently,
\begin{equation} \label{eq:latentSymmetryNormalSymmetry}
    \left[T, \mathcal{R}_S (A,\lambda)\right]_- = 0 \Rightarrow \exists Q= T \oplus \overline{Q} : \left[Q,A\right]_- = 0,
\end{equation}
where $[A,B]_- = A B - B A$ denotes the commutator.
Two points are noteworthy here.
Firstly, there is a strong connection between latent symmetries---symmetries of the isospectral reduction---and symmetries of the underlying Hamiltonian.
In particular, this connection implies that one can use the tools developed for conventional symmetries also to latent ones. A good example is group theory and its connection to spectral degeneracies, which were treated in Ref. \cite{Rontgen2021PRL126180601LatentSymmetryInducedDegeneracies}.
Secondly, the symmetry $Q$ corresponding to a latent symmetry $T$ usually has a rather complicated structure, i.e. it does not have a direct geometrical interpretation in the form of a permutation matrix\footnote{There is one exception to this statement: The ISR shown in \cref{fig:Latentcells}(f) features a reflection symmetry about a vertical axis that goes through the upper of the three sites; the corresponding symmetry $Q$ of the original setup [\cref{fig:Latentcells} (b)] is a permutation matrix.}. 
We shall see examples of this statement throughout the manuscript.

With some additional technical preface, a reasoning similar to \cref{eq:latentSymmetryNormalSymmetry} can be applied to crystalline symmetries.
Let us assume that our unit cell features a latent $T$ symmetry (for instance, $C_n$) that becomes apparent after performing an ISR to the set $S$ of its sites, i.e. $\left[ \mathcal{R}_S (H,E), T \right]_{-} = 0$. 
We then build a lattice such that (i) the ISR performed simultaneously on the sites $S$ in all unit cells is $T$-symmetric, and (ii) intercell-coupling only exists between sites $S$. It is now straightforward to show the following:
If we denote the reduced Bloch-Hamiltonian of the lattice by $\mathfrak{h}(k)$, we find, analogous to Ref. \cite{benalcazar_electric_2017},
\begin{equation}
    T \mathfrak{h}(\mathbf{k}) T^{-1} = \mathfrak{h}(D_T\kk)\,,\label{eq:cryst}
\end{equation}
where $D_T$ is the representation of $T$ on the vector space of reciprocal lattice vectors. For example, mirror symmetry ($T \equiv M$) flips spatial coordinates, such that $D_M k = -k$.
The analogue of \cref{eq:latent} is then 
\begin{equation}
    \left(T \oplus \overline{Q}\right) h(\kk) \left(T \oplus \overline{Q}\right)^{-1} = h(D_T \mathbf{k}).\label{eq:symQ}
\end{equation}

To make things more specific, let us now investigate the latent SSH model introduced in \cref{eq:latSSH}.
This model features a (1D) latent mirror symmetry: $T = M = \sigma_x$ and $D_M k = -k$. The symmetry $Q$ of $h_\text{LSSH}(k)$ is given by (details on the derivation of the matrix Q are given in \cref{app:Qmat})
\begin{equation} \label{eq:fullSymmetryLatentSSH}
    Q = M \oplus \overline{Q} = \begin{pmatrix}
        0 & 1 & 0 & 0 \\
        1 & 0 & 0 & 0 \\
        0 & 0 & \frac{1}{\sqrt{2}} & \frac{1}{\sqrt{2}} \\
        0 & 0 & \frac{1}{\sqrt{2}} & \frac{-1}{\sqrt{2}}
    \end{pmatrix}.
\end{equation}
The existence of $Q$ is central to the analysis considered in this work, as it forms the link between latent and conventional symmetries.
Specifically, and as we shall demonstrate now, it allows us to use the machinery developed for the treatment of geometric HOTIs also for the treatment of latent HOTIs.
In the case of \cref{eq:fullSymmetryLatentSSH}, the matrix $Q$ acts as a permutation on the red sites, but as a general orthogonal transformation on the other sites. Importantly, since $Q$ acts similarly on $h_\text{LSSH}(k)$ as an ordinary mirror symmetry would, that is $Q h_\text{LSSH}(k) Q^{-1} = h_\text{LSSH}(-k)$, it poses the exact same restrictions as mirror symmetry would. For example, the Wilson loop eigenvalues $\nu_\alpha$ of $h_\text{LSSH}(k)$ are restricted to be $0$, $\pi$ or come in pairs $\{-\nu,\nu\}$. Consequently, its Zak phase $\varphi$, which is the sum of Wilson loop eigenvalues, is quantized to $0$ or $\pi$. Both the Wilson loop eigenvalues and the Zak phase at different filling factors are shown in Table~\ref{tab:latssh}.

\begin{table}
\centering
\begin{tabular}{p{0.1\linewidth}>{\centering}p{0.25\linewidth}>{\centering}p{0.15\linewidth}>{\centering}p{0.25\linewidth}>{\centering\arraybackslash}p{0.15\linewidth}}
\toprule
     n & \multicolumn{2}{c}{$|v|>1$} & \multicolumn{2}{c}{$|v|<1$} \\
     \midrule
     {} & $\nu_\alpha$ & $\varphi$ & $\nu_\alpha$ & $\varphi$ \\
     \cmidrule(l){2-3}\cmidrule(l){4-5}
     $1$ & $0$ & $0$ & $\pi$ & $\pi$ \\
     $2$ & $\phantom{\pi, } \{-\nu, \nu\}$ & $0$ & $\phantom{\pi, } \{-\nu, \nu\}$ & $0$ \\
     $3$ & $0, \{-\nu, \nu\}$ & $0$ & $\pi, \{-\nu, \nu\}$ & $\pi$ \\
\bottomrule
\end{tabular}
\caption{Wilson loop eigenvalues $\nu_\alpha$ and Zak phase $\varphi$ for different fillings $n$ of $h_\text{LSSH}(k)$. For $|v|>1$ the model is always trivial, while for $|v|<1$ the model is topological for fillings $n=1$ and $n=3$.}\label{tab:latssh}
\end{table}
Secondly, besides revealing latent symmetries, the ISR also allows for a more straightforward topological characterisation of Bloch Hamiltonians. More concretely, consider two Hamiltonians $h_1(\kk,\mathbf{t}=\{t_1,t_2,\dots,t_n\})$ and $h_2(\kk,\mathbf{g}=\{g_1,g_2,\dots,g_m\})$. The first one, $h_1(\kk,\mathbf{t})$, represents a known model which exhibits a gap closing at $E=E_0$ captured through a condition on the parameters $\mathbf{t}$ that could equivalently be written as $f(\mathbf{t}) = 0$. For example, it could represent the SSH model, introduced in Eq.~\eqref{eq:SSH}. In this case, we have $E_0=0$, $\mathbf{t} = (v)$, and the gap closes at $f(v) = 1 - |v| = 0$; see the example below.  Let us then assume that the ISR of the second model has the same form as $h_1(\kk)$, but with energy-dependent parameters $\{\mathfrak{t}_1(E,\mathbf{g}),\mathfrak{t}_2(E,\mathbf{g}),\dots,\mathfrak{t}_n(E,\mathbf{g})\} \equiv \boldsymbol{\mathfrak{t}}(E,\mathbf{g})$, up to a constant energy-dependent shift $a(E,\mathbf{g})$ \footnote{Recall that through the ISR, the reduced parameters depend on both energy $E$ and the original system parameters $\mathbf{g}$.}, i.e.,
\begin{equation}
    \mathfrak{h}_2(\kk) = h_1\big[\kk,\boldsymbol{\mathfrak{t}}(E,\mathbf{g})\big] + a(E,\mathbf{g})\mathbb{I}. \label{eq:similar}
\end{equation}
For instance, for the latent SSH model described in  Eq.~\eqref{eq:latSSH}, this is the case [c.f. Eq.\eqref{eq:latSSHreduc}]. If this is the case, we can extract the gap closing energies, $E^*$, of $h_2(\kk,\mathbf{g})$ from 
\begin{equation}
    E^* - a(E^*,\mathbf{g}) = E_0. \label{eq:criten}
\end{equation}
This may be understood by rewriting the eigenvalue problem corresponding to Eq.~\eqref{eq:similar}, i.e. $\mathfrak{h}_2(\kk) \ket{\psi} = E \ket{\psi}$ in the form
\begin{equation}
    h_1\left[ \kk \mathfrak{t}(E,\mathbf{g}) \right] \ket{\psi} = \left[E-a(E,\mathbf{g}) \right] \ket{\psi}.
\end{equation}
Imposing $h_1(\kk)$ has $E_0$ as a gap closing energy yields Eq.~\eqref{eq:criten}.
The corresponding values of $\mathbf{g}$ can be obtained by solving $f\big[\boldsymbol{\mathfrak{t}}(E^*,\mathbf{g})\big] = 0$.
When the ISR reduces a system to a known model, the topological characterisation of the system is reduced from calculating multiband topological invariants to solving an algebraic problem \cite{PhysRevB.110.035106, eek_emergent_2024}. As an example, consider again the latent SSH model described by \cref{eq:latSSH}. The `ordinary' SSH model has a gap closing at $E_0 = 0$ for $f(v) = 1- |v| = 0$. The gap closing energies $E^*$ for the latent model are thus determined through
\begin{equation}
    E^* - a(E^*,w) =  E^* - \frac{2w^2}{E^*} = 0.\label{eq:enSSH}
\end{equation}
Equation~\eqref{eq:enSSH} has two solutions $E^* = \pm \sqrt{2}w$, which is corroborated by \cref{fig:1DISR}(d) showing two sets of in-gap modes in red. The energies $E^*$ predict the topological phase transition through $f[s(E^*,w)]=1-|s(E^*,w)| = 0$, which in this case reduces to $|w|=1$,  akin to the SSH model. Analogously, the system is topological for $|s(E^*,w)|<1$, which in this specific case reduces to $|w|<1$. From this we conclude that we can fully predict the topological behavior of the latent SSH model by imposing knowledge of the SSH model itself.

\subsection{Difference between symmetries and \textit{latent} symmetries}

Now, we discuss symmetries, both latent and non-latent, in more detail. It is important to realize that there are no asymmetric Hamiltonians.
Indeed, if one defines symmetry as the commutation of a matrix with a given Hamiltonian $h$, then there is an \emph{infinite} number of symmetries; they can be constructed directly from the eigenvectors of $h$\footnote{For instance, if $\ket{a}$ and $\ket{b}$ are eigenvectors of $h$, then the matrix $\ket{a}\bra{b}$ is a symmetry of $h$.}.
Colloquially, however, only those operations that are sufficiently easy to find are considered to be symmetries; with examples being reflection, rotation, and inversion operations.
The remaining operations commuting with the Hamiltonian are considered to be ``hidden'' symmetries

An example of this (arbitrary) classification into obvious and non-obvious (often called ``hidden") symmetries are spectral degeneracies: In group theory one can make a strong link between spectral degeneracies and the group that is formed by its symmetries. In particular, through Schur's lemma, the number and dimension of the so-called irreducible representations of the group are directly linked to the degeneracies of the Hamiltonian. Importantly, it might be that not all (or even none) of the degeneracies of the Hamiltonian can be explained from the group theoretical analysis of its obvious symmetries. In this case, the degeneracies are said to be \textit{accidental}, though they can be understood by considering the non-obvious symmetries as well.
Specifically, given some $n$-fold degenerate states $\ket{i}$, by considering operators of the form $\sum_{i,j} a_{i,j} \ket{i}\bra{j}$, it is straightforward to construct a set of matrices that commute with the Hamiltonian and which form a suitable group that can explain the degeneracy\footnote{The key is to realize that the coefficients $a_{i,j}$ can be identified with an $n$-dimensional matrix $A$. One then has to choose any suitable group with a $n$-dimensional irreducible representation, and identitfy this representation with the set of matrices $A$. A possible choice would be the so-called standard representation of $S_{n+1}$, that is, the symmetric group of order $n+1$ \cite{Fulton2004129RepresentationTheory}.}.
For two-fold degeneracies, a pedagogical explanation of this approach can be found in Ref.~\cite{Hou2018HiddenDegeneracy}.
Another example for the prevalence of a classification into obvious and non-obvious symmetries is the hydrogen atom.
Geometrically, it posseses a $SO(3)$ symmetry, but additionally it also has a hidden symmetry that leads to the conservation of the Runge-Lenz vector \cite{Fock1935ZP98145ZurTheorieWasserstoffatoms}.

There are good reasons for the division of symmetries into obvious and non-obvious. Usually, the symmetries of the Hamiltonian are used to simplify the eigenvalue problem $h\boldsymbol{\psi} = E \boldsymbol{\psi}$. For instance, if $h$ is reflection symmetric, its eigenvectors have a definite parity, such that the $h$ can be block diagonalized, yielding two smaller eigenvalue problems. Any obvious symmetry, that is: one that can be found before diagonalizing $h$ itself, is valuable in solving eigenvalue problems.

Latent symmetries fill the gap between obvious and non-obvious symmetries.
Firstly, due to \cref{eq:latentSymmetryNormalSymmetry}, every latent symmetry corresponds to a symmetry $Q$ of the full Hamiltonian.
The matrix $Q$ is block-diagonal, with the first block a permutation---an obvious symmetry---, while the other is a general orthogonal transformation, that is, a non-obvious symmetry.
The matrix $Q$ is thus a chimera; a mixture living in both worlds.

 \section{Topological indices from (latent) rotational symmetry}\label{sec:class}
In the above, we have discussed a latent SSH model, which has been previously investigated both in its Hermitian \cite{PhysRevB.110.035106} and its non-Hermitian \cite{eek_emergent_2024} version. In the remainder of this work, we apply a similar reasoning to develop the concept of \textit{latent HOTIs}. To this end, we start by reviewing the existing methods for the classification of usual HOTIs. We focus on two-dimensional insulators that preserve time-reversal symmetry (class AI in the Altland-Zirnbauer classification \cite{altland_nonstandard_1997,ryu_topological_2010}). The introduction of crystalline symmetries allows for a further classification of these materials \cite{fu_topological_2011, fang_bulk_2012}. Here, we will follow Ref.~\cite{benalcazar_quantization_2019}.

\subsection{Recap: Topological indices through conventional geometric rotation symmetries}
The presence of a rotation symmetry $\hat{C}_n$, which rotates sites in a lattice by $2\pi/n$ rad around some point, is represented on the level of the Bloch Hamiltonian by
\begin{equation}
    \hat{C}_n h(\kk) \hat{C}_n^{-1} = h(D_{C_n} \kk),
\end{equation}
where, similar to \cref{eq:cryst}, $D_{C_n}$ is a linear transformation on $\mathbf{k}$ that depends on $C_n$. At high symmetry points (HSPs) in the Brillouin zone, i.e. points that are mapped to themselves (modulo a reciprocal lattice vector), $D_{C_n} \boldsymbol{\Pi^{(n)}} = \boldsymbol{\Pi^{(n)}}$, we have
\begin{equation}
    \left[\hat{C}_n, h\left( \boldsymbol{\Pi^{(n)}}\right)\right]_- = 0\,.
\end{equation}
Because $\hat{C}_n$ and $h(k)$ commute at a HSP $\boldsymbol{\Pi^{(n)}}$, they share an eigenbasis. Thus, the Bloch states $\ket{u(\mathbf{k}=\boldsymbol{\Pi^{(n)}})}$ can be chosen as eigenstates of $\hat{C}_n$, such that
\begin{equation}\label{eq:rotEV}
    \hat{C}_n \ket{u\left( \boldsymbol{\Pi^{(n)}} \right)} = \Pi^{(n)}_p \ket{u\left( \boldsymbol{\Pi^{(n)}} \right)}.
\end{equation}
Since $\left(\hat{C}_n \right)^n = \mathbb{I}$, its eigenvalues are the $n$-th roots of unity:
\begin{equation}
    \Pi^{(n)}_p = e^{\frac{2\pi i}{n}(p-1)}, \qquad \text{with} \qquad  p \in \{1,2, \dots n\}.
\end{equation}
From these eigenvalues, we can construct rotation topological invariants of the form
\begin{equation}
    \left[ \Pi^{(n)}_p \right] \equiv \# \Pi^{(n)}_p - \# \Gamma^{(n)}_p,
\end{equation}
where $\# \Pi^{(n)}_p$ denotes the number of bands below the energy gap with eigenvalue $\Pi^{(n)}_p$ and $\boldsymbol{\Gamma} = \mathbf{0}$ is the gamma point in the Brillouin zone, which is a natural reference point to calculate the rotational invariants\footnote{For example, $[K_2^{(3)}] = \# K^{(3)}_2 - \# \Gamma^{(3)}_2$, where $\# K^{(3)}_2$ represents the number of occupied eigenvectors at $\boldsymbol{\Pi} = \mathbf{K}$ which have $C_3$ eigenvalue $\exp{2 \pi i/3}$. The same holds for $\# \Gamma^{(3)}_2$ but evaluated at the $\boldsymbol{\Gamma}$ point.}. $[\Pi^{(n)}_p]$ characterises the topology of $C_n$-symmetric insulators in a similar way to inversion eigenvalues: A difference of inversion eigenvalues between two HSPs indicates band inversion, i.e. non-trivial topology. In a similar manner, the difference of rotation eigenvalues between HSPs, captured by $[ \Pi^{(n)}_p]$, allows for a comparison of the representations of rotation symmetry. If different representations exist, the energy bands exhibit non-trivial topology. As a consequence of TRS and the fact that the number of bands is constant through the Brillouin zone, one obtains a set of independent topological indices $\chi^{(n)}$ for $C_n$-symmetric materials \cite{benalcazar_quantization_2019},
\begin{figure*}
    \centering
    \includegraphics[width=\textwidth]{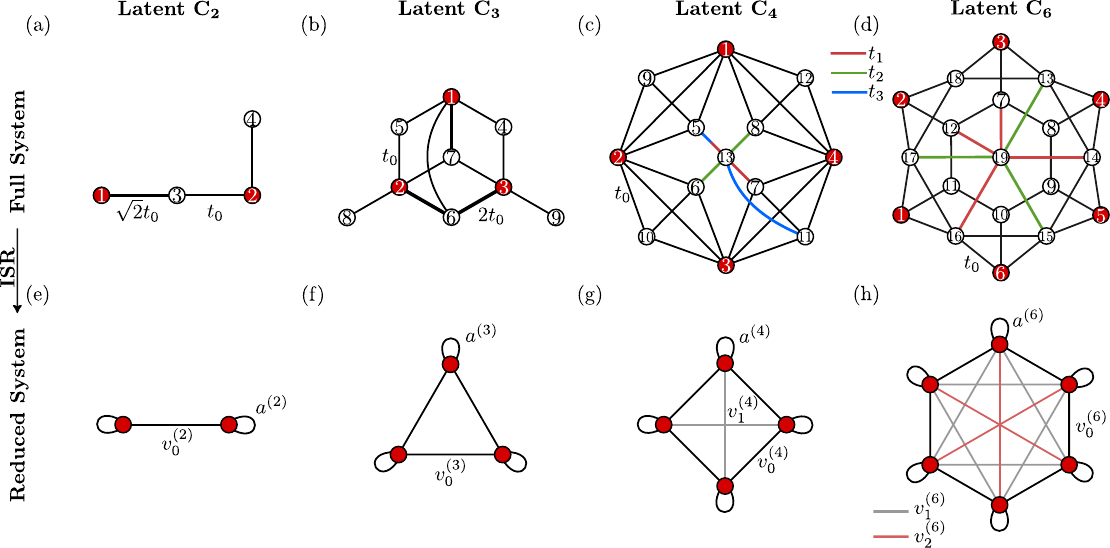}
    \caption{Latently $C_n$ symmetric unit cells. (a-d) Minimal examples of unit cells that show a latent $C_2$, $C_3$, $C_4$, and $C_6$ symmetry without a direct geometric symmetry. Relevant hopping parameters are indicated. (e-h) Cells obtained upon performing an ISR to the red sites of the unit cells in (a-d), the reduced systems show geometric $C_n$ symmetries, revealing the latent symmetries in (a-d). Relevant (energy-dependent) parameters are indicated. The numbers on the sites indicate the order in which they appear in the matrix representations of the Hamiltonians.}\label{fig:Latentcells}
\end{figure*}
\begin{align}
    \chi^{(2)} &= \left( \left[ X_1^{(2)} \right], \left[ Y_1^{(2)} \right], \left[ M_1^{(2)} \right] \right), \notag \\
    \chi^{(4)} &= \left( \left[ X_1^{(2)} \right], \left[ M_1^{(4)} \right], \left[ M_2^{(4)} \right] \right), \notag \\
    \chi^{(3)} &= \left( \left[ K_1^{(3)} \right], \left[ K_2^{(3)} \right] \right), \notag \\ 
    \chi^{(6)} &= \left( \left[ M_1^{(2)} \right], \left[ K_1^{(3)} \right] \right). \label{eq:rotinvar}
\end{align}
Here, $\mathbf{X}$, $\mathbf{Y}$, $\mathbf{M}$, and $\mathbf{K}$ ($X$, $Y$, $M$, and $K$) are high-symmetry points in the Brillouin zone.

\subsection{Topological indices through latent rotation symmetries}
Having reviewed the theory regarding geometric symmetries, let us now discuss how latent symmetries fit into the scheme.
Rather surprisingly, it can be shown that the invariants $\chi^{(n)}$ are still applicable.
The key for this insight is \cref{eq:latentSymmetryNormalSymmetry}.
That is, the fact that to every latent symmetry $\hat{C}_n$ there corresponds a matrix $Q = \hat{C}_n \oplus \overline{Q}$ that commutes with the underlying Hamiltonian.

Equipped with this knowledge, one only needs to replace the symmetry operator $\hat{C}_n$ in \cref{eq:rotEV} by $\hat{C}_n\oplus \overline{Q}$, as defined in \cref{eq:symQ}. Upon doing so, the rotation invariants of latent symmetric models can be evaluated to characterize the topology of the system. This leads to the definition of latent rotation symmetric HOTIs through
\begin{equation}
    \left[ \hat{C}_n \oplus \overline{Q}, h\left( \boldsymbol{\Pi^{(n)}}\right) \right]_- = 0,
\end{equation}
where the total symmetry $\hat{C}_n \oplus \overline{Q}$ is not necessarily geometric. Equivalently, we have
\begin{equation}
    \left[ \hat{C}_n, \mathfrak{h}\left( \boldsymbol{\Pi^{(n)}}\right) \right]_- = 0.
\end{equation}

\subsection{Classification of HOTIs: Dipole moment and corner charges}
The rotation invariants $\chi^{(n)}$ can be related to physical properties of the systems, such as the dipole moment $\mathbf{P}^{(n)}$ and corner charge $Q^{(n)}$. These quantities are derived in Ref.~\cite{benalcazar_quantization_2019} for geometric symmetries and can be shown to hold also for latent symmetries.

Below, we will directly state the results for the dipole moment,
\begin{align}
    \mathbf{P}^{(2)} &= \frac{e}{2} \left(\left[ Y_1^{(2)} \right] + \left[ M_1^{(2)} \right] \right)\mathbf{a_1} \notag \\
    & \qquad \qquad + \frac{e}{2} \left(\left[ X_1^{(2)} \right] + \left[ M_1^{(2)} \right] \right)\mathbf{a_2}, \notag \\ 
    \mathbf{P}^{(4)} &= \frac{e}{2} \left[ X_1^{(2)} \right] (\mathbf{a_1}+\mathbf{a_2}), \notag \\
    \mathbf{P}^{(3)} &= \frac{2e}{3} \left(\left[ K_1^{(3)} \right] + 2\left[ K_2^{(3)} \right]\right) (\mathbf{a_1}+\mathbf{a_2}), \notag \\
    \mathbf{P}^{(6)} &= \mathbf{0}, \label{eq:dipole} 
\end{align}
and for the corner charge,
\begin{align}
    Q^{(2)} &= \frac{e}{4} \left( -\left[ X_1^{(2)} \right] - \left[ Y_1^{(2)} \right] + \left[ M_1^{(2)} \right] \right), \notag \\
    Q^{(4)} &= \frac{e}{4} \left( \left[ X_1^{(2)} \right] + 2 \left[ M_1^{(4)} \right] + 3\left[ M_2^{(4)} \right] \right), \notag \\
    Q^{(3)} &= \frac{e}{3} \left[ K_2^{(3)} \right], \notag \\
    Q^{(6)} &= \frac{e}{4} \left[ M_1^{(2)} \right] + \frac{e}{6} \left[ K_1^{(3)} \right]. \label{eq:charge}
\end{align}
In the expressions for the dipole moments, $\mathbf{a_1}$ and $\mathbf{a_2}$ represent the lattice vectors. The dipole moments are defined modulo $e/A_\text{cell}$ times a lattice vector and the corner charges are defined modulo $e$. Here $A_\text{cell}$ denotes the area of the unit cell. Notice also that we group the dipole moments $\mathbf{P}^{(2)}$ and $\mathbf{P}^{(4)}$ ($\mathbf{P}^{(3)}$ and $\mathbf{P}^{(6)}$) because they share at least one common topological index (the same is done for the corner charges as well).

Before we continue, let us briefly remind the reader about the connection between these two quantities and HOTIs.
To this end, we consider a system under OBC, terminated respecting its (latent or geometric) $C_n$ symmetry.
The system is a HOTI if it is (higher order) topological and insulating. The former requirement translates in a nonzero corner charge.
If the dipole moment $\mathbf{P}^{(n)}$ does not vanish, there will be in-gap states when OBC are imposed. This poses two constraints on a HOTI, be it conventional or latent:
\begin{itemize}
    \item The corner charge $Q^{(n)}$, as given by \cref{eq:charge}, is nonzero.
    \item The dipole moment $\mathbf{P}^{(n)}$, as given by \cref{eq:dipole}, vanishes.
\end{itemize}
In the remainder of this manuscript, we will show how one can construct a latent HOTI, that is, a latently $C_n$-symmetric system fulfilling these two criteria.

\section{Building blocks for latent HOTIs}\label{sec:generators}
A necessary first step to construct a latent HOTI is a sufficiently large set of latently $C_n$ symmetric lattices.
In this section, we start by showing how they can be built in subsection \ref{sec:latent cells}.

Afterwards, we introduce specific examples of unit cells that feature a latent $C_2-$, $C_3-$, $C_4-$, and $C_6-$symmetry, as depicted in \cref{fig:Latentcells}.
Then, in subsection \ref{sec:latticeStructures} we shall use these unit cells to construct lattices such that the total system retains the latent $C_n$ rotation symmetry.
Finally, in subsection \ref{sec:primgen}, we present a systematic way of constructing latently $C_n$-symmetric setups with a pre-defined topological index.
Equipped with all these tools, we will then finally construct latent HOTIs in \cref{sec:corn}.

\subsection{Latently symmetric unit cells}\label{sec:latent cells}
In the following, we will present unit cells featuring a latent $C_2$, $C_3$, $C_4$, or $C_6$ symmetry.
We emphasize that each of them is just a specific example of a large class of unit cells featuring these latent symmetries.
Before we continue, let us explain how these families of latently $C_n$ symmetric unit cells can be designed.

For the design of latently $C_2$-symmetric unit cells, there are different techniques that range from exhaustive search \cite{Rontgen2020PRA101042304DesigningPrettyGoodState} to more sophisticated graph-theoretical results \cite{Godsil1982AM25257ConstructingCospectralGraphs}.
The interested reader is referred to the literature on graphs\footnote{We note that there is a one-to-one mapping between a graph and its adjacency matrix $H$, which for many graphs is Hermitian and could thus be interpreted as a Hamiltonian.} with cospectral vertices \cite{Schwenk1973PTAAC257AlmostAllTreesAre,Godsil1982AM25257ConstructingCospectralGraphs}; every graph with this property has recently been shown to have a latent $C_2$-symmetry \cite{Kempton2020LAIA594226CharacterizingCospectralVerticesIsospectral}.
The four-site latently $C_2$-symmetric cell that we depict in \cref{fig:Latentcells}(a) was found analytically by starting with a four-site long chain with couplings $a,b,c$ and all on-site energies equal to zero.
Then, demanding that two sites $u,v$ in this chain are latently symmetric, analytical expressions for the three couplings were found.

The construction of unit cells with a latent $C_3$, $C_4$, and $C_6$-symmetry is more demanding but can be done using the complement multiplet technique explained in the Supplemental Material of Ref.~\cite{Rontgen2021PRL126180601LatentSymmetryInducedDegeneracies}.
This technique starts with a Hamiltonian that features a geometric $C_n$ symmetry.
It can be shown that such a Hamiltonian likewise features a latent $C_n$ symmetry over some sites $S$ that are mapped onto each other by the geometric symmetry \footnote{This can be shown by performing the isospectral reduction on any $n$ sites that are mapped onto each other by the corresponding symmetry operator $C_n$ commuting with the underlying Hamiltonian.}.
The principle is then to identify changes to the Hamiltonian that keep the latent symmetry.
In many cases, these changes break the initial geometric $C_n$ symmetry.
As a result, one obtains a purely latently $C_n$ symmetric Hamiltonian.
To identify such changes, one must find a set of sites $M \subset \sbar$---with $\sbar$ defined as in \cref{eq:ISR}---such that, for any positive integer $k$ and for any site $s \in S$, the expression
\begin{equation}
    \sum_{m \in M} \left(H \overline{H}^k \right)_{s,m} = c_k
\end{equation}
is fulfilled.
Here, $\overline{H}$ denotes the matrix obtained from the Hamiltonian $H$ by setting the coupling between $S$ and $\sbar$ to zero.
In the language of Ref.~\cite{Rontgen2021PRL126180601LatentSymmetryInducedDegeneracies} , the set $M$ is called a ``complement multiplet''.
Given such a complement multiplet $M$, one can connect each of its sites (with the same coupling) to a new site $c$ without breaking the latent symmetry\footnote{We remark that one could also connect different complement multiplets $M_i$ to this new site $c$. If some of these multiplets share sites, the coupling of these overlap sites to $c$ must be modified; see Ref.~\cite{Rontgen2021PRL126180601LatentSymmetryInducedDegeneracies}.}.
The setups depicted in \cref{fig:Latentcells}(b), (c), and (d) have been found using the complement multiplet technique.

Having discussed how latently $C_n$-symmetric unit cells can be designed, let us now investigate specific examples of these.

\subsubsection{Latent $C_2$ symmetry}
In Sec.~\ref{sec:anomaly}, the notion of a latently mirror symmetric system was introduced in the context of the latent SSH model. In 1D, $C_2$ and mirror symmetry are the same, hence we take the latent SSH unit cell as a building block for latent $C_2$-symmetric HOTIs. The unit cell of the latent SSH model, and its ISR to the red sites, are once more displayed in Figs.~\ref{fig:Latentcells}(a) and \ref{fig:Latentcells}(e), respectively. The Hamiltonian for the single cell in \cref{fig:Latentcells}(a) is given by
\begin{equation}\label{eq:latcellc2}
    H^{(2)}_{L} = t_0\begin{pmatrix}
        0 & 0 & \sqrt{2} & 0 \\
        0 & 0 & 1 & 1 \\
        \sqrt{2} & 1 & 0 & 0 \\
        0 & 1 & 0 & 0
    \end{pmatrix}.
\end{equation}
The parameters in the isospectrally reduced model in \cref{fig:Latentcells}(e) are given by $a^{(2)}=2t_0^2/E$ and $v^{(2)}_0=\sqrt{2}t_0^2/E$. The reduced model obeys a $C_2$-symmetry,
\begin{equation}\label{eq:C2}
    \hat{C}_2 = \begin{pmatrix}
        0 & 1 \\
        1 & 0  \\
    \end{pmatrix},
\end{equation}
which corresponds to the symmetry
\begin{align}
    Q^{(2)} &\equiv \hat{C}_2 \oplus \overline{Q}^{(2)} = 
\begin{pmatrix}
    0 & 1 \\
    1 & 0\\
\end{pmatrix}
\oplus
\begin{pmatrix}
\frac{1}{\sqrt{2}} & \frac{1}{\sqrt{2}}\\
\frac{1}{\sqrt{2}} & -\frac{1}{\sqrt{2}}
\end{pmatrix}
\end{align}
of the full Hamiltonian. 

\subsubsection{Latent $C_3$ symmetry}
An example of a unit cell exhibiting latent $C_3$ symmetry is depicted in \cref{fig:Latentcells}(b). It has 9 sites, with hopping given by $t_0$ (thin black lines) and $2t_0$ (thick black lines). Its Hamiltonian is given by
\begin{equation}\label{eq:latentc3}
    H^{(3)}_{L} = t_0\begin{pmatrix}
        0 & 0 & 0 & 1 & 1 & 1 & 2 & 0 & 0 \\
        0 & 0 & 0 & 0 & 1 & 2 & 1 & 1 & 0 \\
        0 & 0 & 0 & 1 & 0 & 2 & 1 & 0 & 1 \\ 
        1 & 0 & 1 & 0 & 0 & 0 & 0 & 0 & 0 \\
        1 & 1 & 0 & 0 & 0 & 0 & 0 & 0 & 0 \\
        1 & 2 & 2 & 0 & 0 & 0 & 0 & 0 & 0 \\
        2 & 1 & 1 & 0 & 0 & 0 & 0 & 0 & 0 \\ 
        0 & 1 & 0 & 0 & 0 & 0 & 0 & 0 & 0 \\
        0 & 0 & 1 & 0 & 0 & 0 & 0 & 0 & 0
    \end{pmatrix}.
\end{equation}
Upon performing an ISR to the red sites, the cell shown in \cref{fig:Latentcells}(f) is obtained, displaying a $C_3$-symmetry. The reduced model has a single energy-dependent hopping parameter $v_0^{(3)} = 5t_0^2/E$ and onsite potential $a^{(3)}~=~7t_0^2/E$.\footnote{For brevity, we have dropped the energy dependence in the notation of the reduced hopping parameter and onsite potential: $v_i^{(n)} = v_i^{(n)}(E)$ and $a^{(n)} = a^{(n)}(E)$.} The corresponding symmetry matrix of the full Hamiltonian $Q^{(3)}$ [cf. \cref{eq:latentSymmetryNormalSymmetry}] of the full system depicted in \cref{fig:Latentcells}(b) is given by
\begin{align}
    Q^{(3)} &\equiv \hat{C}_3 \oplus \overline{Q}^{(3)}\\
    &= 
\begin{pmatrix}
    0 & 0 & 1 \\
    1 & 0 & 0 \\
    0 & 1 & 0
\end{pmatrix}
\oplus
\begin{pmatrix}
\frac{1}{2} & 0 & \frac{1}{2} & -\frac{1}{2} & \frac{1}{2} & 0 \\
\frac{1}{2} & \frac{1}{2} & 0 & 0 & -\frac{1}{2} & \frac{1}{2} \\
0 & \frac{1}{2} & \frac{1}{2} & \frac{1}{2} & 0 & -\frac{1}{2} \\
0 & -\frac{1}{2} & \frac{1}{2} & \frac{1}{2} & 0 & \frac{1}{2} \\
\frac{1}{2} & 0 & -\frac{1}{2} & \frac{1}{2} & \frac{1}{2} & 0 \\
-\frac{1}{2} & \frac{1}{2} & 0 & 0 & \frac{1}{2} & \frac{1}{2} \\
\end{pmatrix}. \notag
\end{align}

\subsubsection{Latent $C_4$ symmetry}
Next, let us investigate the setup shown in \cref{fig:Latentcells}(c), which has 13-sites and features four different hopping parameters: $t_0$ (black), $t_1$ (green), $t_2$ (red), and $t_3$ (blue). A double-coloured line implies a sum of the two hopping strengths, i.e. the blue-and-red line has hopping parameter $t_2 + t_3$. The matrix form, $H_L^{(4)}$, of \cref{fig:Latentcells}(c) is given in \cref{app:expr4} by \cref{eq:H4L}.
Depending on the choice of couplings, this unit cell has different geometric symmetries.
Firstly, if $t_1 = t_2$ and $t_3 =0$, it enjoys a $C_4$-symmetry.
Keeping $t_3=0$ but breaking the equality of the first two couplings such that $t_1 \neq t_2$, this symmetry is partly broken and only a geometric $C_2$ symmetry is left.
The situation becomes much easier, though, when performing an ISR on the red sites.
The resulting reduced model is depicted in \cref{fig:Latentcells}(g); it has an energy-dependent on-site potential $a^{(4)}$ and hopping parameters $v_0^{(4)}$ (black) and $v_1^{(4)}$ (grey).
It also features a $C_4$ symmetry given by
\begin{equation}\label{eq:C4}
    \hat{C}_4 = \begin{pmatrix}
        0 & 0 & 0 & 1 \\
        1 & 0 & 0 & 0 \\
        0 & 1 & 0 & 0 \\
        0 & 0 & 1 & 0
    \end{pmatrix}.
\end{equation}
Thus, regardless of the choice of couplings $t_i$, the unit cell is latently $C_4$-symmetric.
Once again, this demonstrates that the ISR gives a wider, more comprehensive viewpoint on a system than merely checking its geometric symmetries.
Before continuing, we remark that---again, by \cref{eq:latentSymmetryNormalSymmetry}---this latent $C_4$ symmetry corresponds to a non-geometric symmetry $Q^{(4)} \equiv \hat{C}_4 \oplus \overline{Q}^{(4)}$ of the full unit cell.

Expressions for $\overline{Q}^{(4)}$ and the parameters in \cref{fig:Latentcells}(g) are given in \cref{app:expr4}.

\subsubsection{Latent $C_6$ symmetry}
Finally, we come to the unit cell depicted in \cref{fig:Latentcells}(d).
It has 19 sites and features three different hopping parameters $t_0$ (black), $t_1$ (green), $t_2$ (red). A more complex latent $C_6$-symmetric unit cell is given in \cref{app:expr6}. This model is $C_6$-symmetric only when $t_1=t_2=0$. Furthermore, it is $C_3$ symmetric if only $t_2=0$.
Once again, the picture becomes clearer when performing an ISR to the red sites.
The resulting reduced model is depicted in \cref{fig:Latentcells}(h); it is described in terms of the on-site potential $a^{(6)}$ and hopping parameters $v_0^{(6)}$ (black), $v_1^{(6)}$ (grey), and $v_2^{(6)}$ (red). It can be promptly seen that the reduced model is $C_6$-symmetric.
Thus, irrespective of the choice of coupling parameters $t_i$, the full unit cell is latently $C_6$-symmetric.
This latent symmetry corresponds to a non-geometric symmetry of the full unit cell, given by $Q^{(6)} \equiv \hat{C}_6 \oplus \overline{Q}^{(6)}$, with full expressions for $\overline{Q}^{(6)}$ and the parameters in \cref{fig:Latentcells}(h) given in \cref{app:expr6}.

\subsection{Lattice structures preserving latent $C_n$-symmetries} \label{sec:latticeStructures}
Equipped with a set of latently $C_n$-symmetric unit cells, the next step is to embed these into lattices, such that the total setup keeps this symmetry.
As we now show, this task is rather straightforward.
Let us assume that a given lattice structure is composed of a unit cell with a geometric $C_n$-symmetry, such that the lattice as a whole keeps the symmetry, i.e. the coupling between different cells also respects the $C_n$ symmetry.
Next, let us replace the geometrically $C_n$-symmetric unit cell by one whose ISR on a set $S$ of sites has the same symmetry, i.e. a latently $C_n$-symmetric unit cell.
One can then show by direct calculation, that the lattice's ISR---for clarity, we mean the simultaneous reduction on the union of sites $S$ in each unit cell---is $C_n$-symmetric. In other words, performing the ISR for all of the unit cells in the lattice simultaneously, instead of performing the ISR of a single cell, should yield a $C_n$-symmetric system.

Examples of systems with these features are discussed in more detail in the following section.

\subsection{Primitive generators and their topological classification}\label{sec:primgen}

With the material presented so far, one could construct a large number of latently $C_n$-symmetric lattices.
However, to be HOTIs, they need a vanishing dipole moment and a non-vanishing corner charge.
Via \cref{eq:dipole,eq:charge}, both quantities are connected to the topological indices in \cref{eq:rotinvar}.
In principle, one could then find latent HOTIs by a brute-force search, that is, by computing these quantities for a large number of latently $C_n$-symmetric setups and filtering out the ones that are HOTIs.
However, there is a more elegant and systematic way that is based on the concept of \emph{primitive generators} \cite{benalcazar_quantization_2019}.
Essentially, these are building blocks with certain properties, which can be connected to each other in a specific manner, such that the resulting setup features a well-defined topological index.

To introduce these generators, let us start with an interesting fact on topological indices.
As pointed out in Ref.~\cite{benalcazar_quantization_2019}, two models with the same $C_n$-symmetry, described by Bloch Hamiltonians $h_1$ and $h_2$, characterized by $\chi^{(n)}_1$ and $\chi^{(n)}_2$, respectively, may be combined to form a third model
\begin{equation}
    h_3 = \begin{pmatrix}
        h_1 & \gamma \\
        \gamma^\dagger & h_2 
    \end{pmatrix}.
\end{equation}
In absence of $\gamma$, $h_3$ features the combined gap structure of $h_1$ and $h_2$. The specific form of $\gamma$ should be chosen such that it does not close these gaps and it respects the $C_n$ symmetry of $h_1$ and $h_2$. In this case, the rotation invariant of $h_3$ is given by $\chi^{(n)}_3 = \chi^{(n)}_1 + \chi^{(n)}_2$. Consequently, if for a given symmetry $\chi^{(n)}$ has $N$ components, it is sufficient to have $N$ models with linearly independent $\chi^{(n)}$ to span the whole topological phase space. This sets the basis to the notion of \textit{primitive generators}. The primitive generators form a minimal set from which a setup with an arbitrary topological index can be constructed.
From there, the construction of an actual (latent) HOTI is only a minor step. 

In the following, we will discuss and classify (latent) primitive generators for every one of the four classes $C_2$, $C_3$, $C_4$, and $C_6$ that are compatible with translational invariance of a crystal.
The key results are graphically depicted in \cref{fig:C2full,fig:C3full,fig:C4full,fig:C6full} and are discussed in more detail in the text.
For each of the four classes, we first treat the conventional case of geometric symmetry, and then treat the new case of latent symmetry.
This dual treatment might seem redundant, but it serves two purposes.
Firstly, the characteristics (for instance, the band structure) of the geometric and latent setups have some striking similarity that would otherwise be unnoticed.
Secondly, our treatment of the geometrically symmetric primitive generators represents minor generalisations of the primitive generators introduced in Ref.~\cite{benalcazar_quantization_2019}.

In each of the following examples, the primitive generators are the Bloch-Hamiltonians of a crystal obtained by inserting a specific unit cell (with either a geometric or latent $C_n$ symmetry) into a specific lattice structure. 
We remark that a given $C_n$ might support different lattice structures.

\subsubsection{$C_2$-symmetry}\label{sec:PC2}
For a $C_2$ symmetry, we investigate only one lattice structure, which we call the ``stacked SSH model''.

\paragraph{Geometric stacked SSH.} We consider the primitive generator given by the Hamiltonian
\begin{equation}\label{eq:ham_c2}
    h_1^{(2)}(\kk) = \begin{pmatrix}
        2w \cos k_y & t + e^{ik_x}\\
        t + e^{-ik_x} & 2w \cos k_y
    \end{pmatrix}.
\end{equation}
\begin{figure}
    \centering
    \includegraphics[width=\linewidth]{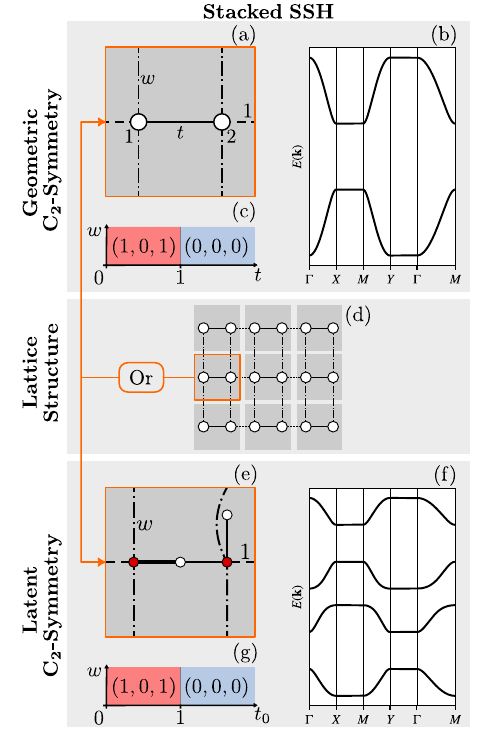}
    \caption{(a) Unit cell of the geometric $C_2$-symmetric primitive generator $h_1^{(2)}(\kk)$, with intracell hopping $t$. Horizontal intercell hopping is indicated by dashed lines and is fixed to $1$; vertical intercell hopping is represented by a dot-dashed line and is given by $w$. (b) Energy spectrum of $h_1^{(2)}(\kk)$ for $t = 1/2$ and $w=0$. (c) Phase diagram in the $(w,t)$-plane, in which the rotation invariants $\chi^{(2)}$ are shown for gapped phases. Since $w$ enters the Hamiltonian as an identity term, it does no alter the phase diagram. (d) Lattice structure of the $C_2$-symmetric primitive generators. The lattice represents stacked 1D SSH chains. (e) Unit cell of the Latent $C_2$-symmetric primitive generator $h^{(2)}_{L,1}(\kk)$. Values of the hopping parameters are indicated in \cref{fig:Latentcells}(a) and horizontal intercell hoppings (dashed lines) are fixed to $1$, while vertical intercell hoppings (dot-dashed lines) are given by $w$. (f) Spectrum of $h^{(2)}_{L,1}(\kk)$ for $t_0 = 1/2$ and $w=0$. (g) Phase diagram in the $(w,t_0)$-plane at one filled band. Rotation invariants $\chi^{(2)}$ are displayed for gapped phases.}
    \label{fig:C2full}
\end{figure}
It is the Bloch-Hamiltonian of a system obtained by inserting the unit cell depicted in \cref{fig:C2full}(a)  into the stacked SSH lattice structure of \cref{fig:C2full}(d).
The system corresponds to multiple SSH chains stacked in the $y$-direction.
Hopping within a chain occurs with an intracell hopping given by $t$ (solid black line) and a horizontal intercell hopping of $1$ (dashed black line).
The chains are connected with a vertical intercell hopping of $w$ (dot-dashed black line). The lattice vectors are given by $\mathbf{a_1}=(1,0)$ and $\mathbf{a_2}=(0,1)$.
\cref{eq:ham_c2} admits a $C_2$-symmetry given by
\begin{equation}
    \hat{C}_2 h_1^{(2)}(k_x,k_y) \hat{C}_2^{-1} = h_1^{(2)}(-k_x,-k_y),
\end{equation}
where $\hat{C}_2$ is given by \cref{eq:C2}.
Consequently, the topology of the system may be characterized using $\chi^{(2)}$. Since the value of $w$ does not affect the gap structure, we set $w=0$. For general values of $t$, the spectrum is gapped at half filling, as shown in \cref{fig:C2full}(b), giving rise to two distinct phases. These phases are separated by a gap closing at $t=+1$ ($-1$), taking place along the $\mathbf{XM}$ ($\mathbf{Y\Gamma}$) path in the Brillouin zone. For $|t|<1$, the system is in its topological phase, corresponding to $\chi^{(2)} = (1,0,1)$ and $\mathbf{P}^{(2)} = (e/2) \mathbf{a_1}$. For $|t| > 1$, the system is trivial with $\chi^{(2)} = (0,0,0)$ and $\mathbf{P}^{(2)} = \mathbf{0}$. The different topological phases are depicted in \cref{fig:C2full}(c).
We note that a second, independent, generator for $C_2$ can be obtained by rotating \cref{fig:C2full}(d) by 90 degrees. This would correspond by letting $k_x \to k_y$ and $k_y \to -k_x$ in \cref{eq:ham_c2}. As a result, the topological phase would now have $\chi^{(2)} = (0,1,1)$ and $\mathbf{P}^{(2)} = (e/2)\mathbf{a_2}$. This would yield $2$ generators for $C_2$-symmetric systems. A third generator can be obtained by taking one of the $C_4$ generators in the next section and making the hopping in the $x-$ and $y-$direction different.

\paragraph{Latent stacked SSH.} If we insert the unit cell depicted in \cref{fig:C2full}(e) into the lattice structure of \cref{fig:C2full}(d), we obtain a system with a Bloch-Hamiltonian given by
\begin{equation} \label{eq:c2LatentGenerator}
    h^{(2)}_{L,1}(\kk) = H^{(2)}_{L} + \begin{pmatrix}
        2w \cos k_y & e^{ik_x} & 0 & 0 \\
        e^{-ik_x} & 2w \cos k_y & 0 & 0 \\
        0 & 0 & 0 & 0 \\
        0 & 0 & 0 & 0
    \end{pmatrix},
\end{equation}
where $H^{(2)}_{L}$ has been defined in \cref{eq:latcellc2}. In \cref{fig:C2full}(e), the horizontal intercell hopping (dashed black line) is fixed to $1$ and the vertical intercell hopping (dot-dashed black line) is given by $w$.
Figure~\ref{fig:C2full}(f) depicts the spectrum of \cref{eq:c2LatentGenerator} for $t_0=1/2$ and $w=0$. Notice that the spectrum resembles two copies of the one in \cref{fig:C2full}(b). Since this primitive generator represents stacked (latent) SSH chains, its phase diagram is the same as that of the SSH chain\footnote{Notice that $w=0$ represents the case of uncoupled SSH chains, which poses a trivial example. Nevertheless, $w$ enters the (reduced) Hamiltonian as an identity term [see \cref{eq:ham_c2}]. Consequently, it will never close the gap and all phases with $w\neq 0$ will be adiabatically connected to $w=0$. Therefore, they will be in the same topological phase.}. At the end of Sec.~\ref{sec:anomaly}, we showed that for this specific latent SSH model, phase transitions occur at $|t|= 1$, just like for the SSH model (again, the value of $w$ does not affect the gap structure). Following this reasoning, $h^{(2)}_{L,1}(\kk)$ has a gap closing at $|t_0| = 1$, separating the trivial phase $\chi^{(2)} = (0,0,0)$ ($|t_0|>1$) from the topological phase $\chi^{(2)} = (1,0,1)$ ($|t_0|<1$), as shown in the phase diagram in \cref{fig:C2full}(g). 

\subsubsection{$C_4$-symmetry}\label{sec:PC4}
For a $C_4$-symmetry, we consider three lattice structures: a ``2D SSH'', a ``breathing square-octagon'', and a ``stacked breathing square-octagon''.
\begin{figure*}
    \centering
    \includegraphics[width=\linewidth]{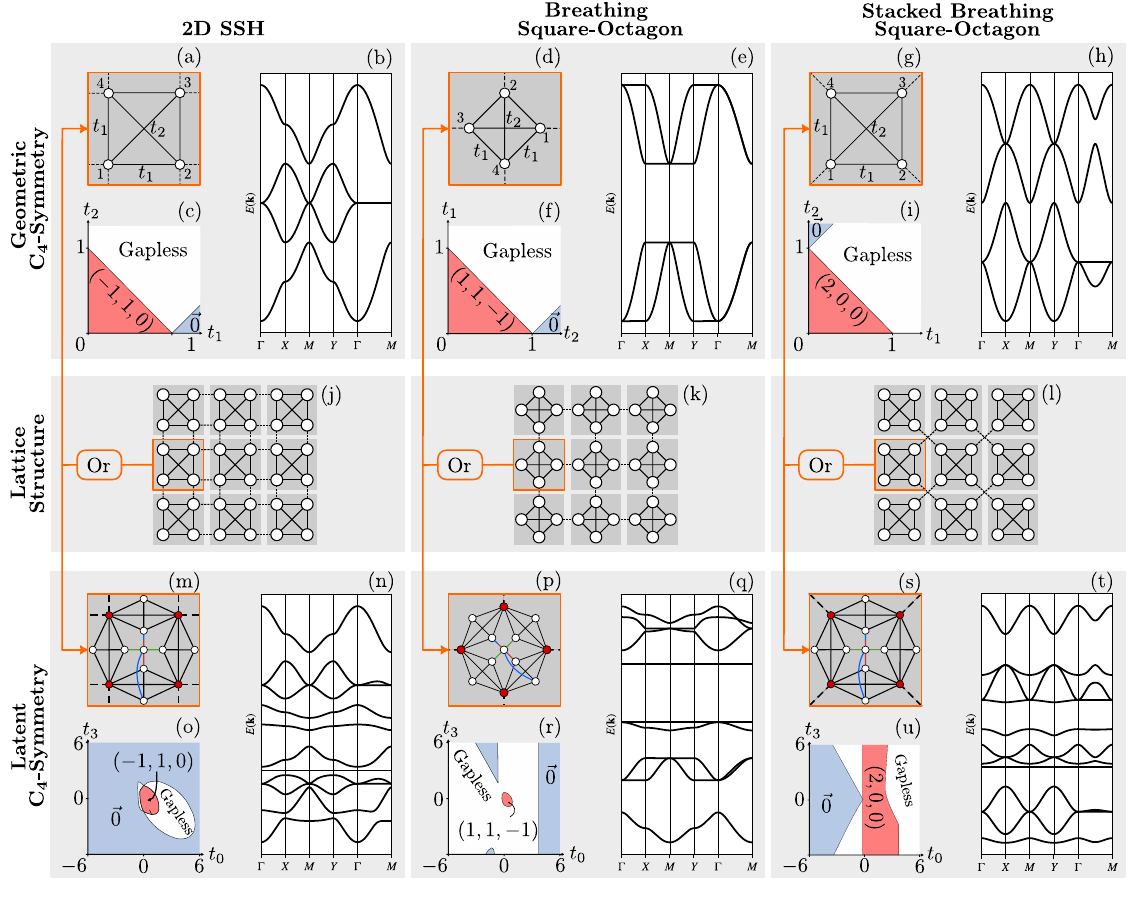}
    \caption{(a,d,g) Unit cells of the geometric $C_4$-symmetric primitive generators $h^{(4)}_1(\kk)$, $h^{(4)}_2(\kk)$, and $h^{(4)}_3(\kk)$, respectively. The nearest-neighbour hoppings are represented by $t_1$; the next-nearest-neighbour hoppings are represented by $t_2$; the intercell hoppings are indicated by dashed lines and are fixed to be equal to $1$. (b,e,h) Energy spectra of $h_1^{(4)}(\kk)$ at $t_1 = 1/2$ and $t_2 = 0$, $h_2^{(4)}(\kk)$ at $t_1 = 1/2$ and $t_2 =0$, and $h_3^{(4)}(\kk)$ at $t_1 = 0$ and $t_2 =1/2$. (c,f,i) Phase diagrams of the geometric primitive generators. For gapped phases, the rotation invariants $\chi^{(4)}$ are shown. (j,k,l) Lattice structures of the $C_4-$symmetric primitive generators. The lattices represent a 2D SSH, a breathing square-octagon, and a stacked breathing square octagon lattice, respectively. (m,p,s) Unit cells of the latent $C_4$-symmetric primitive generators $h^{(4)}_{L,1}(\kk)$, $h^{(4)}_{L,2}(\kk)$, and $h^{(4)}_{L,3}(\kk)$, respectively. Values of the hopping parameters are indicated in \cref{fig:Latentcells}(c) and intercell hoppings (dashed lines) are fixed to $1$. (n,q,t) Energy spectra of $h^{(4)}_{L,1}(\kk)$ for $t_0 = t_1 = t_2 = 1$, $t_3 = 1/2$, $h^{(4)}_{L,2}(\kk)$ for $t_0 = t_1 = t_2 = 1/2$, $t_3 = 1$, and $h^{(4)}_{L,3}(\kk)$ for $t_0 = t_1 = t_2 = -1$, $t_3 = 0$. (o,r,u) Phase diagrams of $h_{L,1}^{(4)}(\kk)$ at 1 filled band, $h_{L,2}^{(4)}(\kk)$ at 2 filled bands, and $h_{L,3}^{(4)}(\kk)$ at 3 filled bands, as a function of $t_0$ and $t_3$, $t_1=t_2=t_0$. Rotation invariants $\chi^{(4)}$ are shown for gapped phases.}
    \label{fig:C4full}
\end{figure*}

\paragraph{Geometric 2D SSH.} Inserting the unit cell of \cref{fig:C4full}(a) into a lattice structure, we obtain \cref{fig:C4full}(j), which corresponds to a system described by the Bloch-Hamiltonian
\begin{equation}\label{eq:C4_1}
    h_1^{(4)}(\mathbf{k}) = \begin{pmatrix}
        0 & t_1 + e^{ik_x} & t_2 & t_1 + e^{ik_y}\\
        t_1 + e^{-ik_x} & 0 & t_1 + e^{ik_y}& t_2 \\
        t_2 & t_1 + e^{-ik_y} & 0 & t_1 + e^{-ik_x} \\
        t_1 + e^{-ik_y} & t_2 & t_1 + e^{ik_x} & 0
    \end{pmatrix}.
\end{equation}
The system has 4 sites per unit cell, as depicted in \cref{fig:C4full}(a), and is also known as the 2D SSH model \cite{Liu2019R20191TopologicallyProtectedEdgeState,Coutant2021JoAP129125108TopologicalTwodimensionalSuSchriefferHeegerAnalog,PhysRevB.108.245140}. The sites are connected with hopping parameters $t_1$ and $t_2$ for nearest-neighbour and next-nearest-neighbour hopping, respectively. The intercell hopping (dashed lines) is fixed to unity. The lattice vectors are given by $\mathbf{a_1}=(1,0)$ and $\mathbf{a_2}=(0,1)$. Equation~\ref{eq:C4_1} has a $C_4$-symmetry, that is:
\begin{equation}
    \hat{C}_4 h_1^{(4)}(k_x,k_y) \hat{C}_4^{-1} = h_1^{(4)}(k_y,-k_x),
\end{equation}
where $\hat{C}_4$ is given by \cref{eq:C4}.
Figure~\ref{fig:C4full}(b) shows the spectrum of $h_1^{(4)}(\kk)$ for $t_{1} = 1/2$ and $t_{2}= 0$, which is gapped at $1/4$- and $3/4$-filling. For this parameter choice, we have $[ X_1^{(2)} ] = -1$, $[ M_1^{(4)} ] = 1$, $[ M_2^{(4)} ] = 0$, hence $\chi^{(4)} = (-1,1,0)$, which corresponds to $\mathbf{P}^{(4)}= e (\mathbf{a_1}+\mathbf{a_2})/2$. The spectrum closes its gap for $|t_{1}| + |t_{2}| = 1$, which occurs simultaneously at the $\mathbf{X}$ and $\mathbf{Y}$ point, owing to the $C_4$-symmetry. For finite $t_{2}$, the system remains gapless till $|t_{1}| = 1+|t_{2}|$. At this point, the system becomes gapped once again with $\chi^{(4)} = (0,0,0)$, corresponding to $\mathbf{P}^{(4)}=\mathbf{0}$. This behavior, together with the corresponding symmetry indicators $\chi^{(4)}$, is depicted in  \cref{fig:C4full}(c). A third primitive generator for $C_2$ is obtained by letting the hoppings in the $x-$ and $y-$dirrection be different. In that case, the $C_4$-symmetry in \cref{fig:C4full}(a) gets broken into a $C_2$ symmetry given by
\begin{equation}
    \mathcal{I} h_1^{(4)}(k_x,k_y) \mathcal{I}^{-1} = h_1^{(4)}(-k_x,-k_y),
\end{equation}
where $\mathcal{I}\equiv \hat{C}_4^2$. 

\paragraph{Geometric breathing square-octagon.}
The second generator for $C_4$ is given by the Bloch-Hamiltonian
\begin{align}\label{eq:C4_2}
    h_2^{(4)}(\mathbf{k}) = \begin{pmatrix}
        0 & t_1 & t_2 + e^{ik_x} & t_1 \\
        t_1 & 0 & t_1 & t_2 + e^{ik_y} \\
        t_2 + e^{-ik_x} & t_1 & 0 & t_1  \\
        t_1 & t_2 + e^{-ik_y} & t_1 & 0
    \end{pmatrix}.
\end{align}
The underlying system is obtained by inserting the unit cell of \cref{fig:C4full}(d) into a lattice to form the structure shown in \cref{fig:C4full}(k).
The internal structure of the unit cell is rotated with regard to $h_1^{(4)}(\kk)$.
If the next-nearest-neighbour hopping is set to zero, this lattice represents (breathing) T-Graphene; otherwise, we call it the square-octagon lattice. The lattice has attracted much attention recently, with results ranging from topological phases to flat-band superconductivity \cite{dai_electronic_2014,Nunes2020PRB101224514FlatbandSuperconductivityTightbindingElectrons,Yan_2023}.
The spectrum is gapped for $|t_{1}| + |t_{2}| < 1$ at half filling, accompanied with rotation invariants $\chi^{(4)} = (1,1,-1)$ and $\mathbf{P}^{(4)} = e(\mathbf{a_1}+\mathbf{a_2})/2$. An example of the spectrum for $t_{1} = 0$ and $t_{2} = 1/2$ is shown in  \cref{fig:C4full}(e). When $|t_{1}| + |t_{2}| = 1$, the gap generally closes at the $\mathbf{M}$ point. However, when $t_{1} = 0$ (and therefore $|t_{2}| = 1$), it closes at both the $\mathbf{M}$ and $\boldsymbol{\Gamma}$ points. Moreover, when $t_{2} = 0$ ($|t_{1}|=1$), the gap closes over the full $\mathbf{XM}$ and $\mathbf{YM}$ lines in the Brillouin zone. For $|t_{2}| > 1 + |t_{1}|$, a new gap opens at the $\mathbf{M}$ point (if $|t_{1}| = 0$, this gap opens along the whole $\mathbf{XM}$ and $\mathbf{YM}$ lines). This gap is trivial and characterized by $\chi^{(4)} = (0,0,0)$, $\mathbf{P}^{(4)}= \mathbf{0}$. For other parameter choices, the system is gapless, as shown in \cref{fig:C4full}(f).

\paragraph{Geometric stacked breathing square-octagon.}
Finally, the third $C_4$-symmetric generator is governed by the Hamiltonian
\begin{align} \label{eq:C4_3}
    h_3^{(4)}(\mathbf{k}) &= \begin{pmatrix}
        0 & t_1 & t_2 & t_1 \\
        t_1 & 0 & t_1 & t_2 \\
        t_2 & t_1 & 0 & t_1  \\
        t_1  & t_2 & t_1  & 0
    \end{pmatrix} \\
    &+ \begin{pmatrix}
        0 & 0 & e^{i(k_x+k_y)} & 0 \\
        0 & 0 & 0 & e^{-i(k_x-k_y)} \\
        e^{-i(k_x+k_y)} & 0 & 0 & 0  \\
        0 & e^{i(k_x-k_y)} & 0 & 0
    \end{pmatrix}. \notag
\end{align}
The underlying system is obtained by inserting the unit cell of \cref{fig:C4full}(g) into a lattice to form the structure in \cref{fig:C4full}(l).
The system has the same internal structure as $h_1^{(4)}(\kk)$, as can be seen in  \cref{fig:C4full}(a), but has different intercell hopping. The lattice is formed by overlapping two breathing square-octagon lattices. The spectrum of $h_3^{(4)}(\kk)$ [ \cref{fig:C4full}(h)] is gapped at half filling for $|t_{1}|+|t_{2}| < 1$, characterized by the rotation invariants $\chi^{(4)} = (2,0,0)$. This corresponds to $\mathbf{P}^{(4)}=\mathbf{0}$. For $|t_{1}|+|t_{2}| = 1$, the gap closes at the $\mathbf{X}$ and $\mathbf{Y}$ points (if $t_{1} = 0$, it also closes at the $\boldsymbol{\Gamma}$ and $\mathbf{M}$ points). For $|t_{2}| > 1+|t_{1}|$, a trivial gap opens at the $\boldsymbol{\Gamma}$ and $\mathbf{M}$ points with $\chi^{(4)} = (0,0,0)$, $\mathbf{P}^{(4)}=\mathbf{0}$. For other values of $t_{1}$ and $t_{2}$, the system is gapless, as shown in \cref{fig:C4full}(i). 

\paragraph{Latent $C_4$-symmetric structures.} The latent $C_4$-symmetric cell in \cref{fig:Latentcells}(c) can be inserted in the three different lattice structures of \cref{fig:C4full}(j), \cref{fig:C4full}(k), or \cref{fig:C4full}(l).
This yields systems with Bloch-Hamiltonians of the form
\begin{equation}
    h^{(4)}_{L,i}(\kk) = H^{(4)}_{L} + \begin{pmatrix}
        \Tilde{h}^{(4)}_i(\kk) & \emptyset_{4\times 9}\\
        \emptyset_{9\times 4} & \emptyset_{9\times 9}
    \end{pmatrix}.
\end{equation}
Here, $H^{(4)}_{L}$ has been defined in \cref{eq:H4L}, the symbol $\emptyset_{m \times n}$ represents a zero matrix of size $m\times n$, and $\Tilde{h}^{(4)}_i (\kk)$ are the Hamiltonians given in \cref{eq:C4_1}, \cref{eq:C4_2}, and \cref{eq:C4_3}  with all intracell hoppings set to zero, i.e. only intercell hopping. This is because the second term is only there to connect the latently symmetric cells on a lattice. 

For all $h^{(4)}_{L,i}(\kk)$, we set $t_1 = t_2 = t_0$. Consequently, the Hamiltonians have a geometric $C_4$-symmetry for $t_3 = 0$. For finite values of $t_3$, the geometric $C_4$ symmetry of $h^{(4)}_{L,i}(\kk)$ gets broken. Nevertheless, the latent symmetry of the unit cell is inherited, which becomes clear upon taking the ISR to the red sites
\begin{equation}
    \mathfrak{h}^{(4)}_{L,i}(\kk) = h^{(4)}_{i}\left(\kk, t_{1} = v^{(4)}_0, t_{2} = v^{(4)}_1 \right) + a^{(4)} \mathbb{I}, \label{eq:c4reduced}
\end{equation}
which commutes with $\hat{C}_4$ given in \cref{eq:C4}. The parameters $a^{(4)}$, $v_0^{(4)}$ and $v_1^{(4)}$ are polynomials of a degree larger than $5$ [See \cref{app:expr4}]. Consequently, it is not possible to analytically resolve the gap closing conditions as outlined at the end of Sec.~\ref{sec:anomaly}. 
Figure \ref{fig:C4full}(n) shows the spectrum of $h^{(4)}_{L,1}(\kk)$ for $t_0=t_1=t_2 = 1$ and $t_3 = 1/2$. The spectrum is gapped at multiple fillings. For simplicity we consider a single filled band. Figure \ref{fig:C4full}(o) shows the numerically obtained phase diagram of $h_{L,1}^{(4)}(\kk)$ for a single filled band for $t_1 = t_2 = t_0$. The rotation invariants $\chi^{(4)}$ are depicted for the gapped phases.
Similarly, the spectrum of $h^{(4)}_{L,2}(\kk)$ for $t_0 = t_1=t_2 = -1$ and $t_3 = 0$ is shown in \cref{fig:C4full}(q). The corresponding phase diagram for $2$ filled bands is shown in \cref{fig:C4full}(r).
Finally, \cref{fig:C4full}(t) shows the spectrum of $h^{(4)}_{L,3}(\kk)$ for $t_0 = t_1=t_2 = 1/2$ and $t_3 = 1$. The phase diagram shown in \cref{fig:C4full}(u) is (numerically) obtained for 3 filled bands. The rotation invariants $\chi^{(4)}$ are depicted for the gapped phases. The topological and trivial phases are represented by the red and blue regions, respectively. The white regions denote metallic gapless states. 

\subsubsection{$C_3$-symmetry}\label{sec:PC3}
For $C_3$-symmetry, we consider two lattice strucures, namely a ``breathing kagome'' and a ``bearded breathing kagome''.

\begin{figure}[h]
    \centering
    \includegraphics[width=\linewidth]{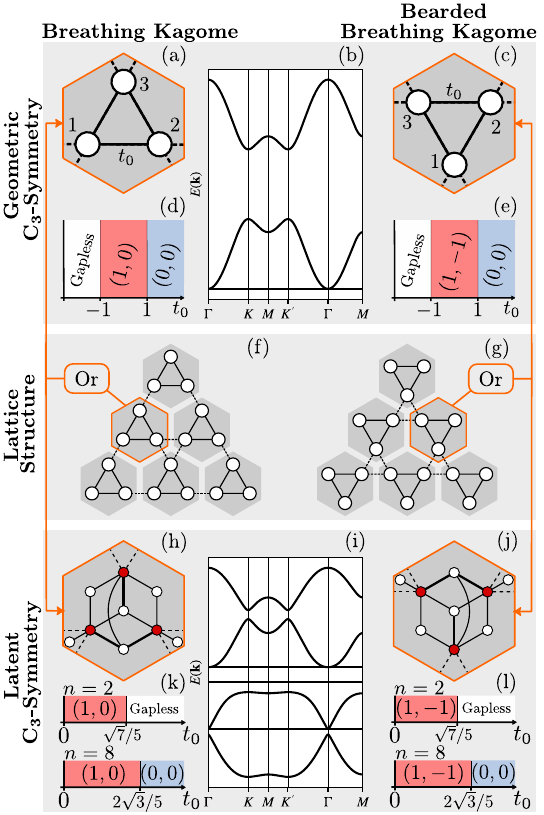}
    \caption{(a,c) Unit cells of the geometric $C_3-$symmetric primitive generators $h^{(3)}_1(\kk)$ and $h^{(3)}_2(\kk)$, respectively. Intracell hopping (black line) has a strength of $t_0$, while intercell hopping (dashed lines) is fixed to $1$. (b) Spectrum of $h^{(3)}_1(\kk)$ and $h^{(3)}_2(\kk)$ at $t_0 = 1/2$. (d,e) Phase diagrams of $h^{(3)}_1(\kk)$ and $h^{(3)}_2(\kk)$ at $2/3$-filling as a funciton of $t_0$. Rotation invariants $\chi^{(3)}$ are indicated for the gapped phases. (f,g) Lattice structures of the $C_3-$-symmetric primitive generators. The lattices represent a breathing kagome and a bearded breathing kagome lattice, respectively. (h,j) Unit cells of the latent $C_3-$symmetric primitive generators $h^{(3)}_{L,1}(\kk)$ and $h^{(3)}_{L,2}(\kk)$, respectively. Values of the hopping parameters are indicated in \cref{fig:Latentcells}(b) and intercell hoppings (dashed lines) are fixed to $1$. (i) Spectrum of $h^{(3)}_{L,1}(\kk)$ and $h^{(3)}_{L,2}(\kk)$ for $t_0 = 0.6$. (k,l) Phase diagrams at filling $n=2$ and $n=8$ for $h^{(3)}_{L,1}$ and $h^{(3)}_{L,2}$, respectively. The rotation invariants $\chi^{(3)}$ are indicated in the different gapped phases.}
    \label{fig:C3full}
\end{figure}

\paragraph{Geometric $C_3$-symmetric structures.} The internal structure of our geometrically $C_3$-symmetric unit cells are shown in Figs.~\ref{fig:C3full}(a) and (c). They consist of 3 sites connected with an intracell hopping $t_0$ and an intercell hopping fixed to $1$. The two lattice structures for $C_3$ symmetry are presented in Figs.~\ref{fig:C3full}(f)-(g). 
Since the resulting systems correspond to different terminations of the same Kagome lattice, we will treat them together.
The Hamiltonian corresponding to the unit cell shown in \cref{fig:C3full}(a) inserted into a lattice to form the structure in \cref{fig:C3full}(f) is given by
\begin{equation}
    h^{(3)}_1(\kk) = \begin{pmatrix}
        0 & t_0 + e^{i\kk\cdot \mathbf{a_1}} & t_0 + e^{i\kk\cdot \mathbf{a_2}} \\
        t_0 + e^{-i\kk\cdot \mathbf{a_1}} & 0 & t_0 + e^{-i\kk\cdot \mathbf{a_3}} \\
        t_0 + e^{-i\kk\cdot \mathbf{a_2}} & t_0 + e^{i\kk\cdot \mathbf{a_3}} & 0
    \end{pmatrix}, \label{eq:C3_1}
\end{equation}
while the Bloch-Hamiltonian corresponding to the unit cell in \cref{fig:C3full}(c), which forms the lattice structure in \cref{fig:C3full}(g), is given by
\begin{equation}
    h^{(3)}_2(\kk) = \begin{pmatrix}
        0 & t_0 + e^{i\kk\cdot \mathbf{a_2}} & t_0 + e^{-i\kk\cdot \mathbf{a_3}} \\
        t_0 + e^{-i\kk\cdot \mathbf{a_2}} & 0 & t_0 + e^{-i\kk\cdot \mathbf{a_1}} \\
        t_0 + e^{i\kk\cdot \mathbf{a_3}} & t_0 + e^{i\kk\cdot \mathbf{a_1}} & 0
    \end{pmatrix}. \label{eq:C3_2}
\end{equation}
The lattice vectors are defined through $\mathbf{a_1} = (1,0)$, $\mathbf{a_2} = (1/2, \sqrt{3}/2)$, and $\mathbf{a_3} = \mathbf{a_1}-\mathbf{a_2}$. $h^{(3)}_1(\kk)$ and $h^{(3)}_2(\kk)$ represent a (bearded) breathing Kagome lattice \cite{ kempkes_robust_2019,van_miert_topological_2020,herrera_corner_2022} which exhibits a $C_3$ symmetry of the form
\begin{equation}
    \hat{C}_3 h_i^{(3)}(k_x,k_y) \hat{C}_3^{-1} = h_i^{(3)}(D_{C_3}\kk),
\end{equation}
with $ D_{C_3}\kk = \left( -k_x-\sqrt{3}k_y,-k_y+\sqrt{3} k_x \right)/2$ and
\begin{equation}
    \hat{C}_3 = \begin{pmatrix}
        0 & 0 & 1 \\
        1 & 0 & 0 \\
        0 & 1 & 0
    \end{pmatrix}.
\end{equation}
The spectrum for both Hamiltonians is shown in \cref{fig:C3full}(b) for $t_0 = 1/2$. For $|t_0|<1$, the spectrum of $h^{(3)}_1(\kk)$ [$h^{(3)}_2(\kk)$] is gapped at $2/3$ filling and is described by a topological invariant $\chi^{(3)} = (1,0)$ [$\chi^{(3)} = (1,-1)$] with $\mathbf{P}^{(3)}= (2e/3)(\mathbf{a_1}+\mathbf{a_2})$ [$\mathbf{P}^{(3)}= (e/3)(\mathbf{a_1}+\mathbf{a_2})$]. At $t_0 = 1$, the gap closes at $\mathbf{K}$ and $\mathbf{K^\prime}$ and opens again in a trivial phase with $\chi^{(3)} = (0,0)$ and $\mathbf{P}^{(3)}=\mathbf{0}$ for $t_0>1$. At $t_0 = -1$, the gap closes at $\boldsymbol{\Gamma}$ and the spectrum remains gapless for $t_0<-1$. The phase diagrams of $h^{(3)}_1(\kk)$ and $h^{(3)}_2(\kk)$ are shown in Figs.~\ref{fig:C3full}(d) and (e), respectively.

\paragraph{Latent $C_3$-symmetric structures.} Inserting the latently $C_3$-symmetric unit cell \cref{fig:Latentcells}(b) into a lattice to form the structures in Figs.~\ref{fig:C3full}(f) and (g) yields two latent $C_3$-symmetric primitive generators with Bloch-Hamiltonians
\begin{equation}
    h^{(3)}_{L,i}(\kk) = H^{(3)}_{L} + \begin{pmatrix}
        \Tilde{h}^{(3)}_i(\kk) & \emptyset_{3\times 6}\\
        \emptyset_{6\times 3} & \emptyset_{6\times 6}
    \end{pmatrix},
\end{equation}
with $i=1,2$. $\Tilde{h}^{(3)}_i (\kk)$ are the Hamiltonians given in \cref{eq:C3_1} and \cref{eq:C3_2} with all intracell hoppings set to zero, i.e. only intercell hopping (once again, just to connect the larger cells on a lattice), and $H^{(3)}_{L}$ has been defined in \cref{eq:latentc3}. Figures \ref{fig:C3full}(f) and (g) show lattices corresponding to $h^{(3)}_{L,1}$ and $h^{(3)}_{L,2}$, respectively. Analogous to the non-latent primitive generators for $C_3$ symmetry, $h^{(3)}_{L,1}$ and $h^{(3)}_{L,2}$ share the same spectrum, as depicted in \cref{fig:C3full}(i).  

The ISR of $h^{(3)}_{L,i}(\kk)$ to the red sites is given by
\begin{equation}
    \mathfrak{h}^{(3)}_{L,i}(\kk) = h^{(3)}_{i}\left(\kk, t_0 = v^{(3)}_0\right) + a^{(3)}\mathbb{I},
\end{equation}

Owing to the simplicity of the parameters $v^{(3)}_0$~$(=5t_0^2/E)$ and $a^{(3)}$~$(=7t_0^2/E)$, it is possible to analytically derive the phase diagram of this model. In Sec.~\ref{sec:generators} we showed that $h^{(3)}_i(\kk)$ has a gap closing at $E=1$ for $t_0=1$, and at $E=0$ for $t_0=-1$. Therefore, we obtain the energies $E_1^*$ and $E_2^*$ at which the latent models have a gap closing
\begin{equation}
    E_1^*-a^{(3)}(E_1^*) = 1 \quad \text{and} \quad E_2^*-a^{(3)}(E_2^*) = 0.
\end{equation}
Solving the above equations yields
\begin{equation}
    E_1^* = \frac{1}{2}\left(1\pm \sqrt{1+28 t_0^2}\right) \quad \text{and} \quad E_2^* = \pm \sqrt{7}t_0.\label{eq:critC3}
\end{equation}
From \cref{eq:critC3}, we extract the phase transitions through $v^{(3)}_0(E_1^*) = 1$ and $v^{(3)}_0(E_2^*) = -1$, resulting in 
\begin{equation}
    t_0 = \pm \frac{2\sqrt{3}}{5}, \pm \frac{\sqrt{7}}{5}.
\end{equation}
The gap closing energies correspond to a filling of $n=2$ and $n=8$ bands out of $9$. Using the above derived constraints, we obtain the phase diagrams depicted in Figs.~\ref{fig:C3full}(k) and (l) for $h_{L,1}^{(3)}(\kk)$ and $h_{L,2}^{(3)}(\kk)$, respectively. The rotation invariants $\chi^{(3)}$ are shown for the different phases. The invariants are the same as those obtained for $h^{(3)}_i(\kk)$.

\subsubsection{$C_6$-symmetry}\label{sec:PC6}
For $C_6$-symmetry, we consider two lattice structures, namely, the ``breathing ruby lattice'', and the ``Kekulé'' structure.

\begin{figure*}
    \centering
    \includegraphics[width=\linewidth]{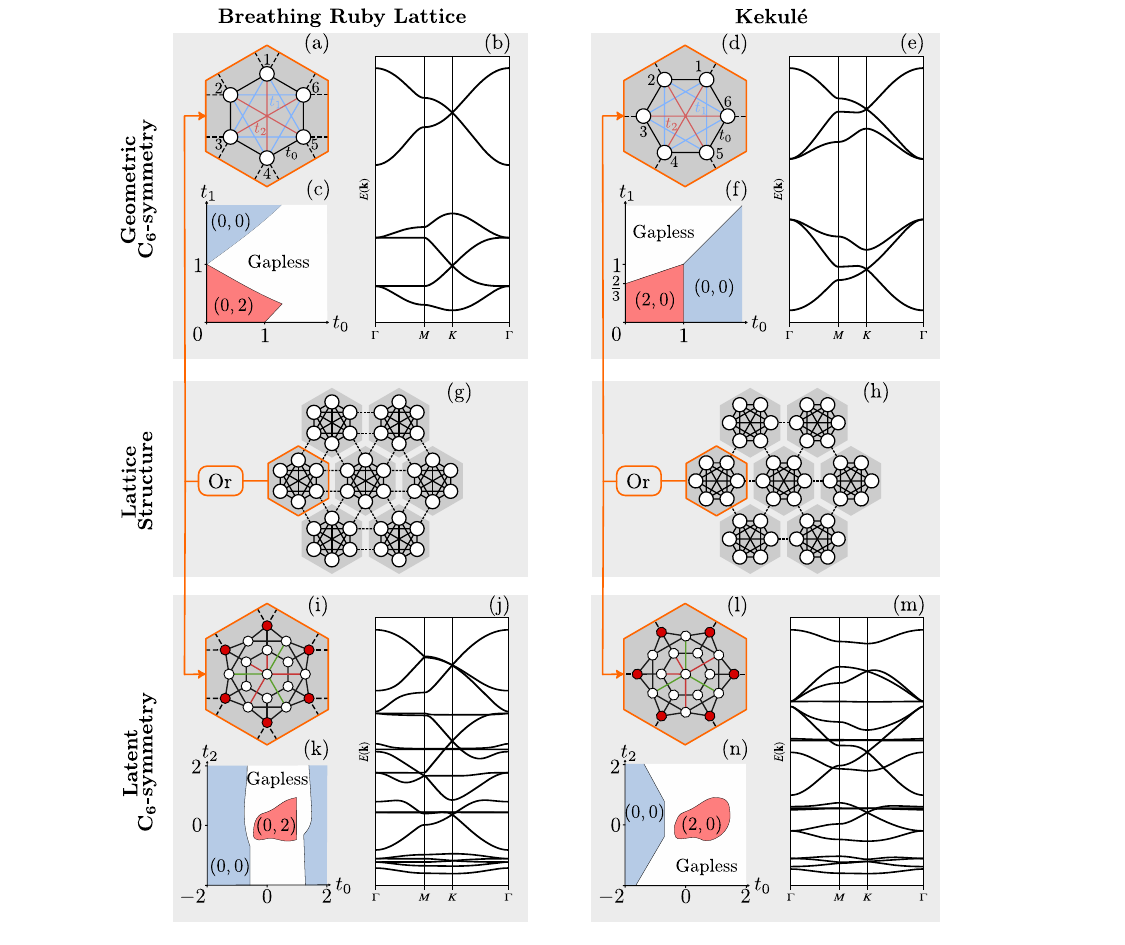}
    \caption{(a,d) Unit cells of the geometric $C_6$-symmetric primitive generators $h_1^{(6)}(\kk)$ and $h_2^{(6)}(\kk)$. Three $C_6$ preserving hoppings $t_0$ (black), $t_1$ (blue) and $t_2$ (red) are indicated. Intercell hopping (dashed) is fixed to $1$. (b,e) Spectra of $h^{(6)}_1(\kk)$ and $h^{(6)}_2(\kk)$, for $t_0 = 1/2$, $t_1 = t_2 = 0$. (c,f) Phase diagrams of $h^{(6)}_1(\kk)$, and $h^{(6)}_2(\kk)$ in the $(t_0, t_1)$-plane with $t_2 = 0$. Rotation invariants $\chi^{(6)}$ are indicated for gapped phases. (g,h) Lattice structures of the $C_6-$symmetric primitive generators. The lattices represent a breathing ruby lattice and a kekulé lattice, respectively. (i,l) Unit cells of the latent $C_6$-symmetric primitive generators $h_{L,1}^{(6)}(\kk)$ and $h_{L,2}^{(6)}(\kk)$. Values of the hopping parameters are indicated in \cref{fig:Latentcells}(d) and intercell hoppings (dashed lines) are fixed to $1$. (j,m) Spectra of $h_{L,1}^{(6)}(\kk)$ and $h_{L,2}^{(6)}(\kk)$, for $t_0 = 3/4$, $t_1 = 7/8$, $t_2 = 1/8$. (k,n) Phase diagrams of $h_{L,1}^{(6)}(\kk)$ for $n = 4$ filled bands and $h_{L,2}^{(6)}(\kk)$ for $n = 3$ filled bands, with $t_1 = 1/4$. Rotation invariants $\chi^{(6)}$ are indicated in the gapped phases.}
    \label{fig:C6full}
\end{figure*}

\paragraph{Geometric breathing ruby lattice} The Hamiltonian for the first $C_6$-symmetric primitive generator is given by
\begin{align}
    h^{(6)}_1&(\kk) = \begin{pmatrix}
        0 & t_0 & t_1 & t_2 & t_1 & t_0 \\
        t_0 & 0 & t_0 & t_1 & t_2 & t_1 \\
        t_1 & t_0 & 0 & t_0 & t_1 & t_2 \\
        t_2 & t_1 & t_0 & 0 & t_0 & t_1 \\
        t_1 & t_2 & t_1 & t_0 & 0 & t_0 \\
        t_0 & t_1 & t_2 & t_1 & t_0 & 0 \\
    \end{pmatrix} \label{eq:C6_1}\\
    &+
    \begin{pmatrix}
        0 & 0 & e^{-i \kk \cdot \mathbf{a_2}} & 0 & e^{i \kk \cdot \mathbf{a_3}} & 0 \\
        0 & 0 & 0 & e^{i \kk \cdot \mathbf{a_3}} & 0 & e^{i \kk \cdot \mathbf{a_1}} \\
        e^{i \kk \cdot \mathbf{a_2}} & 0 & 0 & 0 & e^{i \kk \cdot \mathbf{a_1}} & 0 \\
        0 & e^{-i \kk \cdot \mathbf{a_3}} & 0 & 0 & 0 & e^{i \kk \cdot \mathbf{a_2}} \\ 
        e^{-i \kk \cdot \mathbf{a_3}} & 0 & e^{-i \kk \cdot \mathbf{a_1}} & 0 & 0 & 0 \\
        0 & e^{-i \kk \cdot \mathbf{a_1}} & 0 & e^{-i \kk \cdot \mathbf{a_2}} & 0 & 0 
    \end{pmatrix} . \notag 
\end{align}
It is obtained by inserting the unit cell from \cref{fig:C6full}(a) into a lattice to obtain the structure in \cref{fig:C6full}(g).
This lattice is a breathing version of a Ruby lattice. The lattice vectors are given by $\mathbf{a_1} = (1,0)$, $\mathbf{a_2} = (1/2, \sqrt{3}/2)$ and $\mathbf{a_3} = \mathbf{a_1}-\mathbf{a_2}$. $h^{(6)}_1(\kk)$ exhibits a $C_6$ symmetry given by
\begin{equation}
    \hat{C}_6 h_1^{(6)}(k_x,k_y) \hat{C}_6^{-1} = h_1^{(6)}(D_{C_6}\kk), \label{eq:C6sym}
\end{equation}
with $D_{C_6}\kk = (k_x-\sqrt{3}k_y,+\sqrt{3}k_x+ky)/2$ and
\begin{equation}\label{eq:C6}
    \hat{C}_6 = \begin{pmatrix}
        0 & 0 & 0 & 0 & 0 & 1 \\
        1 & 0 & 0 & 0 & 0 & 0 \\
        0 & 1 & 0 & 0 & 0 & 0 \\
        0 & 0 & 1 & 0 & 0 & 0 \\
        0 & 0 & 0 & 1 & 0 & 0 \\
        0 & 0 & 0 & 0 & 1 & 0
    \end{pmatrix}.
\end{equation}
First, consider the case $t_1 = t_2 = 0$. The spectrum is gapped at $2/3$-filling if $|t_0|<1$. The gap is characterised by $\chi^{(6)} = (0,2)$, which corresponds to $\mathbf{P^{(6)}} = \mathbf{0}$ and $Q^{(6)} = e/3$. This can be seen in the spectrum in \cref{fig:C6full}(b), which is calculated for $t_0 = 1/2$. At $t_0 = 1$, the gap closes at the $\boldsymbol{\Gamma}$ point and upon further increase of $t_0$, the spectrum remains gapless. If we only impose $t_2 = 0$, the phase diagram in \cref{fig:C6full}(c) is obtained. For values of $t_0$ and $t_1$ in the red region, the system is in a topological phase with $\chi^{(6)} = (0,2)$. Leaving the topological phase (red) by increasing $t_1$, the gap closes at the $\mathbf{K}$ and $\mathbf{K'}$ points and the spectrum is gapless. Upon further increasing $t_1$, a trivial gap [$\chi^{(6)} = (0,0)$] reopens at the $\mathbf{K}$ and $\mathbf{K'}$ points.

\paragraph{Geometric Kekulé.}
The Hamiltonian for the second $C_6$-symmetric primitive generator is given by
\begin{align}
    h^{(6)}_2&(\kk) = \begin{pmatrix}
        0 & t_0 & t_1 & t_2 & t_1 & t_0 \\
        t_0 & 0 & t_0 & t_1 & t_2 & t_1 \\
        t_1 & t_0 & 0 & t_0 & t_1 & t_2 \\
        t_2 & t_1 & t_0 & 0 & t_0 & t_1 \\
        t_1 & t_2 & t_1 & t_0 & 0 & t_0 \\
        t_0 & t_1 & t_2 & t_1 & t_0 & 0 \\
    \end{pmatrix}\label{eq:C6_2}\\
    &+
    \begin{pmatrix}
        0 & 0 & 0 & e^{-i \kk \cdot \mathbf{a_2}} & 0 & 0 \\
        0 & 0 & 0 & 0 & e^{i \kk \cdot \mathbf{a_3}} & 0 \\
        0 & 0 & 0 & 0 & 0 & e^{i \kk \cdot \mathbf{a_1}} \\
        e^{i \kk \cdot \mathbf{a_2}} & 0 & 0 & 0 & 0 & 0 \\
        0 & e^{-i \kk \cdot \mathbf{a_3}} & 0 & 0 & 0 & 0 \\
        0 & 0 & e^{-i \kk \cdot \mathbf{a_1}} & 0 & 0 & 0 
    \end{pmatrix}.\notag
\end{align}
The corresponding system is obtained by using the unit cell of \cref{fig:C6full}(d) to form the lattice structure in \cref{fig:C6full}(h).
Note that it has the same unit cell as in the Breathing Ruby Lattice case, but rotated and with different intercell hopping. The lattice vectors $\mathbf{a_1}$, $\mathbf{a_2}$, and $\mathbf{a_3}$ are the same as in the breathing ruby lattice case.
We note that the different intercell hopping structure forms a breathing honeycomb or Kekulé lattice \cite{wu_topo_photonic_dielectric_2015,freeney_edge-dependent_2020}. Just as \cref{eq:C6_1}, the primitive generator \cref{eq:C6_2} has a $C_6$ symmetry given by \cref{eq:C6sym}. Again, first consider $t_1 = t_2 = 0$. The spectrum is gapped at $1/2$-filling for $-1/2<t_0<1$, as shown in \cref{fig:C6full}(e), where $t_0 = 1/2$. This gap is characterised by $\chi^{(6)} = (2,0)$, which corresponds to $\mathbf{P^{(6)}} = \mathbf{0}$ and $Q^{(6)} = e/2$. For fixed $t_2 = 0$, the phase diagram in \cref{fig:C6full}(f) is obtained. The system is in its topological phase [$\chi^{(6)} = (2,0)$] for $-1/2<t_0<1$ and $|t_1|< (t_0+2)/3$. At $t_0=1$ or $|t_1|= (t_0+2)/3$, the gap closes at the $\boldsymbol{\Gamma}$ point. For $t_0>1$ and $|t_1|<|t_0|$, the gap reopens in a trivial phase with $\chi^{(6)} = (0,0)$.\\

\noindent \textit{Latent $C_6$-symmetric structures.} The latent primitive generators for $C_6$-symmetry are given by 
\begin{equation}
    h^{(6)}_{L,i}(\kk) = H^{(6)}_{L} + \begin{pmatrix}
        \Tilde{h}^{(6)}_i(\kk) & \emptyset_{6\times 13}\\
        \emptyset_{13\times 6} & \emptyset_{13\times 13}
    \end{pmatrix} \,.
\end{equation}
Here, $\Tilde{h}^{(6)}_i (\kk)$ are the Hamiltonians given in \cref{eq:C6_1}, \cref{eq:C6_2} with all intracell hoppings set to zero, i.e. only intercell hopping (to connect the large unit cells on a lattice).
The two systems corresponding to $h^{(6)}_{L,1}(\kk)$ and $h^{(6)}_{L,2}(\kk)$ are obtained by inserting the latently $C_6$-symmetric unit cell shown in \cref{fig:Latentcells}(d) into a lattice structure to form \cref{fig:C6full}(g) and \cref{fig:C6full}(h), respectively.
The intercell hopping (dashed) is fixed to $1$.
The ISR of $\mathfrak{h}^{(6)}_{L,i}(\kk)$ is given by
\begin{align}
    \mathfrak{h}^{(6)}_{L,i}(\kk) = h^{(6)}_{i}\left(\kk, t_0 = v^{(6)}_0, t_1 = v^{(6)}_1, t_2 = v^{(6)}_2\right)\notag\\ + a^{(6)}\mathbb{I},\label{eq:c6reduced}
\end{align}
which is symmetric under the action of $\hat{C}_6$, i.e. \cref{eq:C6sym}. Since the parameters $a^{(6)}$ and $v^{(6)}_i$ are ratios of large order polynomials in $E$, it is not possible to analytically obtain the phase diagrams of $h^{(6)}_{L,i}(\kk)$. Figure~\ref{fig:C6full}(j) [\cref{fig:C6full}(m)] show the spectrum of $h^{(6)}_{L,1}(\kk)$ [$h^{(6)}_{L,2}(\kk)$] for $t_0 = t_1 = t_2 = 1/4$. A phase diagram of $h^{(6)}_{L,1}(\kk)$ for $4$ filled bands is shown in \cref{fig:C6full}(k), revealing a topological phase characterised by $\chi^{(6)} = (0,2)$ separated by a gapless phase from the trivial phase $\chi^{(6)}=(0,0)$. Moreover, \cref{fig:C6full}(n) shows a phase diagram of $h^{(6)}_{L,2}(\kk)$ for $3$ filled bands. Both phase diagrams reveal the presence of the same topological phases as in $h^{(6)}_i(\kk)$.\\

We once more emphasize that the presence of a hidden (latent) symmetry in the lattices protects these higher-order topological phases. Since these latent symmetries behave much like their geometric counterparts, they also give rise to very similar topological phases. 

\section{Construction of latent HOTIs}\label{sec:corn}
We are now finally equipped with all the necessary tools to allow an efficient design of latent HOTIs.

As mentioned above, any HOTI (geometric or latent) must fulfill the following two constraints:
\begin{itemize}
    \item The corner charge $Q^{(n)}$, as given by \cref{eq:charge}, is nonzero.
    \item The dipole moment $\mathbf{P}^{(n)}$, as given by \cref{eq:dipole}, vanishes.
\end{itemize}
In the following, we will construct latent HOTIs featuring a $C_n$ symmetry. 
Although there are many possibilities, for brevity, we only show a single example of $C_3-$, $C_4-$, and $C_6-$symmetry each. $C_2-$symmetry is left out as constructing a lattice with only two `corners' would correspond to a 1D chain, which is already treated in Ref.~\cite{PhysRevB.110.035106}. Alternatively, a square latent $C_2-$symmetric lattice could be considered, which only shows pairwise equal corner charges.

\begin{figure*}
    \centering
    \includegraphics[width=\textwidth]{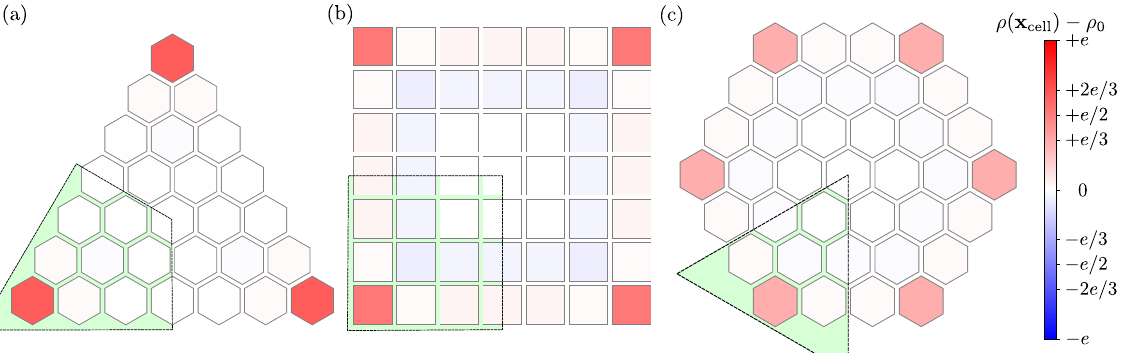}
    \caption{Latent higher-order topological insulators. (a) Triangular OBC flake of $h^{(3)}_L(\kk)$ for $t_0=0.2$ and $g=0.1$ at filling $N_f = 18 N_\text{cells} + 2$. Each of the corners shows a quantized excess corner charge of $Q_\text{corner} = 2e/3$. (b) Square OBC flake of $h^{(4)}_{L,2}(\kk)$ for $t_0=t_1=t_2 = 1/2$ and $t_3 = 1$ at filling $N_f = 3 N_\text{cells} + 2$. Each of the corners shows a quantized excess corner charge of $Q_\text{corner} = e/2$. (c) Hexagonal OBC flake of $h^{(6)}_{L,1}(\kk)$ for $t_0=t_1=t_2 = 1/4$ at filling $N_f = 17 N_\text{cells} + 2$. Each of the corners shows a quantized excess corner charge of $Q_\text{corner} = e/3$.}\label{fig:corner}
\end{figure*}

\paragraph{Latent $C_3$-symmetric HOTI.}
The latent $C_3$-symmetric generators $h^{(3)}_{L,i}(\kk)$ have topological phases with $\chi^{(3)}=(0,2)$ and $\chi^{(3)}=(2,0)$, which corresponds to $\mathbf{P}^{(3)} = (e/3)(\mathbf{a_1}+\mathbf{a_2})$ and $\mathbf{P}^{(3)} = (2e/3)(\mathbf{a_1}+\mathbf{a_2})$, respectively. For neither of these models the dipole moment vanishes. In Sec.~\ref{sec:primgen} we argued that the generators may be stacked to obtain models with arbitrary rotational invariant, and, thus arbitrary $\mathbf{P}^{(n)}$ and $Q^{(n)}$. By `stacking' $h^{(3)}_{L,1}(\kk)$ and $h^{(3)}_{L,2}(\kk)$, we obtain
\begin{equation}
    h^{(3)}_L(\kk) = \begin{pmatrix}
        h^{(3)}_{L,1}(\kk) & \mathcal{T} \\
        \mathcal{T}^\dagger & h^{(3)}_{L,2}(\kk)
    \end{pmatrix},    
\end{equation}
where $\mathcal{T}$ couples the two models without breaking $C_3$-symmetry and without closing the gap. Here, we choose
\begin{equation}
    \mathcal{T} = \begin{pmatrix}
        T & \emptyset_{3\times 6}\\
        \emptyset_{6\times 3} & \emptyset_{6\times 6}
    \end{pmatrix}, \quad T = \begin{pmatrix}
        g & 0 & g \\
        g & g & 0 \\
        0 & g & g
    \end{pmatrix}.
\end{equation}
From \cref{fig:C3full}, we observe that both $h^{(3)}_{L,i}(\kk)$ are topological for $t_0=0.2$ at 8 filled bands. Taking $g=0.1$ does not close the gap, such that $h^{(3)}_L(\kk)$ is characterised by $\chi^{(3)} = (0,2)+(2,0)=(2,2)$. This translates to $\mathbf{P}^{(3)} = \mathbf{0}$ and $Q^{(3)} = 2e/3$. Figure~\ref{fig:corner}(a) shows a triangular flake with OBC corresponding to $h^{(3)}_L(\kk)$, in which every hexagon represents a (stacked) unit cell. The colour of the cells represents $\rho(\mathbf{x}_\text{cell}) - \rho_0$ where
\begin{equation}
    \rho(\mathbf{x}_\text{cell}) = e \sum_{\mathbf{x} \in \mathbf{x}_\text{cell}}   \sum_i^{N_f} |\psi_i(\mathbf{x})|^2,
\end{equation}
is the electronic charge density per unit cell and $N_f$ is the amount of filled states. $\psi_i(\mathbf{x})$ is the wavefunction of the $i^{\text{th}}$ energy eigenstate of the electron. $\rho_0$ is the (ionic) background charge density of the unit cells ($\rho_0 = e \times \# \text{filled bands} \times \# \text{cells}$). In \cref{fig:corner}(a), we take a filling of $16$ bands ($8$ filled bands of each model). There is a clear localisation of excess charge in the three corners of the flake. Adding up the excess charge within a single sector (indicated in green), reveals that the corner charge is quantized and equal to $Q_\text{corner} = 2e/3$. 

\paragraph{Latent $C_4$-symmetric HOTI.}
As an example of $C_4$-symmetric latent HOTI, we may take $h^{(4)}_{L,2}(\kk)$, which in its topological phase [$\chi^{(4)}=(2,0,0)$] at 3 filled bands has $\mathbf{P}^{(4)}=\mathbf{0}$ and $Q^{(4)} = e/2$, i.e., it is not necessary to stack two models. Figure~\ref{fig:corner}(b) shows the charge density in a square flake with OBC, described by $h^{(4)}_{L,2}(\kk)$ for $t_0=t_1=t_2 = 1/2$ and $t_3 = 1$. Summing over a single sector (green) reveals a total corner charge of $Q_\text{corner} = e/2$, predominantly localized at the corners.

\paragraph{Latent $C_6$-symmetric HOTI.}
Finally, from \cref{eq:dipole}, it follows that the dipole moment always vanishes for any $C_6$-symmetric system. Consequently, both primitive generators $h^{(6)}_{L,i}(\kk)$ will represent a HOTI in their topological phase. Here, we consider $h^{(6)}_{L,1}(\kk)$ for $t_0 = t_1 = t_2 = 1/4$ for $17$ filled bands. For these parameters, the system is gapped and characterised by $\chi^{(6)}=(0,2)$, corresponding to $\mathbf{P}^{(6)} = \mathbf{0}$ and $Q^{(6)} = e/3$. Figure~\ref{fig:corner}(c) shows a hexagonal flake of this system, with OBC. It displays corner charges $Q_\text{corner} = e/3$ in each of the six corners of the flake. \\

\begin{figure*}[!hbt]
    \centering
    \includegraphics[width=.75\textwidth]{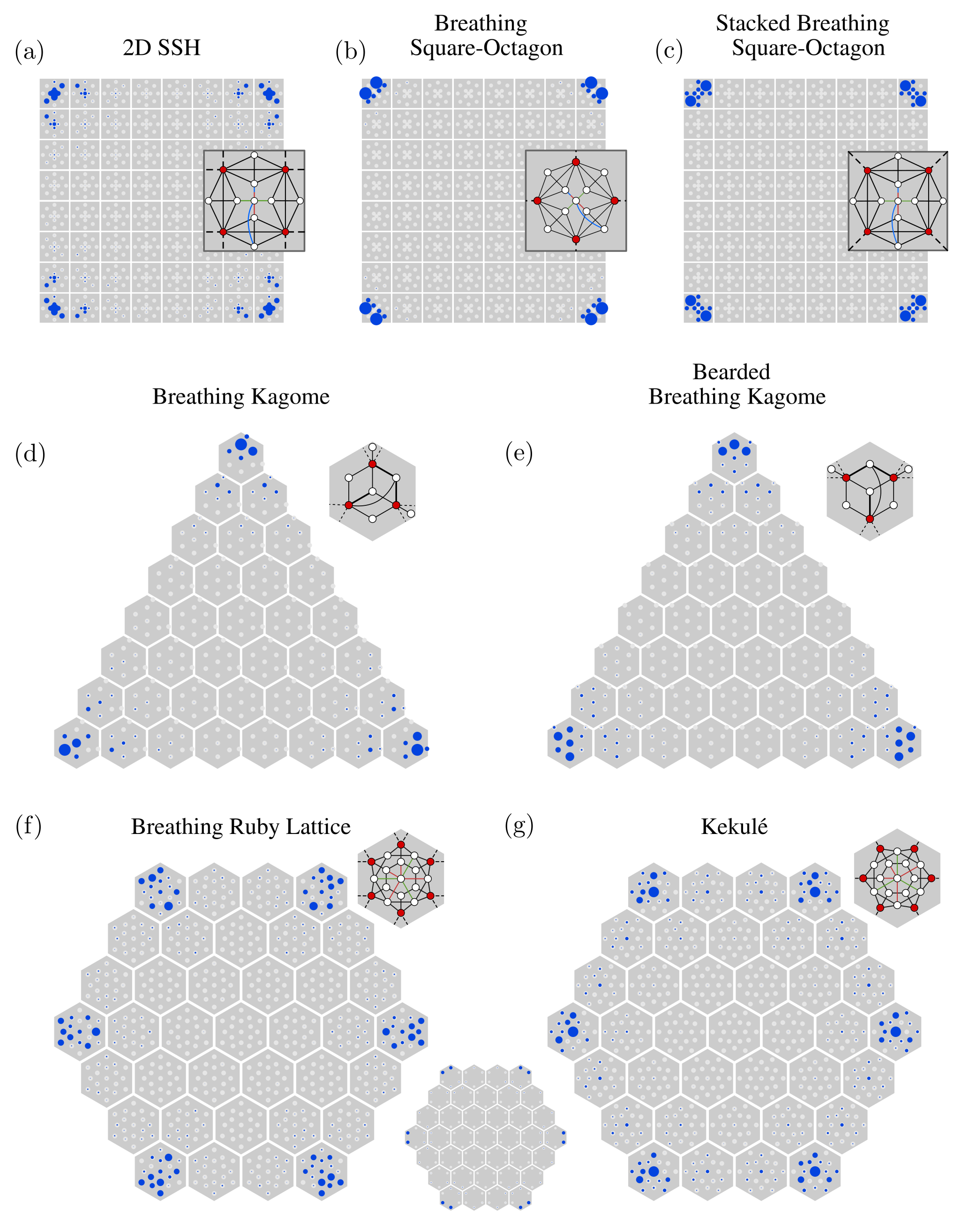}
    \caption{Corner states in the latent lattices considered in Figs.~\ref{fig:C4full}, \ref{fig:C3full}, and \ref{fig:C6full}. Blue circles represent the wavefunction amplitude of the corner states as defined through \cref{eq:cornerWF} while grey circles denote the sites in the unit cell. States are calculated for parameters: (a) $t_0 = t_1 = t_2 = 1$, $t_3 = 0.5$, (b) $t_0 = t_1 = t_2 = 0.25$, $t_3 = 0$, (c) $t_0 = t_1 = t_2 = 0.5$, $t_3 = 1$, (d,e) $t_0 = 0.2$, (f,g) $t_0 = t_1 = t_2 = 0.25$.}
    \label{fig:corner_states}
\end{figure*}

In addition to the charge density distributions of the three examples shown in \cref{fig:corner}, we further present in \cref{fig:corner_states} the corner modes in each of the lattices that we treated. 
In \cref{fig:corner}, we displayed the quantized fractional corner charges through the quantity $\rho(\mathbf{x}_\text{cell}) - \rho_0$. We now show the states themselves. For a $C_n$-symmetric system, there are $n$ corner states. In \cref{fig:corner_states}, the size of the blue circles represents the combined amplitude of one set of corner states, i.e.
\begin{equation}\label{eq:cornerWF}
    v(\mathbf{x}) = \sum_{i = 1}^n |\psi^\text{corner}_i(\mathbf{x})|^2,
\end{equation}
where $n = 2, 3, 4, 6$ for $C_2-, C_3-, C_4-, C_6-$symmetry, respectively.\\

A close inspection of the latent $C_3-$ and $C_6-$symmetric lattices [Figs.~\ref{fig:corner_states}(d-g)] leads to the conclusion that these states do not have a rotation symmetry. Indeed, this can be explained by the fact that these states are protected by a latent symmetry instead of a conventional symmetry. Moreover, when only considering the support of the wavefunction on the $S$ sites (red sites in the insets), the corner states look indeed symmetric. This is verified by the smaller plot next to \cref{fig:corner_states}(f), which shows the same wavefunction but only on $S$.

\section{Topological robustness}\label{sec:robust}
\begin{figure*}[!hbt]
    \centering
    \includegraphics[width=0.9\linewidth]{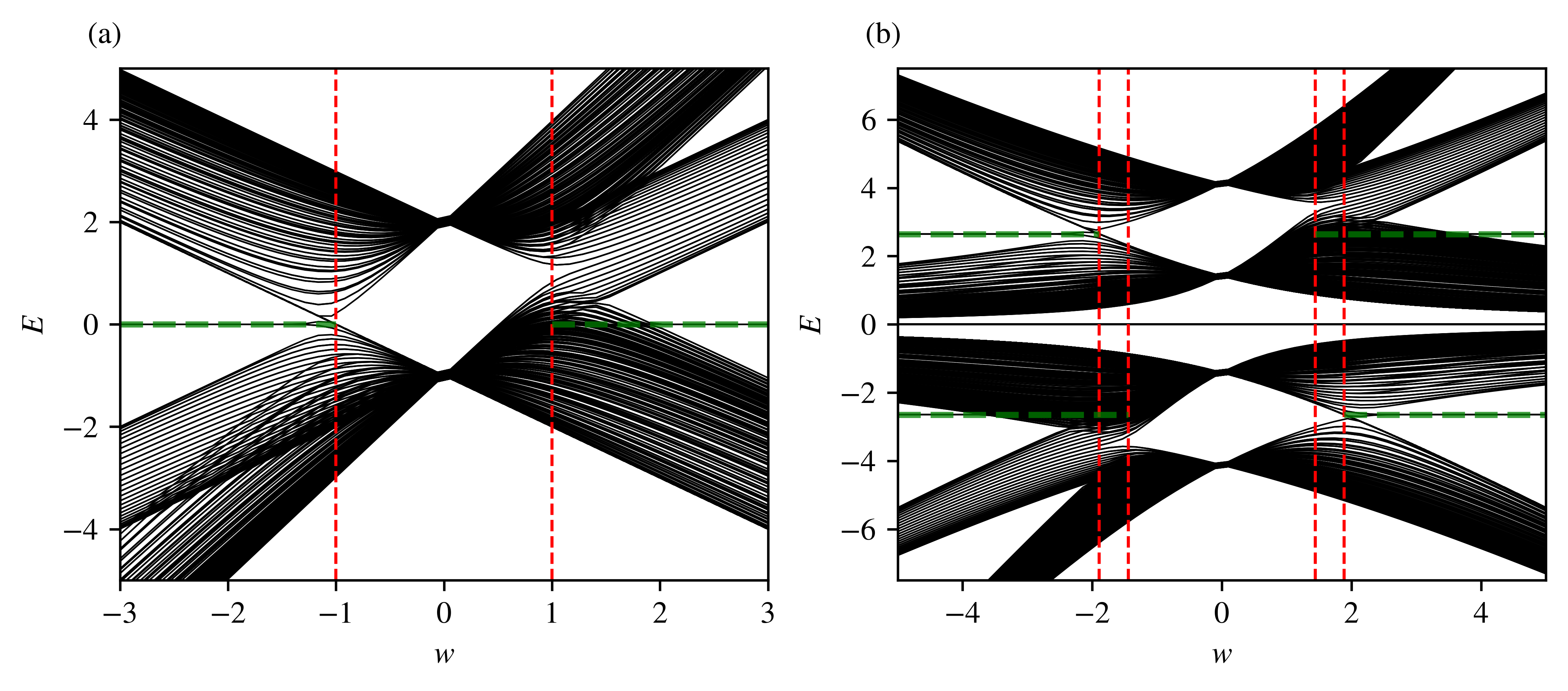}
    \caption{OBC spectra of (a) the geometric and (b) latent breathing kagome model on a triangular flake geometry as a function of intercell hopping $w$ (intracell hopping $t_0 = 1$). Red lines indicate values of $w$ for which a topological phase transition occurs, while green lines mark the topological corner states. The flat band at $E=0$ is not due to latent symmetries; its origin is discussed in \cref{App:flatBand}.
    }
    \label{fig:OBCKagome}
\end{figure*}

In this section, we further investigate the robustness of the latent higher-order topological phases compared to their conventional counterparts. As a working example, we choose the (latent) breathing Kagome lattice. Before including disorder, we show in Figs.~\ref{fig:OBCKagome}(a) and \ref{fig:OBCKagome}(b) the OBC spectra of a triangular flake of the geometric and latent breathing Kagome lattice, respectively. The spectrum is plotted as a function of the intercell hopping parameter $w$, which was previously fixed to $1$.
The vertical red dashed lines indicate the value of $w$ for which a topological phase transition occurs through a bulk gap closing. The horizontal green dashed lines indicate the presence of a topological corner mode. Notice that in \cref{fig:OBCKagome}(b), the corner states do not lie at zero energy, and can sometimes be hidden by the bulk-bands. 

\subsection{Onsite disorder}

\begin{figure*}[!hbt]
    \centering
    \includegraphics[width=0.8\textwidth]{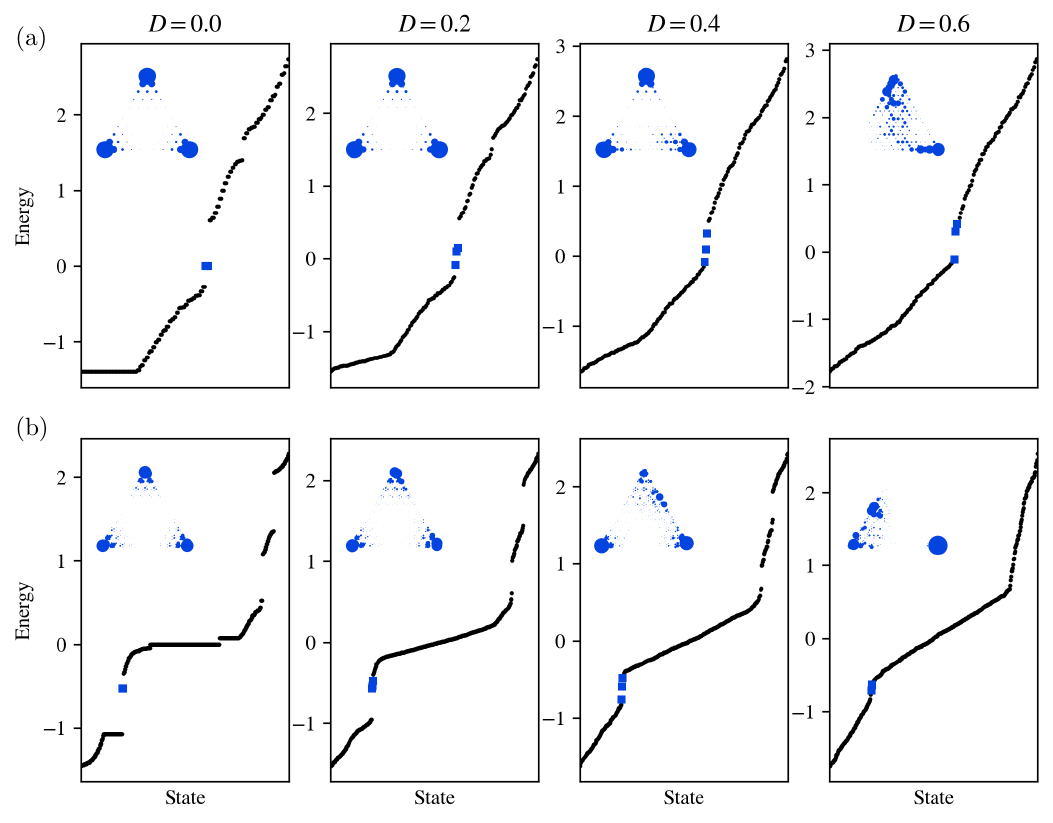}
    \caption{Open boundary condition flake of (a) the geometric Kagome model for $t_0 = 0.4$ and (b) the latent Kagome model for $t_0 = 0.2$, at different disorder strength $D$. Insets show the (combined) wavefunction amplitudes of the three corner states indicated by the blue squares.}
    \label{fig:geokagodisorder}
\end{figure*}

The systems considered in this work are protected by latent/geometric symmetries, which offer weaker protection than spectral symmetries. Consequently, most perturbations break the (latent) $C_n$ symmetries that protect the crystalline topological phases. If the perturbation is sufficiently small, the corner modes may persist, although their degeneracy will be lifted by a small amount. To demonstrate this, we consider random onsite disorder, i.e., a perturbation Hamiltonian of the form 
\begin{equation}
    H_\text{disorder} = \sum_{i} D_i^{} c^\dagger_i c_i^{},
\end{equation}
where the disorder strength $D_i$ is sampled from a uniform distribution $[-D,D]$. Figure~\ref{fig:geokagodisorder} shows the spectra and corner-state wavefunctions for an open boundary conditions flake of the geometric and latent kagome model at different disorder strengths $D$.
The introduction of onsite disorder immediately lifts the degeneracy of the corner states. However, if the disorder is sufficiently small, they will still be pinned to the corners. If the disorder strength becomes large enough, in this case at $D\approx 0.6$, the corner states cease to exist, as seen in the right panels of Figs.~\ref{fig:geokagodisorder}(a) and \ref{fig:geokagodisorder}(b). 

\subsection{Orientational disorder}
\begin{figure*}
    \centering
    \includegraphics[width=0.9\linewidth]{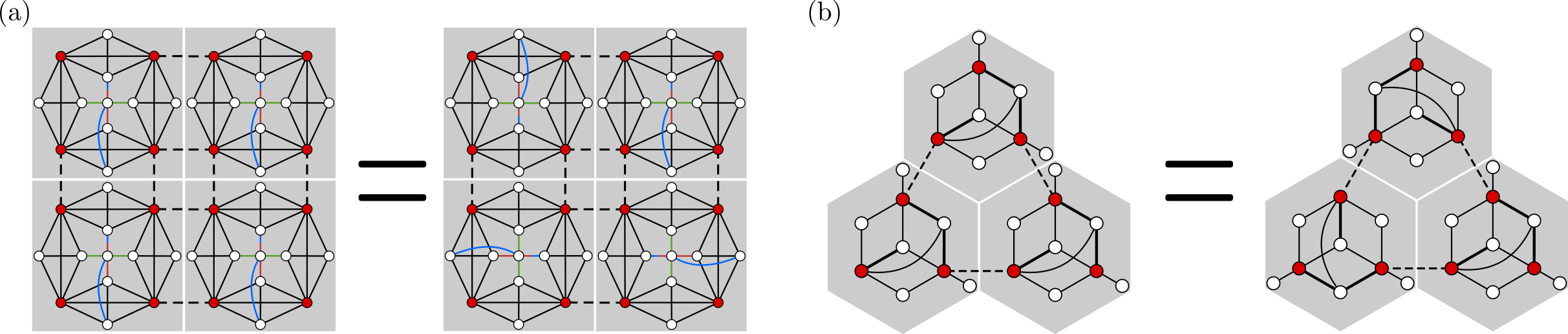}
    \caption{Examples of a latently symmetric lattice with (a) latent $C_4$ and (b) latent $C_6$ symmetry together with examples of orientational disorder.}
    \label{fig:orientation}
\end{figure*}

The latent lattices allow for a type of disorder exclusive to systems with latent symmetries, namely, orientational disorder \cite{PhysRevB.110.035106}. It is created by taking the latently symmetric unit cells in a lattice and rotating them by their respective symmetry, as seen in \cref{fig:orientation}. Since rotating the unit cell does not affect the isospectral reduction of a lattice, a lattice with orientational disorder has the same spectral properties as a lattice with no orientational disorder. Consequently, the behavior of such a lattice is unperturbed.

\section{Conclusion} \label{Sec: Conclusion}
In this work, we have demonstrated that the presence of non-trivial fractional corner charges in 2D systems does not require the explicit preservation of \(C_n\)-symmetry. Instead, these features can arise due to a latent symmetry, which manifests through the operator \(\hat{C}_n \oplus \overline{Q}^{(n)}\). While the protection of the topological phase is based on the existence of the symmetry $Q$, the $C_n$ component of this symmetry is able to entirely characterize the topological classification of these systems. At high-symmetry points, this latent symmetry behaves identically to a conventional \(C_n\)-symmetry, satisfying $[h(\boldsymbol{\Pi}), \hat{C}_n \oplus \overline{Q}^{(n)}]_- = 0$. As a result, any 2D system possessing this form of latent symmetry can support non-trivial corner modes and be described in terms of rotation invariants. To our knowledge, this is the first extension of higher-order topological phases to non-geometrical discrete symmetries.

In Section~\ref{sec:generators}, we explored examples of primitive generators that can be used to construct models with higher-order topological phases protected by latent symmetries. A significant advantage of such latent symmetries lies in their ability to simplify the classification of topological phases through symmetry indicators. This revealed another strength of our framework: if a Hamiltonian, under isospectral reduction, transforms into an energy-dependent version of a well-studied model, the properties of the latter can serve to characterize the former. This concept was formalized at the conclusion of \cref{sec:anomaly}.

We provided explicit examples of unit cells exhibiting latent $C_n$-symmetry for the four primary rotational symmetries that tile two-dimensional space. These serve as the foundation for constructing latent HOTIs. However, our approach is not limited to these unit cells. Any cell obeying a latent $C_n$-symmetry can be employed to construct a latent HOTI. The unit cells highlighted in this work were chosen merely to illustrate specific instances of latent HOTIs. These examples, while illustrative, do not encompass the entire scope of potential latent HOTIs. For instance, two primitive generators based on unit cells—whether geometric- or latent-symmetric—can be combined to create a new generator. This opens the possibility of constructing HOTIs by mixing geometric and latent generators.

The specific models discussed in this work—namely the 2D SSH model, the breathing square-octagon lattice, T-graphene, the breathing kagome lattice, the breathing ruby lattice, and the Kekulé lattice—are all models previously studied in the literature in various contexts \cite{Liu2019R20191TopologicallyProtectedEdgeState,Coutant2021JoAP129125108TopologicalTwodimensionalSuSchriefferHeegerAnalog,PhysRevB.108.245140,dai_electronic_2014,Nunes2020PRB101224514FlatbandSuperconductivityTightbindingElectrons,Yan_2023, kempkes_robust_2019,van_miert_topological_2020,herrera_corner_2022,freeney_edge-dependent_2020}. We believe that the results presented here are directly relevant to these systems. The references include prior investigations into topologically protected edge states, flat-band phenomena, and corner states. These connections suggest that the latent symmetry framework might find immediate applications in interpreting known phenomena within these systems.

While this work focuses on HOTIs protected by $C_n$-symmetry, it is important to note that this constitutes only a subset of all crystalline topological phases. The methods outlined here can be extended to systems with other crystalline symmetries. We believe this approach can be used as a starting point for investigating latent generalizations of topological phases protected by different crystalline symmetries. A particularly intriguing direction for future work is the study of latent versions of topological systems that lack translational symmetry but retain global space-group symmetries. For instance, this could involve generalizing recent findings on 1D topological quasicrystalline insulators \cite{moustaj2024anomalous} or crystalline extensions of topological states observed in finite fractal systems \cite{canyellas2023topological, frachald}.

Finally, the generalization of this framework to third-order topological insulators in 3D presents an exciting challenge. Although there is no direct 3D analogue to the theory discussed in Ref.~\cite{benalcazar_quantization_2019}, our approach suggests a potential avenue. Specifically, by employing similar methods to construct unit cells as those used here, it may be possible to identify a latent version of the unit cell in the pyrochlore lattice \cite{ezawa_higher-order_2018}. The latter is already known to host third-order topological phases with quantized fractional charges. Exploring such systems could significantly expand the reach of latent symmetry-based topological classifications.

\begin{acknowledgments}
L.E. and C.M.S. acknowledge the research program “Materials for the Quantum Age” (QuMat) for financial support. This program (registration number 024.005.006) is part of the Gravitation program financed by the Dutch Ministry of Education, Culture and Science (OCW).
\end{acknowledgments}
		
\bibliography{referencesZotD.bib,referencesD.bib,Bibtex}

\begin{thebibliography}{47}%
\makeatletter
\providecommand \@ifxundefined [1]{%
 \@ifx{#1\undefined}
}%
\providecommand \@ifnum [1]{%
 \ifnum #1\expandafter \@firstoftwo
 \else \expandafter \@secondoftwo
 \fi
}%
\providecommand \@ifx [1]{%
 \ifx #1\expandafter \@firstoftwo
 \else \expandafter \@secondoftwo
 \fi
}%
\providecommand \natexlab [1]{#1}%
\providecommand \enquote  [1]{``#1''}%
\providecommand \bibnamefont  [1]{#1}%
\providecommand \bibfnamefont [1]{#1}%
\providecommand \citenamefont [1]{#1}%
\providecommand \href@noop [0]{\@secondoftwo}%
\providecommand \href [0]{\begingroup \@sanitize@url \@href}%
\providecommand \@href[1]{\@@startlink{#1}\@@href}%
\providecommand \@@href[1]{\endgroup#1\@@endlink}%
\providecommand \@sanitize@url [0]{\catcode `\\12\catcode `\$12\catcode `\&12\catcode `\#12\catcode `\^12\catcode `\_12\catcode `\%12\relax}%
\providecommand \@@startlink[1]{}%
\providecommand \@@endlink[0]{}%
\providecommand \url  [0]{\begingroup\@sanitize@url \@url }%
\providecommand \@url [1]{\endgroup\@href {#1}{\urlprefix }}%
\providecommand \urlprefix  [0]{URL }%
\providecommand \Eprint [0]{\href }%
\providecommand \doibase [0]{https://doi.org/}%
\providecommand \selectlanguage [0]{\@gobble}%
\providecommand \bibinfo  [0]{\@secondoftwo}%
\providecommand \bibfield  [0]{\@secondoftwo}%
\providecommand \translation [1]{[#1]}%
\providecommand \BibitemOpen [0]{}%
\providecommand \bibitemStop [0]{}%
\providecommand \bibitemNoStop [0]{.\EOS\space}%
\providecommand \EOS [0]{\spacefactor3000\relax}%
\providecommand \BibitemShut  [1]{\csname bibitem#1\endcsname}%
\let\auto@bib@innerbib\@empty
\bibitem [{\citenamefont {Altland}\ and\ \citenamefont {Zirnbauer}(1997)}]{altland_nonstandard_1997}%
  \BibitemOpen
  \bibfield  {author} {\bibinfo {author} {\bibfnamefont {A.}~\bibnamefont {Altland}}\ and\ \bibinfo {author} {\bibfnamefont {M.~R.}\ \bibnamefont {Zirnbauer}},\ }\bibfield  {title} {\bibinfo {title} {Nonstandard symmetry classes in mesoscopic normal-superconducting hybrid structures},\ }\href {https://doi.org/10.1103/PhysRevB.55.1142} {\bibfield  {journal} {\bibinfo  {journal} {Phys. Rev. B}\ }\textbf {\bibinfo {volume} {55}},\ \bibinfo {pages} {1142} (\bibinfo {year} {1997})}\BibitemShut {NoStop}%
\bibitem [{\citenamefont {Ryu}\ \emph {et~al.}(2010)\citenamefont {Ryu}, \citenamefont {Schnyder}, \citenamefont {Furusaki},\ and\ \citenamefont {Ludwig}}]{ryu_topological_2010}%
  \BibitemOpen
  \bibfield  {author} {\bibinfo {author} {\bibfnamefont {S.}~\bibnamefont {Ryu}}, \bibinfo {author} {\bibfnamefont {A.~P.}\ \bibnamefont {Schnyder}}, \bibinfo {author} {\bibfnamefont {A.}~\bibnamefont {Furusaki}},\ and\ \bibinfo {author} {\bibfnamefont {A.~W.~W.}\ \bibnamefont {Ludwig}},\ }\bibfield  {title} {{\selectlanguage {en}\bibinfo {title} {Topological insulators and superconductors: tenfold way and dimensional hierarchy}},\ }\href {https://doi.org/10.1088/1367-2630/12/6/065010} {\bibfield  {journal} {\bibinfo  {journal} {New J. Phys.}\ }\textbf {\bibinfo {volume} {12}},\ \bibinfo {pages} {065010} (\bibinfo {year} {2010})}\BibitemShut {NoStop}%
\bibitem [{\citenamefont {Roy}\ and\ \citenamefont {Harper}(2017)}]{roy_periodic_2017}%
  \BibitemOpen
  \bibfield  {author} {\bibinfo {author} {\bibfnamefont {R.}~\bibnamefont {Roy}}\ and\ \bibinfo {author} {\bibfnamefont {F.}~\bibnamefont {Harper}},\ }\bibfield  {title} {\bibinfo {title} {Periodic table for {Floquet} topological insulators},\ }\href {https://doi.org/10.1103/PhysRevB.96.155118} {\bibfield  {journal} {\bibinfo  {journal} {Phys. Rev. B}\ }\textbf {\bibinfo {volume} {96}},\ \bibinfo {pages} {155118} (\bibinfo {year} {2017})}\BibitemShut {NoStop}%
\bibitem [{\citenamefont {Li}\ \emph {et~al.}(2009)\citenamefont {Li}, \citenamefont {Chu}, \citenamefont {Jain},\ and\ \citenamefont {Shen}}]{li_topological_2009}%
  \BibitemOpen
  \bibfield  {author} {\bibinfo {author} {\bibfnamefont {J.}~\bibnamefont {Li}}, \bibinfo {author} {\bibfnamefont {R.-L.}\ \bibnamefont {Chu}}, \bibinfo {author} {\bibfnamefont {J.~K.}\ \bibnamefont {Jain}},\ and\ \bibinfo {author} {\bibfnamefont {S.-Q.}\ \bibnamefont {Shen}},\ }\bibfield  {title} {\bibinfo {title} {Topological {Anderson} {Insulator}},\ }\href {https://doi.org/10.1103/PhysRevLett.102.136806} {\bibfield  {journal} {\bibinfo  {journal} {Phys. Rev. Lett.}\ }\textbf {\bibinfo {volume} {102}},\ \bibinfo {pages} {136806} (\bibinfo {year} {2009})}\BibitemShut {NoStop}%
\bibitem [{\citenamefont {Kawabata}\ \emph {et~al.}(2019)\citenamefont {Kawabata}, \citenamefont {Shiozaki}, \citenamefont {Ueda},\ and\ \citenamefont {Sato}}]{kawabata_symmetry_2019}%
  \BibitemOpen
  \bibfield  {author} {\bibinfo {author} {\bibfnamefont {K.}~\bibnamefont {Kawabata}}, \bibinfo {author} {\bibfnamefont {K.}~\bibnamefont {Shiozaki}}, \bibinfo {author} {\bibfnamefont {M.}~\bibnamefont {Ueda}},\ and\ \bibinfo {author} {\bibfnamefont {M.}~\bibnamefont {Sato}},\ }\bibfield  {title} {\bibinfo {title} {Symmetry and {Topology} in {Non}-{Hermitian} {Physics}},\ }\href {https://doi.org/10.1103/PhysRevX.9.041015} {\bibfield  {journal} {\bibinfo  {journal} {Phys. Rev. X}\ }\textbf {\bibinfo {volume} {9}},\ \bibinfo {pages} {041015} (\bibinfo {year} {2019})}\BibitemShut {NoStop}%
\bibitem [{\citenamefont {Fu}(2011)}]{fu_topological_2011}%
  \BibitemOpen
  \bibfield  {author} {\bibinfo {author} {\bibfnamefont {L.}~\bibnamefont {Fu}},\ }\bibfield  {title} {{\selectlanguage {en}\bibinfo {title} {Topological {Crystalline} {Insulators}}},\ }\href {https://doi.org/10.1103/PhysRevLett.106.106802} {\bibfield  {journal} {\bibinfo  {journal} {Phys. Rev. Lett.}\ }\textbf {\bibinfo {volume} {106}},\ \bibinfo {pages} {106802} (\bibinfo {year} {2011})}\BibitemShut {NoStop}%
\bibitem [{\citenamefont {Fang}\ and\ \citenamefont {Fu}(2015)}]{fang_new_2015}%
  \BibitemOpen
  \bibfield  {author} {\bibinfo {author} {\bibfnamefont {C.}~\bibnamefont {Fang}}\ and\ \bibinfo {author} {\bibfnamefont {L.}~\bibnamefont {Fu}},\ }\bibfield  {title} {\bibinfo {title} {New classes of three-dimensional topological crystalline insulators: {Nonsymmorphic} and magnetic},\ }\href {https://doi.org/10.1103/PhysRevB.91.161105} {\bibfield  {journal} {\bibinfo  {journal} {Phys. Rev. B}\ }\textbf {\bibinfo {volume} {91}},\ \bibinfo {pages} {161105} (\bibinfo {year} {2015})}\BibitemShut {NoStop}%
\bibitem [{\citenamefont {Slager}\ \emph {et~al.}(2013)\citenamefont {Slager}, \citenamefont {Mesaros}, \citenamefont {Juričić},\ and\ \citenamefont {Zaanen}}]{slager_space_2013}%
  \BibitemOpen
  \bibfield  {author} {\bibinfo {author} {\bibfnamefont {R.-J.}\ \bibnamefont {Slager}}, \bibinfo {author} {\bibfnamefont {A.}~\bibnamefont {Mesaros}}, \bibinfo {author} {\bibfnamefont {V.}~\bibnamefont {Juričić}},\ and\ \bibinfo {author} {\bibfnamefont {J.}~\bibnamefont {Zaanen}},\ }\bibfield  {title} {{\selectlanguage {en}\bibinfo {title} {The space group classification of topological band-insulators}},\ }\href {https://doi.org/10.1038/nphys2513} {\bibfield  {journal} {\bibinfo  {journal} {Nat. Phys.}\ }\textbf {\bibinfo {volume} {9}},\ \bibinfo {pages} {98} (\bibinfo {year} {2013})}\BibitemShut {NoStop}%
\bibitem [{\citenamefont {Zhang}\ \emph {et~al.}(2023)\citenamefont {Zhang}, \citenamefont {Chen}, \citenamefont {Zhang},\ and\ \citenamefont {Zhao}}]{zhang_general_2023}%
  \BibitemOpen
  \bibfield  {author} {\bibinfo {author} {\bibfnamefont {C.}~\bibnamefont {Zhang}}, \bibinfo {author} {\bibfnamefont {Z.~Y.}\ \bibnamefont {Chen}}, \bibinfo {author} {\bibfnamefont {Z.}~\bibnamefont {Zhang}},\ and\ \bibinfo {author} {\bibfnamefont {Y.~X.}\ \bibnamefont {Zhao}},\ }\bibfield  {title} {\bibinfo {title} {General {Theory} of {Momentum}-{Space} {Nonsymmorphic} {Symmetry}},\ }\href {https://doi.org/10.1103/PhysRevLett.130.256601} {\bibfield  {journal} {\bibinfo  {journal} {Phys. Rev. Lett.}\ }\textbf {\bibinfo {volume} {130}},\ \bibinfo {pages} {256601} (\bibinfo {year} {2023})}\BibitemShut {NoStop}%
\bibitem [{\citenamefont {Benalcazar}\ \emph {et~al.}(2017)\citenamefont {Benalcazar}, \citenamefont {Bernevig},\ and\ \citenamefont {Hughes}}]{benalcazar_electric_2017}%
  \BibitemOpen
  \bibfield  {author} {\bibinfo {author} {\bibfnamefont {W.~A.}\ \bibnamefont {Benalcazar}}, \bibinfo {author} {\bibfnamefont {B.~A.}\ \bibnamefont {Bernevig}},\ and\ \bibinfo {author} {\bibfnamefont {T.~L.}\ \bibnamefont {Hughes}},\ }\bibfield  {title} {{\selectlanguage {en}\bibinfo {title} {Electric multipole moments, topological multipole moment pumping, and chiral hinge states in crystalline insulators}},\ }\href {https://doi.org/10.1103/PhysRevB.96.245115} {\bibfield  {journal} {\bibinfo  {journal} {Phys. Rev. B}\ }\textbf {\bibinfo {volume} {96}},\ \bibinfo {pages} {245115} (\bibinfo {year} {2017})}\BibitemShut {NoStop}%
\bibitem [{\citenamefont {Vanderbilt}\ and\ \citenamefont {King-Smith}(1993)}]{vanderbilt_electric_1993}%
  \BibitemOpen
  \bibfield  {author} {\bibinfo {author} {\bibfnamefont {D.}~\bibnamefont {Vanderbilt}}\ and\ \bibinfo {author} {\bibfnamefont {R.~D.}\ \bibnamefont {King-Smith}},\ }\bibfield  {title} {\bibinfo {title} {Electric polarization as a bulk quantity and its relation to surface charge},\ }\href {https://doi.org/10.1103/PhysRevB.48.4442} {\bibfield  {journal} {\bibinfo  {journal} {Phys. Rev. B}\ }\textbf {\bibinfo {volume} {48}},\ \bibinfo {pages} {4442} (\bibinfo {year} {1993})}\BibitemShut {NoStop}%
\bibitem [{\citenamefont {King-Smith}\ and\ \citenamefont {Vanderbilt}(1993)}]{king-smith_theory_1993}%
  \BibitemOpen
  \bibfield  {author} {\bibinfo {author} {\bibfnamefont {R.~D.}\ \bibnamefont {King-Smith}}\ and\ \bibinfo {author} {\bibfnamefont {D.}~\bibnamefont {Vanderbilt}},\ }\bibfield  {title} {\bibinfo {title} {Theory of polarization of crystalline solids},\ }\href {https://doi.org/10.1103/PhysRevB.47.1651} {\bibfield  {journal} {\bibinfo  {journal} {Phys. Rev. B}\ }\textbf {\bibinfo {volume} {47}},\ \bibinfo {pages} {1651} (\bibinfo {year} {1993})}\BibitemShut {NoStop}%
\bibitem [{\citenamefont {van Miert}\ and\ \citenamefont {Ortix}(2018)}]{MiertRot}%
  \BibitemOpen
  \bibfield  {author} {\bibinfo {author} {\bibfnamefont {G.}~\bibnamefont {van Miert}}\ and\ \bibinfo {author} {\bibfnamefont {C.}~\bibnamefont {Ortix}},\ }\bibfield  {title} {\bibinfo {title} {Higher-order topological insulators protected by inversion and rotoinversion symmetries},\ }\href {https://doi.org/10.1103/PhysRevB.98.081110} {\bibfield  {journal} {\bibinfo  {journal} {Phys. Rev. B}\ }\textbf {\bibinfo {volume} {98}},\ \bibinfo {pages} {081110} (\bibinfo {year} {2018})}\BibitemShut {NoStop}%
\bibitem [{\citenamefont {Benalcazar}\ \emph {et~al.}(2019)\citenamefont {Benalcazar}, \citenamefont {Li},\ and\ \citenamefont {Hughes}}]{benalcazar_quantization_2019}%
  \BibitemOpen
  \bibfield  {author} {\bibinfo {author} {\bibfnamefont {W.~A.}\ \bibnamefont {Benalcazar}}, \bibinfo {author} {\bibfnamefont {T.}~\bibnamefont {Li}},\ and\ \bibinfo {author} {\bibfnamefont {T.~L.}\ \bibnamefont {Hughes}},\ }\bibfield  {title} {\bibinfo {title} {Quantization of fractional corner charge in {C} n -symmetric higher-order topological crystalline insulators},\ }\href {https://doi.org/10.1103/PhysRevB.99.245151} {\bibfield  {journal} {\bibinfo  {journal} {Phys. Rev. B}\ }\textbf {\bibinfo {volume} {99}},\ \bibinfo {pages} {245151} (\bibinfo {year} {2019})}\BibitemShut {NoStop}%
\bibitem [{\citenamefont {Smith}\ and\ \citenamefont {Webb}(2019)}]{Smith2019PA514855HiddenSymmetriesRealTheoretical}%
  \BibitemOpen
  \bibfield  {author} {\bibinfo {author} {\bibfnamefont {D.}~\bibnamefont {Smith}}\ and\ \bibinfo {author} {\bibfnamefont {B.}~\bibnamefont {Webb}},\ }\bibfield  {title} {\bibinfo {title} {Hidden symmetries in real and theoretical networks},\ }\href {https://doi.org/10.1016/j.physa.2018.09.131} {\bibfield  {journal} {\bibinfo  {journal} {Physica A}\ }\textbf {\bibinfo {volume} {514}},\ \bibinfo {pages} {855} (\bibinfo {year} {2019})}\BibitemShut {NoStop}%
\bibitem [{\citenamefont {R{\"o}ntgen}(2021)}]{Rontgen2021GraphtheoreticalAnalysisLocalLatent}%
  \BibitemOpen
  \bibfield  {author} {\bibinfo {author} {\bibfnamefont {M.}~\bibnamefont {R{\"o}ntgen}},\ }\emph {\bibinfo {title} {Graph-Theoretical Analysis of Local and Latent Symmetries in Physical Systems}},\ \href {https://ediss.sub.uni-hamburg.de/handle/ediss/9579} {Ph.D. thesis},\ \bibinfo  {school} {Staats- und Universit{\"a}tsbibliothek Hamburg Carl von Ossietzky} (\bibinfo {year} {2021})\BibitemShut {NoStop}%
\bibitem [{\citenamefont {Bunimovich}\ and\ \citenamefont {Webb}(2014)}]{Bunimovich2014IsospectralTransformationsNewApproach}%
  \BibitemOpen
  \bibfield  {author} {\bibinfo {author} {\bibfnamefont {L.}~\bibnamefont {Bunimovich}}\ and\ \bibinfo {author} {\bibfnamefont {B.}~\bibnamefont {Webb}},\ }\href {http://ebookcentral.proquest.com/lib/subhh/detail.action?docID=1967797} {\emph {\bibinfo {title} {Isospectral Transformations: A New Approach to Analyzing Multidimensional Systems and Networks}}},\ \bibinfo {edition} {1st}\ ed.,\ Springer {{Monographs}} in {{Mathematics}}\ (\bibinfo  {publisher} {Springer},\ \bibinfo {address} {New York, NY, United States},\ \bibinfo {year} {2014})\BibitemShut {NoStop}%
\bibitem [{\citenamefont {Kempton}\ \emph {et~al.}(2020)\citenamefont {Kempton}, \citenamefont {Sinkovic}, \citenamefont {Smith},\ and\ \citenamefont {Webb}}]{Kempton2020LAIA594226CharacterizingCospectralVerticesIsospectral}%
  \BibitemOpen
  \bibfield  {author} {\bibinfo {author} {\bibfnamefont {M.}~\bibnamefont {Kempton}}, \bibinfo {author} {\bibfnamefont {J.}~\bibnamefont {Sinkovic}}, \bibinfo {author} {\bibfnamefont {D.}~\bibnamefont {Smith}},\ and\ \bibinfo {author} {\bibfnamefont {B.}~\bibnamefont {Webb}},\ }\bibfield  {title} {\bibinfo {title} {Characterizing cospectral vertices via isospectral reduction},\ }\href {https://doi.org/10.1016/j.laa.2020.02.020} {\bibfield  {journal} {\bibinfo  {journal} {Linear Algebra Its Appl.}\ }\textbf {\bibinfo {volume} {594}},\ \bibinfo {pages} {226} (\bibinfo {year} {2020})}\BibitemShut {NoStop}%
\bibitem [{\citenamefont {Morfonios}\ \emph {et~al.}(2021)\citenamefont {Morfonios}, \citenamefont {Röntgen}, \citenamefont {Pyzh},\ and\ \citenamefont {Schmelcher}}]{morfonios_flat_2021}%
  \BibitemOpen
  \bibfield  {author} {\bibinfo {author} {\bibfnamefont {C.~V.}\ \bibnamefont {Morfonios}}, \bibinfo {author} {\bibfnamefont {M.}~\bibnamefont {Röntgen}}, \bibinfo {author} {\bibfnamefont {M.}~\bibnamefont {Pyzh}},\ and\ \bibinfo {author} {\bibfnamefont {P.}~\bibnamefont {Schmelcher}},\ }\bibfield  {title} {\bibinfo {title} {Flat bands by latent symmetry},\ }\href {https://doi.org/10.1103/PhysRevB.104.035105} {\bibfield  {journal} {\bibinfo  {journal} {Phys. Rev. B}\ }\textbf {\bibinfo {volume} {104}},\ \bibinfo {pages} {035105} (\bibinfo {year} {2021})}\BibitemShut {NoStop}%
\bibitem [{\citenamefont {R{\"o}ntgen}\ \emph {et~al.}(2021)\citenamefont {R{\"o}ntgen}, \citenamefont {Pyzh}, \citenamefont {Morfonios}, \citenamefont {Palaiodimopoulos}, \citenamefont {Diakonos},\ and\ \citenamefont {Schmelcher}}]{Rontgen2021PRL126180601LatentSymmetryInducedDegeneracies}%
  \BibitemOpen
  \bibfield  {author} {\bibinfo {author} {\bibfnamefont {M.}~\bibnamefont {R{\"o}ntgen}}, \bibinfo {author} {\bibfnamefont {M.}~\bibnamefont {Pyzh}}, \bibinfo {author} {\bibfnamefont {C.~V.}\ \bibnamefont {Morfonios}}, \bibinfo {author} {\bibfnamefont {N.~E.}\ \bibnamefont {Palaiodimopoulos}}, \bibinfo {author} {\bibfnamefont {F.~K.}\ \bibnamefont {Diakonos}},\ and\ \bibinfo {author} {\bibfnamefont {P.}~\bibnamefont {Schmelcher}},\ }\bibfield  {title} {\bibinfo {title} {Latent symmetry induced degeneracies},\ }\href {https://doi.org/10.1103/PhysRevLett.126.180601} {\bibfield  {journal} {\bibinfo  {journal} {Phys. Rev. Lett.}\ }\textbf {\bibinfo {volume} {126}},\ \bibinfo {pages} {180601} (\bibinfo {year} {2021})}\BibitemShut {NoStop}%
\bibitem [{\citenamefont {Su}\ \emph {et~al.}(1979)\citenamefont {Su}, \citenamefont {Schrieffer},\ and\ \citenamefont {Heeger}}]{Su1979PRL421698SolitonsPolyacetylene}%
  \BibitemOpen
  \bibfield  {author} {\bibinfo {author} {\bibfnamefont {W.~P.}\ \bibnamefont {Su}}, \bibinfo {author} {\bibfnamefont {J.~R.}\ \bibnamefont {Schrieffer}},\ and\ \bibinfo {author} {\bibfnamefont {A.~J.}\ \bibnamefont {Heeger}},\ }\bibfield  {title} {\bibinfo {title} {Solitons in polyacetylene},\ }\href {https://doi.org/10.1103/PhysRevLett.42.1698} {\bibfield  {journal} {\bibinfo  {journal} {Phys. Rev. Lett.}\ }\textbf {\bibinfo {volume} {42}},\ \bibinfo {pages} {1698} (\bibinfo {year} {1979})}\BibitemShut {NoStop}%
\bibitem [{\citenamefont {R\"ontgen}\ \emph {et~al.}(2024)\citenamefont {R\"ontgen}, \citenamefont {Chen}, \citenamefont {Gao}, \citenamefont {Pyzh}, \citenamefont {Schmelcher}, \citenamefont {Pagneux}, \citenamefont {Achilleos},\ and\ \citenamefont {Coutant}}]{PhysRevB.110.035106}%
  \BibitemOpen
  \bibfield  {author} {\bibinfo {author} {\bibfnamefont {M.}~\bibnamefont {R\"ontgen}}, \bibinfo {author} {\bibfnamefont {X.}~\bibnamefont {Chen}}, \bibinfo {author} {\bibfnamefont {W.}~\bibnamefont {Gao}}, \bibinfo {author} {\bibfnamefont {M.}~\bibnamefont {Pyzh}}, \bibinfo {author} {\bibfnamefont {P.}~\bibnamefont {Schmelcher}}, \bibinfo {author} {\bibfnamefont {V.}~\bibnamefont {Pagneux}}, \bibinfo {author} {\bibfnamefont {V.}~\bibnamefont {Achilleos}},\ and\ \bibinfo {author} {\bibfnamefont {A.}~\bibnamefont {Coutant}},\ }\bibfield  {title} {\bibinfo {title} {Topological states protected by hidden symmetry},\ }\href {https://doi.org/10.1103/PhysRevB.110.035106} {\bibfield  {journal} {\bibinfo  {journal} {Phys. Rev. B}\ }\textbf {\bibinfo {volume} {110}},\ \bibinfo {pages} {035106} (\bibinfo {year} {2024})}\BibitemShut {NoStop}%
\bibitem [{\citenamefont {Eek}\ \emph {et~al.}(2024)\citenamefont {Eek}, \citenamefont {Moustaj}, \citenamefont {Röntgen}, \citenamefont {Pagneux}, \citenamefont {Achilleos},\ and\ \citenamefont {Smith}}]{eek_emergent_2024}%
  \BibitemOpen
  \bibfield  {author} {\bibinfo {author} {\bibfnamefont {L.}~\bibnamefont {Eek}}, \bibinfo {author} {\bibfnamefont {A.}~\bibnamefont {Moustaj}}, \bibinfo {author} {\bibfnamefont {M.}~\bibnamefont {Röntgen}}, \bibinfo {author} {\bibfnamefont {V.}~\bibnamefont {Pagneux}}, \bibinfo {author} {\bibfnamefont {V.}~\bibnamefont {Achilleos}},\ and\ \bibinfo {author} {\bibfnamefont {C.~M.}\ \bibnamefont {Smith}},\ }\bibfield  {title} {\bibinfo {title} {Emergent non-{Hermitian} models},\ }\href {https://doi.org/10.1103/PhysRevB.109.045122} {\bibfield  {journal} {\bibinfo  {journal} {Phys. Rev. B}\ }\textbf {\bibinfo {volume} {109}},\ \bibinfo {pages} {045122} (\bibinfo {year} {2024})}\BibitemShut {NoStop}%
\bibitem [{\citenamefont {Röntgen}\ \emph {et~al.}(2021)\citenamefont {Röntgen}, \citenamefont {Pyzh}, \citenamefont {Morfonios},\ and\ \citenamefont {Schmelcher}}]{rontgen_symmetries_2021}%
  \BibitemOpen
  \bibfield  {author} {\bibinfo {author} {\bibfnamefont {M.}~\bibnamefont {Röntgen}}, \bibinfo {author} {\bibfnamefont {M.}~\bibnamefont {Pyzh}}, \bibinfo {author} {\bibfnamefont {C.~V.}\ \bibnamefont {Morfonios}},\ and\ \bibinfo {author} {\bibfnamefont {P.}~\bibnamefont {Schmelcher}},\ }\href {http://arxiv.org/abs/2105.12579} {\bibinfo {title} {On symmetries of a matrix and its isospectral reduction}} (\bibinfo {year} {2021}),\ \bibinfo {note} {arXiv:2105.12579 [quant-ph]}\BibitemShut {NoStop}%
\bibitem [{\citenamefont {Fulton}\ and\ \citenamefont {Harris}(2004)}]{Fulton2004129RepresentationTheory}%
  \BibitemOpen
  \bibfield  {author} {\bibinfo {author} {\bibfnamefont {W.}~\bibnamefont {Fulton}}\ and\ \bibinfo {author} {\bibfnamefont {J.}~\bibnamefont {Harris}},\ }\href {https://doi.org/10.1007/978-1-4612-0979-9} {\emph {\bibinfo {title} {Representation {{Theory}}}}},\ \bibinfo {series} {Graduate {{Texts}} in {{Mathematics}}}, Vol.\ \bibinfo {volume} {129}\ (\bibinfo  {publisher} {Springer},\ \bibinfo {address} {New York, NY},\ \bibinfo {year} {2004})\BibitemShut {NoStop}%
\bibitem [{\citenamefont {Hou}\ and\ \citenamefont {Chen}(2018)}]{Hou2018HiddenDegeneracy}%
  \BibitemOpen
  \bibfield  {author} {\bibinfo {author} {\bibfnamefont {J.-M.}\ \bibnamefont {Hou}}\ and\ \bibinfo {author} {\bibfnamefont {W.}~\bibnamefont {Chen}},\ }\bibfield  {title} {\bibinfo {title} {{Hidden antiunitary symmetry behind “accidental” degeneracy and its protection of degeneracy}},\ }\href {https://doi.org/10.1007/s11467-017-0712-8} {\bibfield  {journal} {\bibinfo  {journal} {Front. Phys.}\ }\textbf {\bibinfo {volume} {13}},\ \bibinfo {pages} {130301} (\bibinfo {year} {2018})}\BibitemShut {NoStop}%
\bibitem [{\citenamefont {Fock}(1935)}]{Fock1935ZP98145ZurTheorieWasserstoffatoms}%
  \BibitemOpen
  \bibfield  {author} {\bibinfo {author} {\bibfnamefont {V.}~\bibnamefont {Fock}},\ }\bibfield  {title} {\bibinfo {title} {{Zur Theorie des Wasserstoffatoms}},\ }\href {https://doi.org/10.1007/BF01336904} {\bibfield  {journal} {\bibinfo  {journal} {Z. Physik}\ }\textbf {\bibinfo {volume} {98}},\ \bibinfo {pages} {145} (\bibinfo {year} {1935})}\BibitemShut {NoStop}%
\bibitem [{\citenamefont {Fang}\ \emph {et~al.}(2012)\citenamefont {Fang}, \citenamefont {Gilbert},\ and\ \citenamefont {Bernevig}}]{fang_bulk_2012}%
  \BibitemOpen
  \bibfield  {author} {\bibinfo {author} {\bibfnamefont {C.}~\bibnamefont {Fang}}, \bibinfo {author} {\bibfnamefont {M.~J.}\ \bibnamefont {Gilbert}},\ and\ \bibinfo {author} {\bibfnamefont {B.~A.}\ \bibnamefont {Bernevig}},\ }\bibfield  {title} {{\selectlanguage {en}\bibinfo {title} {Bulk topological invariants in noninteracting point group symmetric insulators}},\ }\href {https://doi.org/10.1103/PhysRevB.86.115112} {\bibfield  {journal} {\bibinfo  {journal} {Phys. Rev. B}\ }\textbf {\bibinfo {volume} {86}},\ \bibinfo {pages} {115112} (\bibinfo {year} {2012})}\BibitemShut {NoStop}%
\bibitem [{\citenamefont {R{\"o}ntgen}\ \emph {et~al.}(2020)\citenamefont {R{\"o}ntgen}, \citenamefont {Palaiodimopoulos}, \citenamefont {Morfonios}, \citenamefont {Brouzos}, \citenamefont {Pyzh}, \citenamefont {Diakonos},\ and\ \citenamefont {Schmelcher}}]{Rontgen2020PRA101042304DesigningPrettyGoodState}%
  \BibitemOpen
  \bibfield  {author} {\bibinfo {author} {\bibfnamefont {M.}~\bibnamefont {R{\"o}ntgen}}, \bibinfo {author} {\bibfnamefont {N.~E.}\ \bibnamefont {Palaiodimopoulos}}, \bibinfo {author} {\bibfnamefont {C.~V.}\ \bibnamefont {Morfonios}}, \bibinfo {author} {\bibfnamefont {I.}~\bibnamefont {Brouzos}}, \bibinfo {author} {\bibfnamefont {M.}~\bibnamefont {Pyzh}}, \bibinfo {author} {\bibfnamefont {F.~K.}\ \bibnamefont {Diakonos}},\ and\ \bibinfo {author} {\bibfnamefont {P.}~\bibnamefont {Schmelcher}},\ }\bibfield  {title} {\bibinfo {title} {Designing pretty good state transfer via isospectral reductions},\ }\href {https://doi.org/10.1103/PhysRevA.101.042304} {\bibfield  {journal} {\bibinfo  {journal} {Phys. Rev. A}\ }\textbf {\bibinfo {volume} {101}},\ \bibinfo {pages} {042304} (\bibinfo {year} {2020})}\BibitemShut {NoStop}%
\bibitem [{\citenamefont {Godsil}\ and\ \citenamefont {McKay}(1982)}]{Godsil1982AM25257ConstructingCospectralGraphs}%
  \BibitemOpen
  \bibfield  {author} {\bibinfo {author} {\bibfnamefont {C.~D.}\ \bibnamefont {Godsil}}\ and\ \bibinfo {author} {\bibfnamefont {B.~D.}\ \bibnamefont {McKay}},\ }\bibfield  {title} {\bibinfo {title} {Constructing cospectral graphs},\ }\href {https://doi.org/10.1007/BF02189621} {\bibfield  {journal} {\bibinfo  {journal} {Aequ. Math.}\ }\textbf {\bibinfo {volume} {25}},\ \bibinfo {pages} {257} (\bibinfo {year} {1982})}\BibitemShut {NoStop}%
\bibitem [{\citenamefont {Schwenk}(1973)}]{Schwenk1973PTAAC257AlmostAllTreesAre}%
  \BibitemOpen
  \bibfield  {author} {\bibinfo {author} {\bibfnamefont {A.~J.}\ \bibnamefont {Schwenk}},\ }\bibfield  {title} {\bibinfo {title} {Almost all trees are cospectral},\ }in\ \href@noop {} {\emph {\bibinfo {booktitle} {Proceedings of the {{Third Annual Arbor Conference}}}}}\ (\bibinfo  {publisher} {Academic Press},\ \bibinfo {address} {New York},\ \bibinfo {year} {1973})\ pp.\ \bibinfo {pages} {257--307}\BibitemShut {NoStop}%
\bibitem [{\citenamefont {Liu}\ \emph {et~al.}(2019)\citenamefont {Liu}, \citenamefont {Gao}, \citenamefont {Zhang}, \citenamefont {Ma}, \citenamefont {Zhang}, \citenamefont {Liu}, \citenamefont {Xiang}, \citenamefont {Cui},\ and\ \citenamefont {Zhang}}]{Liu2019R20191TopologicallyProtectedEdgeState}%
  \BibitemOpen
  \bibfield  {author} {\bibinfo {author} {\bibfnamefont {S.}~\bibnamefont {Liu}}, \bibinfo {author} {\bibfnamefont {W.}~\bibnamefont {Gao}}, \bibinfo {author} {\bibfnamefont {Q.}~\bibnamefont {Zhang}}, \bibinfo {author} {\bibfnamefont {S.}~\bibnamefont {Ma}}, \bibinfo {author} {\bibfnamefont {L.}~\bibnamefont {Zhang}}, \bibinfo {author} {\bibfnamefont {C.}~\bibnamefont {Liu}}, \bibinfo {author} {\bibfnamefont {Y.~J.}\ \bibnamefont {Xiang}}, \bibinfo {author} {\bibfnamefont {T.~J.}\ \bibnamefont {Cui}},\ and\ \bibinfo {author} {\bibfnamefont {S.}~\bibnamefont {Zhang}},\ }\bibfield  {title} {\bibinfo {title} {Topologically {{Protected Edge State}} in {{Two-Dimensional Su}}--{{Schrieffer}}--{{Heeger Circuit}}},\ }\href {https://doi.org/10.34133/2019/8609875} {\bibfield  {journal} {\bibinfo  {journal} {Research}\ }\textbf {\bibinfo {volume} {2019}},\ \bibinfo {pages} {1} (\bibinfo {year} {2019})}\BibitemShut {NoStop}%
\bibitem [{\citenamefont {Coutant}\ \emph {et~al.}(2021)\citenamefont {Coutant}, \citenamefont {Achilleos}, \citenamefont {Richoux}, \citenamefont {Theocharis},\ and\ \citenamefont {Pagneux}}]{Coutant2021JoAP129125108TopologicalTwodimensionalSuSchriefferHeegerAnalog}%
  \BibitemOpen
  \bibfield  {author} {\bibinfo {author} {\bibfnamefont {A.}~\bibnamefont {Coutant}}, \bibinfo {author} {\bibfnamefont {V.}~\bibnamefont {Achilleos}}, \bibinfo {author} {\bibfnamefont {O.}~\bibnamefont {Richoux}}, \bibinfo {author} {\bibfnamefont {G.}~\bibnamefont {Theocharis}},\ and\ \bibinfo {author} {\bibfnamefont {V.}~\bibnamefont {Pagneux}},\ }\bibfield  {title} {\bibinfo {title} {Topological two-dimensional {{Su}}--{{Schrieffer}}--{{Heeger}} analog acoustic networks: Total reflection at corners and corner induced modes},\ }\href {https://doi.org/10.1063/5.0042406} {\bibfield  {journal} {\bibinfo  {journal} {J. Appl. Phys.}\ }\textbf {\bibinfo {volume} {129}},\ \bibinfo {pages} {125108} (\bibinfo {year} {2021})}\BibitemShut {NoStop}%
\bibitem [{\citenamefont {Liu}(2023)}]{PhysRevB.108.245140}%
  \BibitemOpen
  \bibfield  {author} {\bibinfo {author} {\bibfnamefont {F.}~\bibnamefont {Liu}},\ }\bibfield  {title} {\bibinfo {title} {Analytic solution of the $n$-dimensional su-schrieffer-heeger model},\ }\href {https://doi.org/10.1103/PhysRevB.108.245140} {\bibfield  {journal} {\bibinfo  {journal} {Phys. Rev. B}\ }\textbf {\bibinfo {volume} {108}},\ \bibinfo {pages} {245140} (\bibinfo {year} {2023})}\BibitemShut {NoStop}%
\bibitem [{\citenamefont {Dai}\ \emph {et~al.}(2014)\citenamefont {Dai}, \citenamefont {Yan}, \citenamefont {Xiao},\ and\ \citenamefont {Guo}}]{dai_electronic_2014}%
  \BibitemOpen
  \bibfield  {author} {\bibinfo {author} {\bibfnamefont {C.~J.}\ \bibnamefont {Dai}}, \bibinfo {author} {\bibfnamefont {X.~H.}\ \bibnamefont {Yan}}, \bibinfo {author} {\bibfnamefont {Y.}~\bibnamefont {Xiao}},\ and\ \bibinfo {author} {\bibfnamefont {Y.~D.}\ \bibnamefont {Guo}},\ }\bibfield  {title} {{\selectlanguage {en}\bibinfo {title} {Electronic and transport properties of {T}-graphene nanoribbon: {Symmetry}-dependent multiple {Dirac} points, negative differential resistance and linear current-bias characteristics}},\ }\href {https://doi.org/10.1209/0295-5075/107/37004} {\bibfield  {journal} {\bibinfo  {journal} {Europhys. Lett.}\ }\textbf {\bibinfo {volume} {107}},\ \bibinfo {pages} {37004} (\bibinfo {year} {2014})}\BibitemShut {NoStop}%
\bibitem [{\citenamefont {Nunes}\ and\ \citenamefont {Smith}(2020)}]{Nunes2020PRB101224514FlatbandSuperconductivityTightbindingElectrons}%
  \BibitemOpen
  \bibfield  {author} {\bibinfo {author} {\bibfnamefont {L.~H. C.~M.}\ \bibnamefont {Nunes}}\ and\ \bibinfo {author} {\bibfnamefont {C.~M.}\ \bibnamefont {Smith}},\ }\bibfield  {title} {\bibinfo {title} {Flat-band superconductivity for tight-binding electrons on a square-octagon lattice},\ }\href {https://doi.org/10.1103/PhysRevB.101.224514} {\bibfield  {journal} {\bibinfo  {journal} {Phys. Rev. B}\ }\textbf {\bibinfo {volume} {101}},\ \bibinfo {pages} {224514} (\bibinfo {year} {2020})}\BibitemShut {NoStop}%
\bibitem [{\citenamefont {Yan}\ \emph {et~al.}(2023)\citenamefont {Yan}, \citenamefont {Zhang}, \citenamefont {Wang},\ and\ \citenamefont {Yan}}]{Yan_2023}%
  \BibitemOpen
  \bibfield  {author} {\bibinfo {author} {\bibfnamefont {L.}~\bibnamefont {Yan}}, \bibinfo {author} {\bibfnamefont {D.}~\bibnamefont {Zhang}}, \bibinfo {author} {\bibfnamefont {X.-J.}\ \bibnamefont {Wang}},\ and\ \bibinfo {author} {\bibfnamefont {J.-Y.}\ \bibnamefont {Yan}},\ }\bibfield  {title} {\bibinfo {title} {Intrinsic topological metal state in {T}-graphene},\ }\href {https://doi.org/10.1088/1367-2630/acccd7} {\bibfield  {journal} {\bibinfo  {journal} {New J. Phys.}\ }\textbf {\bibinfo {volume} {25}},\ \bibinfo {pages} {043020} (\bibinfo {year} {2023})}\BibitemShut {NoStop}%
\bibitem [{\citenamefont {Kempkes}\ \emph {et~al.}(2019)\citenamefont {Kempkes}, \citenamefont {Slot}, \citenamefont {van~den Broeke}, \citenamefont {Capiod}, \citenamefont {Benalcazar}, \citenamefont {Vanmaekelbergh}, \citenamefont {Bercioux}, \citenamefont {Swart},\ and\ \citenamefont {Morais~Smith}}]{kempkes_robust_2019}%
  \BibitemOpen
  \bibfield  {author} {\bibinfo {author} {\bibfnamefont {S.~N.}\ \bibnamefont {Kempkes}}, \bibinfo {author} {\bibfnamefont {M.~R.}\ \bibnamefont {Slot}}, \bibinfo {author} {\bibfnamefont {J.~J.}\ \bibnamefont {van~den Broeke}}, \bibinfo {author} {\bibfnamefont {P.}~\bibnamefont {Capiod}}, \bibinfo {author} {\bibfnamefont {W.~A.}\ \bibnamefont {Benalcazar}}, \bibinfo {author} {\bibfnamefont {D.}~\bibnamefont {Vanmaekelbergh}}, \bibinfo {author} {\bibfnamefont {D.}~\bibnamefont {Bercioux}}, \bibinfo {author} {\bibfnamefont {I.}~\bibnamefont {Swart}},\ and\ \bibinfo {author} {\bibfnamefont {C.}~\bibnamefont {Morais~Smith}},\ }\bibfield  {title} {{\selectlanguage {en}\bibinfo {title} {Robust zero-energy modes in an electronic higher-order topological insulator}},\ }\href {https://doi.org/10.1038/s41563-019-0483-4} {\bibfield  {journal} {\bibinfo  {journal} {Nat. Mat.}\ }\textbf {\bibinfo {volume} {18}},\ \bibinfo {pages} {1292} (\bibinfo {year} {2019})}\BibitemShut {NoStop}%
\bibitem [{\citenamefont {Van~Miert}\ and\ \citenamefont {Ortix}(2020)}]{van_miert_topological_2020}%
  \BibitemOpen
  \bibfield  {author} {\bibinfo {author} {\bibfnamefont {G.}~\bibnamefont {Van~Miert}}\ and\ \bibinfo {author} {\bibfnamefont {C.}~\bibnamefont {Ortix}},\ }\bibfield  {title} {{\selectlanguage {en}\bibinfo {title} {On the topological immunity of corner states in two-dimensional crystalline insulators}},\ }\href {https://doi.org/10.1038/s41535-020-00265-7} {\bibfield  {journal} {\bibinfo  {journal} {npj Quant. Mat.}\ }\textbf {\bibinfo {volume} {5}},\ \bibinfo {pages} {63} (\bibinfo {year} {2020})}\BibitemShut {NoStop}%
\bibitem [{\citenamefont {Herrera}\ \emph {et~al.}(2022)\citenamefont {Herrera}, \citenamefont {Kempkes}, \citenamefont {de~Paz}, \citenamefont {Swart}, \citenamefont {Smith},\ and\ \citenamefont {Bercioux}}]{herrera_corner_2022}%
  \BibitemOpen
  \bibfield  {author} {\bibinfo {author} {\bibfnamefont {M.~A.~J.}\ \bibnamefont {Herrera}}, \bibinfo {author} {\bibfnamefont {S.~N.}\ \bibnamefont {Kempkes}}, \bibinfo {author} {\bibfnamefont {M.~B.}\ \bibnamefont {de~Paz}}, \bibinfo {author} {\bibfnamefont {A.~G.-E.~I.}\ \bibnamefont {Swart}}, \bibinfo {author} {\bibfnamefont {C.~M.}\ \bibnamefont {Smith}},\ and\ \bibinfo {author} {\bibfnamefont {D.}~\bibnamefont {Bercioux}},\ }\bibfield  {title} {{\selectlanguage {en}\bibinfo {title} {Corner modes of the breathing kagome lattice: origin and robustness}},\ }\href {https://doi.org/10.1103/PhysRevB.105.085411} {\bibfield  {journal} {\bibinfo  {journal} {Phys. Rev. B}\ }\textbf {\bibinfo {volume} {105}},\ \bibinfo {pages} {085411} (\bibinfo {year} {2022})}\BibitemShut {NoStop}%
\bibitem [{\citenamefont {Wu}\ and\ \citenamefont {Hu}(2015)}]{wu_topo_photonic_dielectric_2015}%
  \BibitemOpen
  \bibfield  {author} {\bibinfo {author} {\bibfnamefont {L.-H.}\ \bibnamefont {Wu}}\ and\ \bibinfo {author} {\bibfnamefont {X.}~\bibnamefont {Hu}},\ }\bibfield  {title} {\bibinfo {title} {Scheme for achieving a topological photonic crystal by using dielectric material},\ }\href {https://doi.org/10.1103/PhysRevLett.114.223901} {\bibfield  {journal} {\bibinfo  {journal} {Phys. Rev. Lett.}\ }\textbf {\bibinfo {volume} {114}},\ \bibinfo {pages} {223901} (\bibinfo {year} {2015})}\BibitemShut {NoStop}%
\bibitem [{\citenamefont {Freeney}\ \emph {et~al.}(2020)\citenamefont {Freeney}, \citenamefont {van~den Broeke}, \citenamefont {Harsveld van~der Veen}, \citenamefont {Swart},\ and\ \citenamefont {Morais~Smith}}]{freeney_edge-dependent_2020}%
  \BibitemOpen
  \bibfield  {author} {\bibinfo {author} {\bibfnamefont {S.}~\bibnamefont {Freeney}}, \bibinfo {author} {\bibfnamefont {J.}~\bibnamefont {van~den Broeke}}, \bibinfo {author} {\bibfnamefont {A.}~\bibnamefont {Harsveld van~der Veen}}, \bibinfo {author} {\bibfnamefont {I.}~\bibnamefont {Swart}},\ and\ \bibinfo {author} {\bibfnamefont {C.}~\bibnamefont {Morais~Smith}},\ }\bibfield  {title} {\bibinfo {title} {Edge-{Dependent} {Topology} in {Kekulé} {Lattices}},\ }\href {https://doi.org/10.1103/PhysRevLett.124.236404} {\bibfield  {journal} {\bibinfo  {journal} {Phys. Rev. Lett.}\ }\textbf {\bibinfo {volume} {124}},\ \bibinfo {pages} {236404} (\bibinfo {year} {2020})}\BibitemShut {NoStop}%
\bibitem [{\citenamefont {Moustaj}\ \emph {et~al.}(2024)\citenamefont {Moustaj}, \citenamefont {Krebbekx},\ and\ \citenamefont {Smith}}]{moustaj2024anomalous}%
  \BibitemOpen
  \bibfield  {author} {\bibinfo {author} {\bibfnamefont {A.}~\bibnamefont {Moustaj}}, \bibinfo {author} {\bibfnamefont {J.~P.~J.}\ \bibnamefont {Krebbekx}},\ and\ \bibinfo {author} {\bibfnamefont {C.~M.}\ \bibnamefont {Smith}},\ }\href@noop {} {\bibinfo {title} {Anomalous polarization in one-dimensional aperiodic insulators}} (\bibinfo {year} {2024}),\ \Eprint {https://arxiv.org/abs/2404.14916} {arXiv:2404.14916 [cond-mat.dis-nn]} \BibitemShut {NoStop}%
\bibitem [{\citenamefont {Canyellas}\ \emph {et~al.}(2024)\citenamefont {Canyellas}, \citenamefont {Liu}, \citenamefont {Arouca}, \citenamefont {Eek}, \citenamefont {Wang}, \citenamefont {Yin}, \citenamefont {Guan}, \citenamefont {Li}, \citenamefont {Wang}, \citenamefont {Zheng}, \citenamefont {Liu}, \citenamefont {Jia},\ and\ \citenamefont {Smith}}]{canyellas2023topological}%
  \BibitemOpen
  \bibfield  {author} {\bibinfo {author} {\bibfnamefont {R.}~\bibnamefont {Canyellas}}, \bibinfo {author} {\bibfnamefont {C.}~\bibnamefont {Liu}}, \bibinfo {author} {\bibfnamefont {R.}~\bibnamefont {Arouca}}, \bibinfo {author} {\bibfnamefont {L.}~\bibnamefont {Eek}}, \bibinfo {author} {\bibfnamefont {G.}~\bibnamefont {Wang}}, \bibinfo {author} {\bibfnamefont {Y.}~\bibnamefont {Yin}}, \bibinfo {author} {\bibfnamefont {D.}~\bibnamefont {Guan}}, \bibinfo {author} {\bibfnamefont {Y.}~\bibnamefont {Li}}, \bibinfo {author} {\bibfnamefont {S.}~\bibnamefont {Wang}}, \bibinfo {author} {\bibfnamefont {H.}~\bibnamefont {Zheng}}, \bibinfo {author} {\bibfnamefont {C.}~\bibnamefont {Liu}}, \bibinfo {author} {\bibfnamefont {J.}~\bibnamefont {Jia}},\ and\ \bibinfo {author} {\bibfnamefont {C.~M.}\ \bibnamefont {Smith}},\ }\bibfield  {title} {\bibinfo {title} {Topological edge and corner states in bismuth fractal nanostructures},\ }\href {https://doi.org/10.1038/s41567-024-02551-8} {\bibfield  {journal} {\bibinfo  {journal}
  {Nat. Phys.}\ }\textbf {\bibinfo {volume} {20}},\ \bibinfo {pages} {1421} (\bibinfo {year} {2024})}\BibitemShut {NoStop}%
\bibitem [{\citenamefont {Osseweijer}\ \emph {et~al.}(2024)\citenamefont {Osseweijer}, \citenamefont {Eek}, \citenamefont {Moustaj}, \citenamefont {Fremling},\ and\ \citenamefont {Morais~Smith}}]{frachald}%
  \BibitemOpen
  \bibfield  {author} {\bibinfo {author} {\bibfnamefont {Z.~F.}\ \bibnamefont {Osseweijer}}, \bibinfo {author} {\bibfnamefont {L.}~\bibnamefont {Eek}}, \bibinfo {author} {\bibfnamefont {A.}~\bibnamefont {Moustaj}}, \bibinfo {author} {\bibfnamefont {M.}~\bibnamefont {Fremling}},\ and\ \bibinfo {author} {\bibfnamefont {C.}~\bibnamefont {Morais~Smith}},\ }\bibfield  {title} {\bibinfo {title} {Haldane model on the sierpi\ifmmode \acute{n}\else \'{n}\fi{}ski gasket},\ }\href {https://doi.org/10.1103/PhysRevB.110.245405} {\bibfield  {journal} {\bibinfo  {journal} {Phys. Rev. B}\ }\textbf {\bibinfo {volume} {110}},\ \bibinfo {pages} {245405} (\bibinfo {year} {2024})}\BibitemShut {NoStop}%
\bibitem [{\citenamefont {Ezawa}(2018)}]{ezawa_higher-order_2018}%
  \BibitemOpen
  \bibfield  {author} {\bibinfo {author} {\bibfnamefont {M.}~\bibnamefont {Ezawa}},\ }\bibfield  {title} {{\selectlanguage {en}\bibinfo {title} {Higher-{Order} {Topological} {Insulators} and {Semimetals} on the {Breathing} {Kagome} and {Pyrochlore} {Lattices}}},\ }\href {https://doi.org/10.1103/PhysRevLett.120.026801} {\bibfield  {journal} {\bibinfo  {journal} {Phys. Rev. Lett.}\ }\textbf {\bibinfo {volume} {120}},\ \bibinfo {pages} {026801} (\bibinfo {year} {2018})}\BibitemShut {NoStop}%
\bibitem [{\citenamefont {Leykam}\ \emph {et~al.}(2018)\citenamefont {Leykam}, \citenamefont {Andreanov},\ and\ \citenamefont {Flach}}]{Leykam2018AP31473052ArtificialFlatBandSystems}%
  \BibitemOpen
  \bibfield  {author} {\bibinfo {author} {\bibfnamefont {D.}~\bibnamefont {Leykam}}, \bibinfo {author} {\bibfnamefont {A.}~\bibnamefont {Andreanov}},\ and\ \bibinfo {author} {\bibfnamefont {S.}~\bibnamefont {Flach}},\ }\bibfield  {title} {\bibinfo {title} {Artificial flat band systems: From lattice models to experiments},\ }\href {https://doi.org/10.1080/23746149.2018.1473052} {\bibfield  {journal} {\bibinfo  {journal} {Adv. Phys.}\ }\textbf {\bibinfo {volume} {3}},\ \bibinfo {pages} {1473052} (\bibinfo {year} {2018})}\BibitemShut {NoStop}%
\end{thebibliography}%

\appendix
\begin{widetext}
\section{Derivation of the $\overline{Q}$ matrix}\label{app:Qmat} 
Reference~\cite{Rontgen2021PRL126180601LatentSymmetryInducedDegeneracies} gives a derivation for the $Q$ (and therefore also $\overline{Q}$) matrix. Here, we will briefly outline how to construct $Q$ in the specific case of a latent $C_n$-symmetry. Let $H$ be a real symmetric matrix with a latent $C_n$-symmetry, i.e.
\begin{equation}
    \left[ \mathcal{R}_S(H,E), \hat{C_n} \right]_- = 0.
\end{equation}
Then the eigenstates $\{ \ket{\phi}\}$ of $H$ can be chosen to satisfy
\begin{equation}
    \hat{C}_n \ket{\phi}_S = e^{\frac{2 \pi i}{n}(p-1)}\ket{\phi}_S, \quad \text{with} \quad p \in \{1,2,\dots,n\} \qquad \text{OR} \qquad \ket{\phi}_S = \mathbf{0},
\end{equation}
where $\mathbf{0}$ is a zero vector in the $S$ subspace. We label these states by $\ket{\phi^{(p)}}$ and $\ket{\phi^{(0)}}$, respectively. One can then define the projectors
\begin{equation}
    P_p = \sum_i \ket{\phi_i^{(p)}}\bra{\phi_i^{(p)}}, \qquad P_0 = \sum_i \ket{\phi_i^{(0)}}\bra{\phi_i^{(0)}}.
\end{equation}
From these projectors, $Q$ is defined through
\begin{equation}
    Q = \hat{C}_n \oplus \overline{Q} = P_0 + \sum_{p=1}^n e^{\frac{2 \pi i}{n}(p-1)} P_p.
\end{equation}

\section{Hamiltonians and Symmetries}
\subsection{$C_4$}\label{app:expr4}
The Hamiltonian for the unit cell in \cref{fig:Latentcells}(c) of the main text is given by
\begin{equation}
    H_L^{(4)} = \left(
\begin{array}{ccccccccccccc}
 0 & 1 & 0 & 1 & 1 & 0 & 0 & 1 & 1 & 0 & 0 & 1 & 0 \\
 1 & 0 & 1 & 0 & 1 & 1 & 0 & 0 & 1 & 1 & 0 & 0 & 0 \\
 0 & 1 & 0 & 1 & 0 & 1 & 1 & 0 & 0 & 1 & 1 & 0 & 0 \\
 1 & 0 & 1 & 0 & 0 & 0 & 1 & 1 & 0 & 0 & 1 & 1 & 0 \\
 1 & 1 & 0 & 0 & 0 & 0 & 0 & 0 & 1 & 0 & 0 & 0 & t_1+t_3 \\
 0 & 1 & 1 & 0 & 0 & 0 & 0 & 0 & 0 & 1 & 0 & 0 & t_2 \\
 0 & 0 & 1 & 1 & 0 & 0 & 0 & 0 & 0 & 0 & 1 & 0 & t_1 \\
 1 & 0 & 0 & 1 & 0 & 0 & 0 & 0 & 0 & 0 & 0 & 1 & t_2 \\
 1 & 1 & 0 & 0 & 1 & 0 & 0 & 0 & 0 & 0 & 0 & 0 & 0 \\
 0 & 1 & 1 & 0 & 0 & 1 & 0 & 0 & 0 & 0 & 0 & 0 & 0 \\
 0 & 0 & 1 & 1 & 0 & 0 & 1 & 0 & 0 & 0 & 0 & 0 & t_3 \\
 1 & 0 & 0 & 1 & 0 & 0 & 0 & 1 & 0 & 0 & 0 & 0 & 0 \\
 0 & 0 & 0 & 0 & t_1+t_3 & t_2 & t_1 & t_2 & 0 & 0 & t_3 & 0 & 0 \\
\end{array}
\right). \label{eq:H4L}
\end{equation}
The symmetry on the level of the full Hamiltonian $H_L^{(4)}$ is given by
\begin{align}
Q^{(4)} = \hat{C}_4 \oplus \overline{Q}^{(4)} = \left(\begin{array}{cccc}
 0 & 0 & 0 & 1 \\
 1 & 0 & 0 & 0 \\
 0 & 1 & 0 & 0 \\
 0 & 0 & 1 & 0 \\
\end{array}
\right) \oplus
    \left(
\begin{array}{ccccccccc}
 \frac{3}{4} & -\frac{1}{4} & \frac{1}{4} & \frac{1}{4} & -\frac{1}{4} & -\frac{1}{4} & \frac{1}{4} & \frac{1}{4} & 0 \\
 \frac{1}{4} & \frac{3}{4} & -\frac{1}{4} & \frac{1}{4} & \frac{1}{4} & -\frac{1}{4} & -\frac{1}{4} & \frac{1}{4} & 0 \\
 \frac{1}{4} & \frac{1}{4} & \frac{3}{4} & -\frac{1}{4} & \frac{1}{4} & \frac{1}{4} & -\frac{1}{4} & -\frac{1}{4} & 0 \\
 -\frac{1}{4} & \frac{1}{4} & \frac{1}{4} & \frac{3}{4} & -\frac{1}{4} & \frac{1}{4} & \frac{1}{4} & -\frac{1}{4} & 0 \\
 -\frac{1}{4} & -\frac{1}{4} & \frac{1}{4} & \frac{1}{4} & \frac{3}{4} & -\frac{1}{4} & \frac{1}{4} & \frac{1}{4} & 0 \\
 \frac{1}{4} & -\frac{1}{4} & -\frac{1}{4} & \frac{1}{4} & \frac{1}{4} & \frac{3}{4} & -\frac{1}{4} & \frac{1}{4} & 0 \\
 \frac{1}{4} & \frac{1}{4} & -\frac{1}{4} & -\frac{1}{4} & \frac{1}{4} & \frac{1}{4} & \frac{3}{4} & -\frac{1}{4} & 0 \\
 -\frac{1}{4} & \frac{1}{4} & \frac{1}{4} & -\frac{1}{4} & -\frac{1}{4} & \frac{1}{4} & \frac{1}{4} & \frac{3}{4} & 0 \\
 0 & 0 & 0 & 0 & 0 & 0 & 0 & 0 & 1 \\
\end{array}
\right)
\end{align}
The reduced parameters that enter \cref{fig:Latentcells}(g) and \cref{eq:c4reduced} are given by

\begin{equation}
    a^{(4)}=\frac{t_0^2 \left(4 E  \left(E ^2-t_0^2\right)+t_1^2 (t_0-7 E )+2 \left(t_2-3 t_3\right) t_1 (E +t_0)+t_2^2 (t_0-7 E )+t_3^2 (t_0-7 E )+2 t_2 t_3 (E +t_0)\right)}{(t_0-E ) \left(E  (t_0-E ) (E +t_0)+2 \left(E  t_1^2+t_3 t_1 (E +t_0)+E  \left(t_2^2+t_3^2\right)\right)\right)}
\end{equation}
\begin{equation}
    v^{(4)}_0 = -\frac{t_0 (E +t_0) \left(t_1^2 (-(t_0-2 E ))+t_1 \left(2 E  t_3-2 t_0 t_2\right)-t_2^2 (t_0-2 E )-t_3^2 (t_0-2 E )+E  (t_0-E ) (E +t_0)-2 t_0 t_2 t_3\right)}{(t_0-E ) \left(E  (t_0-E ) (E +t_0)+2 \left(E  t_1^2+t_3 t_1 (E +t_0)+E  \left(t_2^2+t_3^2\right)\right)\right)}
\end{equation}
\begin{equation}
    v^{(4)}_1 = \frac{t_0^2 \left(t_1+t_2+t_3\right){}^2 (E +t_0)}{(t_0-E ) \left(E  (t_0-E ) (E +t_0)+2 \left(E  t_1^2+t_3 t_1 (E +t_0)+E  \left(t_2^2+t_3^2\right)\right)\right)}
\end{equation}

\subsection{$C_6$}\label{app:expr6}
The Hamiltonian for the unit cell in \cref{fig:Latentcells}(d) of the main text is a simplified version of the one denoted in \cref{fig:c6app}. 
\begin{figure}[h]
    \centering
    \includegraphics[]{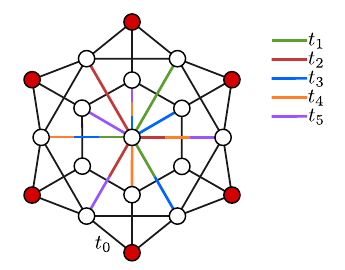}
    \caption{A more complex latent $C_6-$symmetric unit cell. The hopping parameters are indicated by different colors.}
    \label{fig:c6app}
\end{figure}
To obtain the former, we set $t_2 = t_3 = t_4 = 0$ and let $t_5 \to t_2$. The full Hamiltonian is given by
\begin{equation}
    H_L^{(6)}=\left(
\begin{array}{ccccccccccccccccccc}
 0 & 0 & 0 & 0 & 0 & 0 & 0 & 0 & 0 & 0 & t_0 & 0 & 0 & 0 & 0 & t_0 & t_0 & 0 & 0 \\
 0 & 0 & 0 & 0 & 0 & 0 & 0 & 0 & 0 & 0 & 0 & t_0 & 0 & 0 & 0 & 0 & t_0 & t_0 & 0 \\
 0 & 0 & 0 & 0 & 0 & 0 & t_0 & 0 & 0 & 0 & 0 & 0 & t_0 & 0 & 0 & 0 & 0 & t_0 & 0 \\
 0 & 0 & 0 & 0 & 0 & 0 & 0 & t_0 & 0 & 0 & 0 & 0 & t_0 & t_0 & 0 & 0 & 0 & 0 & 0 \\
 0 & 0 & 0 & 0 & 0 & 0 & 0 & 0 & t_0 & 0 & 0 & 0 & 0 & t_0 & t_0 & 0 & 0 & 0 & 0 \\
 0 & 0 & 0 & 0 & 0 & 0 & 0 & 0 & 0 & t_0 & 0 & 0 & 0 & 0 & t_0 & t_0 & 0 & 0 & 0 \\
 0 & 0 & t_0 & 0 & 0 & 0 & 0 & t_0 & 0 & 0 & 0 & t_0 & 0 & 0 & 0 & 0 & 0 & 0 & t_3+t_4+t_5 \\
 0 & 0 & 0 & t_0 & 0 & 0 & t_0 & 0 & t_0 & 0 & 0 & 0 & 0 & 0 & 0 & 0 & 0 & 0 & t_3 \\
 0 & 0 & 0 & 0 & t_0 & 0 & 0 & t_0 & 0 & t_0 & 0 & 0 & 0 & 0 & 0 & 0 & 0 & 0 & 0 \\
 0 & 0 & 0 & 0 & 0 & t_0 & 0 & 0 & t_0 & 0 & t_0 & 0 & 0 & 0 & 0 & 0 & 0 & 0 & t_4 \\
 t_0 & 0 & 0 & 0 & 0 & 0 & 0 & 0 & 0 & t_0 & 0 & t_0 & 0 & 0 & 0 & 0 & 0 & 0 & 0 \\
 0 & t_0 & 0 & 0 & 0 & 0 & t_0 & 0 & 0 & 0 & t_0 & 0 & 0 & 0 & 0 & 0 & 0 & 0 & t_5 \\
 0 & 0 & t_0 & t_0 & 0 & 0 & 0 & 0 & 0 & 0 & 0 & 0 & 0 & t_0 & 0 & 0 & 0 & t_0 & t_1 \\
 0 & 0 & 0 & t_0 & t_0 & 0 & 0 & 0 & 0 & 0 & 0 & 0 & t_0 & 0 & t_0 & 0 & 0 & 0 & t_2+t_4+t_5 \\
 0 & 0 & 0 & 0 & t_0 & t_0 & 0 & 0 & 0 & 0 & 0 & 0 & 0 & t_0 & 0 & t_0 & 0 & 0 & t_1+t_3 \\
 t_0 & 0 & 0 & 0 & 0 & t_0 & 0 & 0 & 0 & 0 & 0 & 0 & 0 & 0 & t_0 & 0 & t_0 & 0 & t_2+t_5 \\
 t_0 & t_0 & 0 & 0 & 0 & 0 & 0 & 0 & 0 & 0 & 0 & 0 & 0 & 0 & 0 & t_0 & 0 & t_0 & t_1+t_3+t_4 \\
 0 & t_0 & t_0 & 0 & 0 & 0 & 0 & 0 & 0 & 0 & 0 & 0 & t_0 & 0 & 0 & 0 & t_0 & 0 & t_2 \\
 0 & 0 & 0 & 0 & 0 & 0 & t_3+t_4+t_5 & t_3 & 0 & t_4 & 0 & t_5 & t_1 & t_2+t_4+t_5 & t_1+t_3 & t_2+t_5 & t_1+t_3+t_4 & t_2 & 0 \\
\end{array}
\right).
\end{equation}
The symmetry on the level of the full Hamiltonian $H_L^{(6)}$ is given by
\begin{equation}
Q^{(6)} = \hat{C}_6 \oplus \overline{Q}^{(6)} =
\left(
\begin{array}{cccccc}
 0 & 0 & 0 & 0 & 0 & 1 \\
 1 & 0 & 0 & 0 & 0 & 0 \\
 0 & 1 & 0 & 0 & 0 & 0 \\
 0 & 0 & 1 & 0 & 0 & 0 \\
 0 & 0 & 0 & 1 & 0 & 0 \\
 0 & 0 & 0 & 0 & 1 & 0 \\
\end{array}
\right)
\oplus
\left(
\begin{array}{ccccccccccccc}
 \frac{3}{8} & \frac{1}{4} & -\frac{1}{8} & \frac{1}{8} & -\frac{1}{4} & \frac{5}{8} & -\frac{3}{8} & \frac{1}{8} & 0 & -\frac{1}{8} & \frac{3}{8} & 0 & 0 \\
 \frac{5}{8} & \frac{3}{8} & \frac{1}{4} & -\frac{1}{8} & \frac{1}{8} & -\frac{1}{4} & 0 & -\frac{3}{8} & \frac{1}{8} & 0 & -\frac{1}{8} & \frac{3}{8} & 0 \\
 -\frac{1}{4} & \frac{5}{8} & \frac{3}{8} & \frac{1}{4} & -\frac{1}{8} & \frac{1}{8} & \frac{3}{8} & 0 & -\frac{3}{8} & \frac{1}{8} & 0 & -\frac{1}{8} & 0 \\
 \frac{1}{8} & -\frac{1}{4} & \frac{5}{8} & \frac{3}{8} & \frac{1}{4} & -\frac{1}{8} & -\frac{1}{8} & \frac{3}{8} & 0 & -\frac{3}{8} & \frac{1}{8} & 0 & 0 \\
 -\frac{1}{8} & \frac{1}{8} & -\frac{1}{4} & \frac{5}{8} & \frac{3}{8} & \frac{1}{4} & 0 & -\frac{1}{8} & \frac{3}{8} & 0 & -\frac{3}{8} & \frac{1}{8} & 0 \\
 \frac{1}{4} & -\frac{1}{8} & \frac{1}{8} & -\frac{1}{4} & \frac{5}{8} & \frac{3}{8} & \frac{1}{8} & 0 & -\frac{1}{8} & \frac{3}{8} & 0 & -\frac{3}{8} & 0 \\
 0 & -\frac{3}{8} & \frac{1}{8} & 0 & -\frac{1}{8} & \frac{3}{8} & \frac{5}{8} & -\frac{1}{4} & \frac{1}{8} & -\frac{1}{8} & \frac{1}{4} & \frac{3}{8} & 0 \\
 \frac{3}{8} & 0 & -\frac{3}{8} & \frac{1}{8} & 0 & -\frac{1}{8} & \frac{3}{8} & \frac{5}{8} & -\frac{1}{4} & \frac{1}{8} & -\frac{1}{8} & \frac{1}{4} & 0 \\
 -\frac{1}{8} & \frac{3}{8} & 0 & -\frac{3}{8} & \frac{1}{8} & 0 & \frac{1}{4} & \frac{3}{8} & \frac{5}{8} & -\frac{1}{4} & \frac{1}{8} & -\frac{1}{8} & 0 \\
 0 & -\frac{1}{8} & \frac{3}{8} & 0 & -\frac{3}{8} & \frac{1}{8} & -\frac{1}{8} & \frac{1}{4} & \frac{3}{8} & \frac{5}{8} & -\frac{1}{4} & \frac{1}{8} & 0 \\
 \frac{1}{8} & 0 & -\frac{1}{8} & \frac{3}{8} & 0 & -\frac{3}{8} & \frac{1}{8} & -\frac{1}{8} & \frac{1}{4} & \frac{3}{8} & \frac{5}{8} & -\frac{1}{4} & 0 \\
 -\frac{3}{8} & \frac{1}{8} & 0 & -\frac{1}{8} & \frac{3}{8} & 0 & -\frac{1}{4} & \frac{1}{8} & -\frac{1}{8} & \frac{1}{4} & \frac{3}{8} & \frac{5}{8} & 0 \\
 0 & 0 & 0 & 0 & 0 & 0 & 0 & 0 & 0 & 0 & 0 & 0 & 1 \\
\end{array}
\right).
\end{equation}
The reduced parameters that enter \cref{fig:Latentcells}(h) and \cref{eq:c6reduced} are of the form
\begin{equation}
    a^{(4)} = \frac{1}{{\Delta}}\sum_{j=0}^9 \alpha^{(j)} E^j , \qquad v_i^{(4)} = \frac{1}{\Delta}\sum_{j=0}^9 \nu_i^{(j)} E^j, \qquad \Delta = \sum_{j=0}^9 \delta^{(j)} E^j.
\end{equation}
The expressions for $\alpha^{(j)}$, $\nu_i^{(j)}$, and $\delta^{(j)}$ are lengthy, but relatively straightforward to calculate; they are not given here. 

\section{Explanation of the flat band of the latent Kagome lattice} \label{App:flatBand}

\begin{figure}[h]
    \centering
    \includegraphics[width=0.5\linewidth]{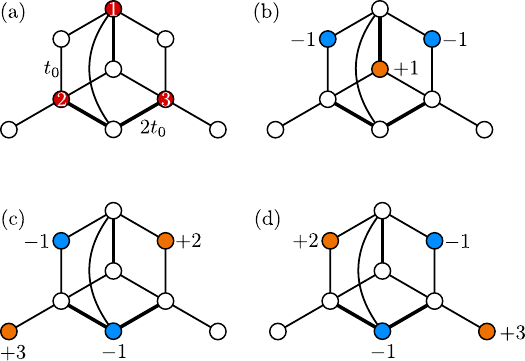}
    \caption{(a) Unit cell of the latent Kagome lattice, corresponding to the band structure depicted in \cref{fig:OBCKagome}(b). The numbers $\{1,2,3\}$ in the red circles mark the sites that are connected to other unit cells when embedding this unit cell in a lattice. (b) to (d): Eigenstates at zero energy. White sites have zero amplitude, blue sites negative, and orange sites positive amplitude. For the sites with non-vanishing amplitude, the amplitude is printed close to the corresponding site.}
    \label{fig:CLS}
\end{figure}

The flat band at zero energy can be explained purely from an analysis of the unit cell, shown in \cref{fig:CLS}(a). For any value of $t_0$, the cell supports three states that have zero energy and vanish on the red sites $\{1,2,3\}$; these states are depicted graphically in Figs.~\ref{fig:CLS}(b) to (d).
States that exactly vanish at some sites are known in the literature as \textit{compact localized states}.
An overview of compact localized states is given in Ref.~\cite{Leykam2018AP31473052ArtificialFlatBandSystems}.
Since these states all vanish on the sites $\{1,2,3\}$ that connect the given unit cell to other unit cells in the lattice, it can be promptly shown that the full lattice also features a zero-energy eigenstate that is at least $3N$-fold degenerate, with $N$ being the number of unit cells.
These $3N$ states can then be superposed to form Bloch states, resulting in three flat bands at zero energy.

\end{widetext}

\end{document}